\documentclass[aps,prd,amsfonts,amssymb,amsmath,mathrsfs,nofootinbib,reprint,showpacs,fixltx2e,subfig,usegraphicx,twocolumn]{revtex4}
\usepackage{graphicx}
\usepackage[font=small,labelfont=bf,textfont=footnotesize,justification=raggedright,singlelinecheck=true]{caption}
\usepackage{subfig}
\usepackage{aas_macros}
\usepackage{mathrsfs}
\usepackage{comment}
\usepackage{appendix}
\usepackage{multirow}
\usepackage{url}
\newcommand{\incgraph}[3]{\includegraphics[angle=#1, width=#2\textwidth]{#3}}

\begin{document}

\title{Cosmology with the lights off: Standard sirens in the Einstein Telescope era}

\author{Stephen R. Taylor}
\email[email: ]{staylor@ast.cam.ac.uk}
\author{Jonathan R. Gair}
\email[email: ]{jgair@ast.cam.ac.uk}
%\thanks{}
\affiliation{Institute of Astronomy, Madingley Road, Cambridge, CB3 0HA, UK }

%\author{Jonathan R. Gair}
%\email[email: ]{jgair@ast.cam.ac.uk}
%\thanks{}
%\affiliation{Institute of Astronomy, Madingley Road, Cambridge, CB3 0HA, UK}


\date{\today}

\begin{abstract}
We explore the prospects for constraining cosmology using gravitational-wave (GW) observations of neutron star binaries by the proposed Einstein Telescope (ET), exploiting the narrowness of the neutron star mass function. This builds on our previous work in the context of advanced-era GW detectors. Double neutron-star (DNS) binaries are expected to be one of the first sources detected after ``first-light'' of Advanced LIGO. DNS systems are expected to be detected at a rate of a few tens per year in the advanced era but the proposed Einstein Telescope (ET) could catalog tens, if not hundreds, of thousands per year. Combining the measured source redshift distributions with GW-network distance determinations will permit not only the precision measurement of background cosmological parameters, but will provide an insight into the astrophysical properties of these DNS systems. Of particular interest will be to probe the distribution of delay times between DNS-binary creation and subsequent merger, as well as the evolution of the star-formation rate density within ET's detection horizon. Keeping $H_0$, $\Omega_{m,0}$ and $\Omega_{\Lambda,0}$ fixed and investigating the precision with which the dark-energy equation-of-state parameters could be recovered, we found that with $10^5$ detected DNS binaries we could constrain these parameters to an accuracy similar to forecasted constraints from future CMB$+$BAO$+$SNIa measurements. Furthermore, modeling the merger delay-time distribution as a power-law ($\propto t^{\alpha}$) and the star-formation rate (SFR) density as a parametrized version of the Porciani and Madau SF2 model, we find that the associated astrophysical parameters are constrained to within $\sim 10\%$. All parameter precisions scaled as $1/\sqrt{N}$, where $N$ is the number of cataloged detections. We also investigated how parameter precisions varied with the intrinsic underlying properties of the Universe and with the distance reach of the network (which is affected, for instance, by the low-frequency cutoff of the detector). We also consider various sources of distance measurement errors in the third-generation era, and how these can be folded into the analysis.
\end{abstract}

%\keywords{}

\maketitle

\section{Introduction}
The era of advanced gravitational-wave (GW) detectors is approaching quickly. The previous decade has seen significant improvements in the sensitivity of GW interferometers, leading to the construction and operation of two Laser Interferometer Gravitational-wave Observatory (LIGO) \citep{ligo2009} detectors in the USA, GEO-$600$ in Germany \citep{geo2008}, Virgo in Italy \citep{virgo2006} and TAMA-$300$ in Japan \citep{tama2004}. The latter detector was designed as a testbed to develop new technologies for the proposed underground, cryogenically cooled KAGRA (formerly LCGT \citep{lcgt2010}) detector \citep{kagra2012}. The LIGO, Virgo and GEO-$600$ detectors have conducted joint searches since $2007$.

The most promising source for the first detection of gravitational waves is the inspiral and merger of a compact-object binary consisting of neutron stars (NSs) and/or black holes \citep{mandel-oshaughnessy2010}. The first joint search for compact binary coalescence signals during the LIGO S5 science run and the Virgo VSR1 data did not result in direct detections \citep{ligo-s5-virgo-vsr1}, nor did the ``enhanced'' detector search during the LIGO S6 science run and the Virgo VSR2+3 data \citep{ligo-s6}. Furthermore, the upper limits placed on compact-binary coalescence rates from the latter search remain two to $3$ orders of magnitude above existing astrophysically predicted rates. However, the LIGO detectors are currently being upgraded to their ``advanced'' configuration \citep{AdvLIGO}, due for completion in $\sim 2015$, for which the horizon distance for NS-NS inspiral detection will be boosted to $\sim 450$ Mpc, giving an almost thousandfold gain in volume sensitivity of the detectors. The advanced detectors are expected to detect double NS inspirals at a rate of $\sim 40$ yr$^{-1}$, although this may vary by approximately $2$ orders of magnitude in either direction \citep{abadie-rate2010}.

Complementing AdLIGO will be a global network of advanced detectors, including AdVirgo \citep{AdvVirgo}, KAGRA \citep{kagra2012} and possibly a third LIGO detector in India, LIGO-India \citep{indigo}. There are currently no prospects for a Southern Hemisphere GW interferometer operating in the advanced era. A global network comprising these detectors will help turn the field from the search for the first detection, into a precise astronomical tool.

The GWs emitted by a compact binary system directly encode the \textit{redshifted} masses and luminosity distance of the system. Double NS (DNS) binary systems are commonly referred to as \textit{self-calibrating standard sirens} because their distance is directly encoded in the waveform, and a means of determining their redshift would mean we could probe the cosmic distance ladder and extract cosmological parameters \citep{Schutz86,hughes-holz-2003,et-cosmography,et-dark-energy}. While the phase evolution of the strain signal in a single interferometer can give precise constraints on the redshifted mass of the system, we require a global network of detectors to constrain the sky location, orbital inclination and GW polarization so that we can break the degeneracies of these angular factors with the luminosity distance.

Unfortunately the redshift and intrinsic mass of the systems enter the waveform only in a combination as the redshifted mass parameter; hence previous techniques for performing GW cosmology using these standard sirens has relied on the association of the GW source with an electromagnetic counterpart, which can provide an independent redshift measurement \citep{HolzHughes:2005,nissanke2010,macleod-hogan,delpozzo2011}. In our previous paper, \citet{hwth2011}, we studied a technique for probing the Hubble constant and NS mass-distribution parameters using only the GWs detected by an advanced-era network. This work relied on the assumption, supported by observations, that the NS mass-distribution is sufficiently narrow, which means that we already have a good idea of the intrinsic masses in the system and the measured redshifted mass parameter then provides a narrow distribution of possible redshifts for each observed source. Combining these redshift distributions with the network-measured luminosity distance for a catalog of observed DNS systems provides constraints on the underlying cosmological parameters, as well as the astrophysical distribution of these systems. This technique (first considered by Markovi\'{c} in Ref.\ \citep{markovic1993} and extended in Refs.\ \citep{chernoff-finn-1993,Finn96}) relies on measurements of the \textit{redshifted chirp mass} (expected to be the best-determined parameter) and luminosity distance for a catalog of detected systems.

In our previous analysis, the cosmological parameters that we could constrain were restricted by the sub-Gpc reach of an advanced-era network. We now extend this technique to a third-generation network, which could have a reach out to tens of Gpcs. Proposed third-generation detectors aim for a broadband factor of $10$ sensitivity improvement with respect to advanced detectors, and to increase sensitivity in the range $\sim 1-10$ Hz, compared to the $\sim 10-20$ Hz lower frequency cutoff of advanced detectors. As a prototypical third-generation detector we use the Einstein Telescope, consisting of three overlapping $10$ km arm-length interferometers in a triangular configuration \citep{triple-michelson2009,sathy-et-potential2011,3rd-gen-science-reach2010}. Each interferometer may actually be two detectors: a cryogenically cooled, underground detector with good low-frequency sensitivity, and a high laser power detector with good high-frequency sensitivity \citep{sensitivity-studies2011}. Keeping $H_0$, $\Omega_{m,0}$ and $\Omega_{\Lambda,0}$ fixed, we find that the sensitivity provided by such a network will be large enough to constrain the dark-energy equation-of-state parameters and NS mass-distribution parameters, as well as the astrophysical distribution of the systems. The latter will inform us about the average time delay between the formation of these compact-binary systems and their merger, as well as the shape of the underlying star-formation-rate density.

Third-generation detectors are unlikely to be online before the mid-2020s, but, if realized, the ambitious and novel design for the Einstein Telescope will have far-reaching scientific advantages. Such a detector could detect as many as hundreds of thousands of DNS inspirals per year, which, along with the distance reach of the detectors, will permit precision GW astronomy. In this paper, we will not consider other methods that have been proposed for using GW observations as cosmological probes. In particular, we do not consider association of GW detections with an electromagnetic (EM) counterpart, which has been studied in Refs.\ \citep{et-cosmography,et-dark-energy}, nor do we consider tidal-coupling corrections to the phase evolution of the strain signal \citep{messenger-read-2011}. The latter method is also a GW-only technique with significant potential, in that these phase-evolution corrections break the mass-redshift degeneracy at $5$PN order and, assuming the NS equation-of-state is well known, will permit the distance-redshift relation to be probed. It may also be possible to apply the method used by Ref.\ \citep{nishizawa-phase-shift-2012}, which was investigated in the context of future space-based detectors, to third-generation ground-based detectors. Their method relies on the measurement of cosmologically-induced shifts in the GW-phase at $4$PN order.

This paper is laid out as follows. Section \ref{sec:detect-characteristics-networks} describes the characteristics of the Einstein Telescope, as well as possible third-generation networks and detection criteria. In Sec.\ \ref{ref:source-systems} we discuss aspects of DNS systems, including the mass distribution of the constituent NSs, and the redshift distribution of DNS mergers. Section \ref{sec:cosmo-models} describes the effect of the dark-energy equation-of-state parameter on cataloged luminosity distances, while Sec.\ \ref{sec:make-analyse-catalogs} describes how we construct and analyze catalogs of detected DNS-system inspirals. Results are shown in Sec.\ \ref{sec:results}, followed by our conclusions in Section \ref{sec:conclusions}.

\section{Detector characteristics and networks} \label{sec:detect-characteristics-networks}
\subsection{The Einstein Telescope} \label{sec:ET-discussion}
\begin{figure}
     \incgraph{270}{0.5}{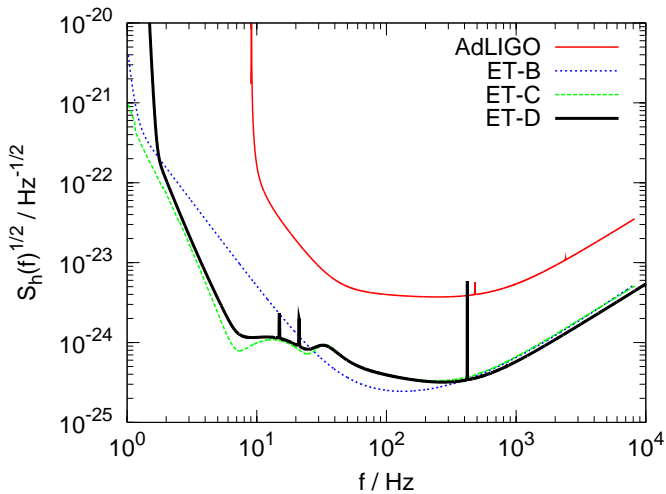}
     \caption{\label{fig:noise_curves}Comparison of the different noise curves for AdLIGO (high-power-zero-detuning) \cite{adligo-noise}, the initial Einstein Telescope noise curve based on conventional techniques, ET-B \citep{conventional-ET-2008}, the initial ``xylophone'' noise curve, ET-C \citep{ET-xylophone-2010}, and the improved, more realistic xylophone noise curve, ET-D \citep{sensitivity-studies2011}. These noise curves are for one 10km right-angled interferometer.}
 \end{figure}
The Einstein Telescope (ET) is a proposed third-generation ground-based interferometer. A recent design study has been carried out within the European Commission's FP7 framework \citep{ET-site} to evaluate the science case for such a detector, and to consider the technological advances required for the science goals to be achieved. Through this three-year design study, some favored designs and configurations have emerged.

The aim for third-generation ground-based detector designs is to achieve a broadband factor of $10$ sensitivity improvement with respect to advanced detectors, and to push the sensitivity down into the $\sim 1-10$ Hz range. Early designs examined the prospects for pushing conventional techniques used in advanced detectors  to their limits to construct a third-generation interferometer \citep{conventional-ET-2008}. This gave the ET-B noise curve in Fig.\ \ref{fig:noise_curves}. Beyond the extension of the arm-length to $10$ km, several techniques were proposed to suppress high- and low-frequency noise, including siting the detector underground.

Crucially, the techniques used to suppress high-frequency noise are not necessarily compatible with the suppression of low-frequency noise. Increasing the laser power will reduce the photon shot noise which dominates the high-frequency range, but this worsens the thermal noise which dominates at low frequencies. The ``xylophone'' design was proposed to address this issue. Instead of having a single broadband instrument, the xylophone design comprises a high-power, high-frequency interferometer (ET-HF) and a cryogenic low-power, low-frequency interferometer (ET-LF) \citep{ET-xylophone-2010}. ET-LF would be an underground instrument, and limited at low frequencies by gravity-gradient noise, while ET-HF would be colocated and co-oriented with ET-LF but surface-sited. ET-HF would employ high-power lasers to suppress high-frequency photon shot noise.

The initial xylophone design gave the ET-C curve \citep{ET-xylophone-2010} in Fig.\ \ref{fig:noise_curves}, which was refined to give the ET-D xylophone design \citep{sensitivity-studies2011}. We will use the ET-D noise curve in the ensuing analysis.

Current and advanced era ground-based interferometers are right-angled interferometers, since an arm opening-angle of $90^{\circ}$ maximizes their sensitivity. However, if both GW polarization states are to be measured at a single site, then two or more colocated nonaligned interferometers are required. Furthermore, at least three colocated interferometers are required to construct a null-stream, i.e.,\ a sum of individual interferometer responses that is insensitive to GWs and can be used to identify noise transients in the data stream. 

Taking these goals into account, the design requiring the shortest total length of tunnels is a triangular configuration with three identical interferometers positioned at each vertex of the triangle, an arm-opening angle of $60^{\circ}$ and rotated relative to each other by $120^{\circ}$ \citep{triple-michelson2009,sathy-et-potential2011,3rd-gen-science-reach2010}. A triangular configuration also provides redundancy, since polarization constraints are still possible with only two vertices operational.

In the following we consider three ET-D interferometers in the triangular configuration, which we denote as a ``single ET''. More than one ET would be very optimistic, so we complement our single ET with a network of individual third-generation right-angled interferometers (also with ET-D sensitivity) to permit source distance determination. While different locations are being mooted, we choose the Virgo location as the reference ET site \citep{eliu-ET}.

\subsection{Signal-to-noise ratio}
The optimal matched filtering signal-to-noise ratio of a GW detection is given by
\begin{equation}
\rho_{\rm opt} = 2\left[\int_0^{\infty}df\frac{|\tilde{h}(f)|^2}{S_h(f)}\right]^{1/2},
\end{equation}
where $S_{h}(f)$ denotes the detector's noise power spectral density. In the quadrupolar approximation, the Fourier transform of the signal amplitude of GWs from an inspiraling binary system takes the following form \citep{Abadie2010,oshaughnessy2010-binary-compact}:
\begin{equation} \label{eq:fourier-waveform}
|\tilde{h}(f)| = \frac{2c}{D_L}\left(\frac{5G\mu}{96c^3}\right)^{1/2}\left(\frac{GM}{\pi^2c^3}\right)^{1/3}\left(\frac{\Theta}{4}\right)f^{-7/6}.
\end{equation}
The function $\Theta$ is defined by
\begin{equation} \label{eq:angle}
{\Theta}\equiv2[F_{+}^{2}(1+\cos^{2}\iota)^{2} + 4F_{\times}^{2}\cos^{2}\iota]^{1/2},
\end{equation}
where $0<\Theta<4$, and
\begin{align}
F_{+}&\equiv{\frac{1}{2}}(1+\cos^{2}\theta)\cos2\phi\cos2\psi-\cos\theta\sin2\phi\sin2\psi, \nonumber\\
F_{\times}&\equiv{\frac{1}{2}}(1+\cos^{2}\theta)\cos2\phi\sin2\psi+\cos\theta\sin2\phi\cos2\psi
\end{align}
are the interferometer's strain responses to the different GW polarizations.

Following Ref.\ \citep{Finn96} we can write the matched filtering signal-to-noise ratio in a single detector as
\begin{equation} \label{eq:snr}
{\rho}=8{\Theta}{\frac{r_{0}}{D_{L}}}{\left({\frac{{\mathcal{M}}_{z}}{1.2M_{\odot}}}\right)}^{5/6}\sqrt{\zeta(f_{\rm{max}})},
\end{equation}
{\noindent}where $\mathcal{M}_{z} = (1+z)\mathcal{M}$ is the redshifted chirp mass,
\begin{align} \label{eq:distance_r0_table}
{r_{0}^{2}}&\equiv{\frac{5}{192\pi}}\left({\frac{3G}{20}}\right)^{5/3}x_{7/3}\frac{M_{\odot}^{2}}{c^3},\nonumber\\
x_{7/3}&\equiv{\int_{0}^{\infty}\frac{df({\pi}M_{\odot})^{2}}{{({\pi}fM_{\odot})^{7/3}}S_{h}(f)}},\nonumber\\
{\zeta(f_{\rm{max}})}&\equiv{\frac{1}{x_{7/3}}}{\int_{0}^{2f_{\rm{max}}}\frac{df({\pi}M_{\odot})^{2}}{{({\pi}fM_{\odot})^{7/3}}S_{h}(f)}},
\end{align}
and $2f_{\rm{max}}$ is the wave frequency at which the inspiral detection template ends \citep{Nutzman2004}. The intrinsic chirp mass, $\mathcal{M}$, is given in terms of the component masses by,
\begin{equation}
\mathcal{M}={\left(\frac{m_{1}m_{2}}{{(m_{1}+m_{2})}^{2}}\right)}^{3/5}(m_{1}+m_{2}).
\end{equation}
The phase evolution of the strain signal in a single interferometer can constrain $\mathcal{M}_{z}$ to subpercent precision \citep{regimbau2012}. In order to measure the luminosity distance, $D_L$, we require a network of separated detectors to break the waveform degeneracy between $D_L$ and $\Theta$ [see Eq.\ (\ref{eq:fourier-waveform})]. Distance measurement errors in a third-generation network will be discussed in Sec.\ \ref{sec:error-section}.

The signal-to-noise (SNR) of a detected system will vary between the individual network sites, as a result of the different $S_{h}(f)$'s and angular dependencies. However, following Refs.\ \citep{Finn96,Nutzman2004}, we assume the network SNR of a detected system is given by the quadrature summation of the individual interferometer SNRs,
\begin{align}\label{eq:network-snr}
{\rho}_{\rm{net}}&=\sqrt{\displaystyle\sum_{k}{\rho}_{k}^{2}}\nonumber\\
&=\frac{8}{D_L}\left(\frac{\mathcal{M}_{z}}{1.2M_{\odot}}\right)^{5/6}\sqrt{\displaystyle\sum_{k}\left(r_{0,k}\Theta_k\zeta_k(f_{\rm{max}})\right)^2}.
\end{align}
where $r_{0,k}$, $\zeta_k(f_{\rm{max}})$ and $\Theta_k$ encode the distance, frequency and angular sensitivities of the $k^{\rm th}$ detector. A comparison of the characteristic distance sensitivities of some second- and third-generation detectors is shown in Table \ref{tab:detector-reach}.

\begin{table}
\caption{\label{tab:detector-reach}The characteristic distance sensitivities [as given by evaluating Eq.\ (\ref{eq:distance_r0_table})] of some advanced-detector configurations and various design studies for the third-generation Einstein Telescope. }
\begin{ruledtabular}
\begin{tabular}{cc}
$\qquad\qquad$Detector & $r_0$ / Mpc$\qquad\qquad$\\
\hline
$\qquad\qquad$AdLIGO\footnotemark & 80$\qquad\qquad$\\
$\qquad\qquad$AdLIGO\footnotemark & 110$\qquad\qquad$\\
$\qquad\qquad$AdLIGO\footnotemark & 119$\qquad\qquad$\\
$\qquad\qquad$AdVirgo\footnotemark & 85$\qquad\qquad$\\
%\hline\hline
$\qquad\qquad$ET\footnotemark & 1527$\qquad\qquad$\\
$\qquad\qquad$ET-B\footnotemark & 1587$\qquad\qquad$\\
$\qquad\qquad$ET-C\footnotemark & 1918$\qquad\qquad$\\
$\qquad\qquad$ET-D\footnotemark & 1591$\qquad\qquad$\\
\end{tabular}
\footnotetext[1]{Low-power zero detuning \cite{adligo-noise}}
\footnotetext[2]{High-power zero detuning \cite{adligo-noise}}
\footnotetext[3]{NS-NS optimised \cite{adligo-noise}}
\footnotetext[4]{Ref.\ \citep{advirgo-noise}}
\footnotetext[5]{Polynomial noise-curve approximation \citep{sathy2009}}
\footnotetext[6]{Conventional technology \citep{ET-site}}
\footnotetext[7]{$3^{\rm rd}$-generation technology, xylophone configuration \citep{ET-site}}
\footnotetext[8]{$3^{\rm rd}$-generation technology, xylophone configuration (updated and more realistic) \citep{ET-site}}
\end{ruledtabular}
\end{table}

\subsection{Network antenna patterns}
The angular dependence of the SNR is encapsulated by the variable $\Theta$. The sky location and binary orientation can be deduced from the network analysis, however in our analysis we will use only $D_L$ and $\mathcal{M}_z$ measurements. We calculate the probability density function for $\Theta$ \citep{Finn96} numerically using Eq.~(\ref{eq:angle}) by choosing $\cos\theta$, $\phi/\pi$, $\cos\iota$ and $\psi/\pi$ to be uncorrelated and distributed uniformly over the range [$-1,1$].

It is unlikely that more than one ET will be constructed. A more likely network configuration will involve a single ET with single third-generation right-angled detectors at other sites. In the interest of verisimilitude we take into account possible detector locations for such a third-generation network. Table \ref{tab:detector-positions} contains the locations and arm-bisector orientations of various current and planned detectors.

\begin{table*}
\caption{\label{tab:detector-positions}A reproduction of the GW-interferometer geographical locations, and arm-bisector orientations from \citet{schutz2011}. We include updated IndIGO information \citep{veitch2012}.\label{tab:detector-locations}}
\begin{ruledtabular}
\begin{tabular}{l c c c c}
Detector & Label & Longitude & Latitude & Orientation\\
\hline
LIGO Livingston, LA, USA & L & $90^{\circ} 46' 27.3''$ W & $30^{\circ} 33' 46.4''$ N & $208.0^{\circ}$(WSW)\\
LIGO Hanford, WA, USA & H & $119^{\circ} 24' 27.6''$ W & $46^{\circ} 27' 18.5''$ N & $279.0^{\circ}$(NW)\\
Virgo, Italy & V & $10^{\circ} 30' 16''$ E & $43^{\circ} 37' 53''$ N & $333.5^{\circ}$(NNW)\\
KAGRA (formerly LCGT), Japan & J & $137^{\circ} 10' 48''$ E & $36^{\circ} 15' 00''$ N & $20.0^{\circ}$(WNW)\\
LIGO-India, India & I & $76^{\circ} 26'$ E & $14^{\circ} 14'$ N & $45.0^{\circ}$(NE)\\
\end{tabular}
\end{ruledtabular}
\end{table*}

To write down the antenna pattern function as a function of the detector position,\footnote{We do not consider modulation of the antenna patterns due to the rotation of the Earth. We justify this in Sec.\ \ref{sec:threshold-change}.} we use the notation and formalism of Ref.\ \citep{schutz2011}.

For a GW source at coordinates $(\theta,\phi)$ on the sky. with polarization angle $\psi$ and a detector with opening angle $\eta$ at latitude $\beta$ and longitude $\lambda$ and such that the bisector of its arms points at an angle $\chi$, counterclockwise from East, the antenna pattern functions are
\begin{equation}
\begin{pmatrix}F_{+}\\F_{\times}\end{pmatrix} = \sin\eta\begin{pmatrix}\cos{(2\psi)}&\sin{(2\psi)}\\-\sin{(2\psi)}&\cos{(2\psi)}\end{pmatrix}\begin{pmatrix}a\\b\end{pmatrix},
\end{equation}
where,
\begin{align}
a =& \frac{1}{16}\sin{(2\chi)}[3-\cos{(2\beta)}][3-\cos{(2\theta)}]\cos{[2(\phi+\lambda)]}\nonumber\\
&+\frac{1}{4}\cos{(2\chi)}\sin\beta[3-\cos{(2\theta)}]\sin{[2(\phi+\lambda)]}\nonumber\\
&+\frac{1}{4}\sin{(2\chi)}\sin{(2\beta)}\sin{(2\theta)}\cos{(\phi+\lambda)}\nonumber\\
&+\frac{1}{2}\cos{(2\chi)}\cos\beta\sin{(2\theta)}\sin{(\phi+\lambda)}\nonumber\\
&+\frac{3}{4}\sin{(2\chi)}\cos^2\beta\sin^2\theta,\nonumber\\
b =& \cos{(2\chi)}\sin\beta\cos\theta\cos{[2(\phi+\lambda)]}\nonumber\\
&{-} \frac{1}{4}\sin{(2\chi)}[3-\cos{(2\beta)}]\cos\theta\sin{[2(\phi+\lambda)]}\nonumber\\
&+ \cos{(2\chi)}\cos\beta\sin\theta\cos{(\phi+\lambda)}\nonumber\\
&{-} \frac{1}{2}\sin{(2\chi)}\sin{(2\beta)}\sin\theta\sin{(\phi+\lambda)}.
\end{align}

As a reference, we use a network comprising three $60^{\circ}$ ET-D sensitivity interferometers at the Virgo location (a single ET), plus right-angled interferometers at the LIGO-Livingston and LIGO-India locations. The characteristic distance reach of all of the interferometers in the network is taken as $1591$ Mpc, corresponding to ET-D sensitivity \citep{sensitivity-studies2011}. This is the sensitivity of a $10$ km right-angle interferometer. We account for the different detector arm-opening angles in the antenna pattern functions.

The network SNR given by Eq.\ (\ref{eq:network-snr}) also depends on ${\zeta}(f_{\rm{max}})$, which describes the overlap of the signal power with the detector bandwidth \citep{Finn96}. The frequency at the end of the inspiral (taken to correspond to the innermost stable circular orbit) is at
\begin{equation}\label{eq:ISCO}
f_{\rm {max}}=\frac{785\text{ Hz}}{1+z}\left({\frac{2.8M_{\odot}}{M}}\right),
\end{equation}
where $M$ is the total mass of the binary system \citep{Abadie2010}. The maximum binary-system mass could conceivably be $\sim 4.2M_{\odot}$.\footnote{Both neutron stars in the binary system would need to have masses $2\sigma$ above the distribution mean at the maximum considered $\mu$ and $\sigma$, where ${\mu}_{\rm{NS}}\in[1.0,1.5]M_{\odot}$, ${\sigma}_{\rm{NS}}\in[0,0.3]M_{\odot}$.} The ET horizon distance for a system with a total mass of $\sim 4M_{\odot}$ is $\sim  25$ Gpc \citep{et-cosmography}. In the $\Lambda$CDM cosmology this corresponds to a redshift of $\sim 2.9$, and from Eq.\ (\ref{eq:ISCO}) this gives $f_{\rm{max}}\sim 134$ Hz. Given the ET-D noise curve \citep{sensitivity-studies2011}, $\sqrt{{\zeta}(f_{\rm{max}}=134 \rm{Hz})}\gtrsim 0.98$. Extending the \textit{redshift} reach out to $z\sim 5$ still gives $\sqrt{{\zeta}(f_{\rm{max}}=87 \rm{Hz})}\gtrsim0.96$. Thus, we feel justified in adopting ${\zeta}(f_{\rm{max}})\simeq1$ for all interferometers in the ensuing analysis.

Using these expressions we were able to numerically estimate the probability distribution for the \textit{effective} $\Theta$,
\begin{equation}
\Theta_{\rm{eff}}=\sqrt{\displaystyle\sum_{k}\Theta_k^2},
\end{equation}
where the sum is over all detectors in the network. We use this $\Theta_{\rm{eff}}$ distribution to choose SNRs for each source in the catalog via Eq.~(\ref{eq:snr}) and then impose a detection criterion. As a reference, we adopt the detection criterion that the network SNR must be greater than $8$. 

\section{DNS systems}\label{ref:source-systems}
\subsection{Neutron-star mass distribution}
For a full discussion of our assumptions and modeling details of the NS mass distribution in DNS systems, see our previous work (Ref.\ \citep{hwth2011} and references therein). We provide here a brief summary of the main assumptions pertinent to the present study.

To lowest order, the GW signal depends on the two neutron-star masses through the chirp mass, $\mathcal{M}$. We assume that the distribution of individual neutron-star masses is normal, as suggested by analysis of Galactic DNS systems \cite{kiziltan2010,valentim2011}, and population synthesis studies (see, e.g., Refs.\ \citep{Belczynski:2008,oshaughnessy2010-binary-compact,MandelOshaughnessy:2010}). For $\sigma_{\rm{NS}} \ll \mu_{\rm{NS}}$, this should also lead to an approximately normal distribution for the chirp mass.

We use a simple ansatz for the relationship between the chirp-mass distribution parameters and the underlying neutron-star mass distribution. The chirp mass distribution is modeled as normal,
\begin{equation}
{\mathcal{M}}\sim N({\mu}_{c}, \sigma_{c}^2), \nonumber
\end{equation}
{\noindent}with mean and standard deviation
\begin{equation} \label{eq:ansatz}
\mu_c\approx 2(0.25)^{3/5}{\mu}_{\rm{NS}},\quad\sigma_c\approx {\sqrt{2}}(0.25)^{3/5}{\sigma}_{\rm{NS}},
\end{equation}
{\noindent}where ${\mu}_{\rm{NS}}$ and ${\sigma}_{\rm{NS}}$ are the mean and standard deviation of the underlying neutron-star mass distribution.

A recent study by \citet{ozel2012} has found that DNS data are consistent with both pulsar and companion having been drawn from the same underlying distribution of masses. The literature indicates an underlying neutron-star mass distribution in DNS systems with ${\sigma}_{\rm{NS}}\lesssim 0.15M_{\odot}$ \cite{kiziltan2010,valentim2011,ozel2012}.\footnote{Indeed, \citet{ozel2012} indicates that the DNS mass distribution is peculiar, since it cannot be explained via electron-capture or core-collapse supernovae mechanisms; rather, its narrow dispersion may be a result of the evolutionary path of these systems.} Hence, we anticipate that Eq.\ (\ref{eq:ansatz}) will be appropriate for generating data sets and we use this in the ensuing analysis. The assumption throughout is that for the volume of the Universe probed by our global network, the neutron-star mass distribution does not change.  

Further population synthesis and observational studies in the following decade will help to shed further light on the nature of the NS mass distribution. The assumption of a unimodal (for DNS systems) Gaussian distribution is an approximation, and if future studies show this to be inappropriate, then a more suitable ansatz could be readily incorporated within the framework described in this paper. 

\subsection{Merger-rate density} \label{sec:merger-rate-density}
In this analysis, we aim to probe not only the background cosmology and NS mass-distribution parameters, but the astrophysical properties of the binary NS population. To this end, we now consider the factors contributing to the coalescence of a binary NS system.

Following several population synthesis studies (e.g., Refs.\ \citep{oshaughnessy2008,oshaughnessy2010-binary-compact}) and Ref.\ \citep{pacheco1997}, we define the merger rate per comoving volume as
\begin{equation} \label{eq:density-convolve}
\dot{n}(t)=\int_{t_*}^t \lambda\frac{dP_m}{dt}(t-t_b)\frac{d\rho_*}{dt}(t_b)dt_b,
\end{equation}
where $\lambda$ is a \textit{mass efficiency}, defined as the number of coalescing DNS binaries per unit star-forming mass \citep{oshaughnessy2010-binary-compact}. $dP_m/dt$ is the probability density distribution of a DNS binary merging at a time $(t-t_b)$ after formation, and $d\rho_*/dt$ is the star-formation rate (SFR) density at cosmological time $t_b$.

Star formation, and the efficiency of double compact-object formation from the progenitor system, may be sensitive to the host-galaxy morphology and environment (e.g.\ metallicity). Furthermore, the distribution of delay times between star formation and the corresponding DNS-system coalescence may have contributions from several different evolutionary paths. However, we are interested here only in a third-generation GW-interferometer network's ability to constrain various astrophysical and cosmological parameters \citep{hwth2011}. As such we consider a single component star-formation distribution, delay-time distribution and mass efficiency, deferring considerations of the other dependencies to a future study. We now discuss the various terms in Eq.\ (\ref{eq:density-convolve}) in more detail.

\subsubsection{Mass efficiency, $\lambda$}
We use values for $\lambda$ obtained from the population synthesis calculations of Ref.\ \citep{oshaughnessy2008}. Smoothed histograms of the mass efficiency are shown in Fig.\ 4 of that paper, with modes at $\sim 10^{-5}M_{\odot}^{-1}$ for DNS systems formed in both elliptical and spiral conditions. However the distribution of $\lambda$ ranges over several orders of magnitude, with $10^{-7}M_{\odot}^{-1}\lesssim\lambda\lesssim 10^{-3}M_{\odot}^{-1}$. We adopt $\lambda=10^{-5}M_{\odot}^{-1}$ as the reference value for our analysis.

\subsubsection{Merger-delay distribution, $dP_m/dt$} \label{sec:merger-delay}
Massive stars in high-mass binary systems evolve into DNS systems on much shorter timescales than typical galaxy evolution or Hubble timescales, such that $dP_m/dt$ is essentially determined by the initial orbital separation, $a$, of the DNS system \citep{totani1997}. The evolutionary time delay between the progenitor formation and the formation of the corresponding DNS system is typically $\lesssim 50$ Myr \citep{schneider2001}, and is therefore negligible compare to the gravitational-wave inspiral timescale, which scales as $\tau_{\rm gr}\propto a^4$. Assuming the number of binaries, $N$, born with separation $a$ scales as $dN/da \propto a^\gamma$~\citep{totani1997}, we obtain
\begin{equation}
\frac{dP_m}{dt}\propto\frac{dN}{d\tau_{\rm gr}}=\frac{dN}{da}\frac{da}{d\tau_{\rm gr}}\propto t^{(\gamma-3)/4} = t^{\alpha}.
\end{equation}
If DNS systems have the same orbital separation distribution as normal-abundance main-sequence stars \citep{piran1992,abt1983}, then $\gamma=\alpha = -1$. However, this scaling is not well constrained and this is discussed in more detail in Appendix~\ref{appsec:merger-delay}. Instead, we adopt the approach of allowing $\alpha$ to be a free parameter that we attempt to fit from our observations and ask with what precision this can be determined. We use the value $\alpha=-1$ for our reference model, which is justified by current (albeit sparse) analysis of Galactic DNS systems \citep{champion2004,guetta2005,guetta2006}, and various population synthesis calculations \citep{schneider2001,oshaughnessy2008,belczynski2006,belczynski-twin-2007,dominik2012,belczynski2010}. For normalisation purposes, we assume a minimum delay-time of $50$ Myr and a maximum delay-time equal to the cosmology-dependent age of the Universe; these choices are discussed in Appendix~\ref{appsec:merger-delay}.

\subsubsection{Star-formation rate density, $d\rho_*/dt$}
The star-formation rate density is also rather uncertain. The SF2 model of \citet{porciani2001} attempts to factor in the uncertainties in the incompleteness of data sets and the amount of dust extinction at early epochs. The SF2 model has the form
\begin{align}\label{eq:SF2}
\frac{d\rho_*}{dt}(z) \approx &\quad0.16\times\left(\frac{\exp{(\beta_1 z)}}{\exp{(\beta_2 z)}+22}\right)\nonumber\\
&\times\frac{E(z)}{(1+z)^{3/2}}\quad\text{$M_{\odot}$Mpc$^{-3}$yr$^{-1}$},
\end{align}
where
\begin{equation}\label{eq:hubble-flow}
E(z)=\sqrt{{\Omega}_{m,0}(1+z)^3+{\Omega}_{k,0}(1+z)^2+{\Omega}_{\Lambda}(z)}
\end{equation}
and $\beta_1=\beta_2=3.4$. In this model, the SFR density remains roughly constant at $z\gtrsim 2$, which may be incompatible with recent observations \citep{bouwens2009,gonzalez2010}. This is discussed in more detail in Appendix~\ref{appsec:sfrdens}. To allow for some uncertainty, we treat $\beta_1$ and $\beta_2$ as free parameters and explore how precisely we can measure them. While this simple ansatz does not cover all possible forms for the SFR density, using it will provide an indication of what GW observations could tell us. The framework is easily adaptable to more complex SFR models.

We must also specify $t_*$, the lower integration bound in Eq. (\ref{eq:density-convolve}), which represents the time of the earliest period of star formation. The highest redshift objects observed are a long gamma-ray burst (GRB) with a photometric redshift of $\sim 9.4$~\citep{cucchiara2011} and a candidate $z\sim10$ galaxy \citep{bouwens2011}. We therefore use $z=10$ as the earliest time of star formation. Future observations, for instance with the \textit{James Webb Space Telescope}, may be able to probe back to the first phases of galaxy formation at $z\sim 15$ and if objects are found at that epoch, this assumption should be revised. However, our results are fairly insensitive to this choice.

\subsubsection{Calculating $\dot{n}(z)$}
\begin{figure}
    \incgraph{270}{0.5}{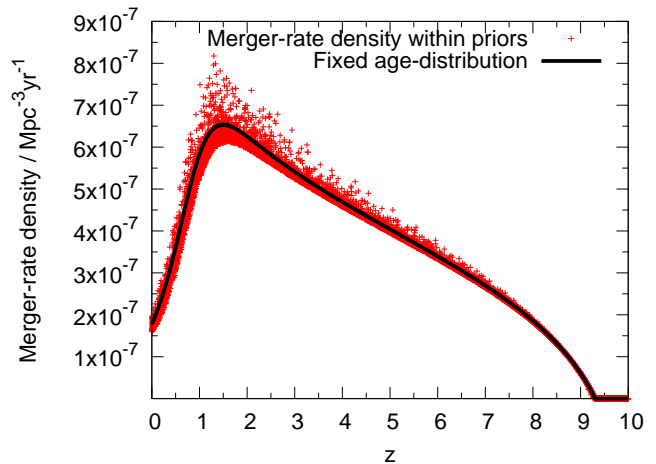}
   \caption{\label{fig:merger-priors}Merger-rate densities computed for the reference cosmology (solid line) and for cosmological parameters chosen randomly from within the prior range (red crosses).}
\end{figure}

Eq.\ (\ref{eq:density-convolve}) can be rewritten as a distribution in redshift using $dt/dz = -1/((1+z) E(z) H_0)$
\begin{align}
\dot{n}(z)=&\int_{10}^z \lambda\frac{dP_m}{dt}(t-t_b)\frac{d\rho_*}{dt}(t_b)\frac{dt_b}{dz_b}dz_b\nonumber\\
=& \frac{0.16\lambda}{H_0}\int_{z}^{10} \frac{dP_m}{dt}(t(z)-t_b(z_b))\nonumber\\
&\times\left(\frac{\exp{(\beta_1z_b)}}{\exp{(\beta_2z_b)}+22}\right)\frac{dz_b}{(1+z_b)^{5/2}}.
\end{align}
Evaluating this expression requires an expensive double integral which created a bottleneck in our analysis. However, because the priors on the cosmological parameters are narrow (see Sec.\ \ref{sec:posterior-calc}), there is little variation in the merger-rate density across this range, as shown in Fig.\ \ref{fig:merger-priors}. We therefore fixed the cosmological parameters at their reference values for the cosmological time calculation, which made the merger-rate density calculation considerably faster. Although this throws away some of the cosmological information, it did not significantly affect the results and made the analysis more tractable.

\section{Cosmological models}\label{sec:cosmo-models}
In our previous analysis~\cite{hwth2011} we considered only a flat cosmology, but here we allow for curvature and an evolving equation of state (EOS) for the dark energy. From the cosmological field equations we have
\begin{equation}
\dot{\rho} + 3\left(\frac{\dot{a}}{a}\right)\left(\rho + \frac{p}{c^2}\right) = 0,
\end{equation}
where $\rho$ and $p$ are the density and pressure of a cosmological fluid respectively, while $a$ is the scale factor of the universe. Derivatives are with respect to physical time. For a perfect fluid ($p=w{\rho}c^2$, where $w$ is the EOS parameter), this reduces to
\begin{equation}
\left(\frac{\dot{\rho}}{\rho}\right) = -3(1+w)\left(\frac{\dot{a}}{a}\right).
\end{equation}
Hence,
\begin{equation}
\rho(a) = \rho(a_{\rm today}=1)\times e^{-3\int_1^a (1+w)(da'/a')}.
\end{equation}
The last decade has seen many proposals for $w$, with different physical motivation. One approach attempts to explain dark energy as a minimally coupled scalar field (``quintessence'') slowly rolling down its potential such that it can exert negative pressure. While it is possible to have a constant EOS in this formalism, this requirement places severe constraints on the potential and so it is natural to expect a time-varying EOS \citep{chevallier2000}.

The simplest approximation is to assume a linear model ($w(z)=w_0+w_1z$), but this is only appropriate for local studies due to the divergence at high redshift. The Shafieloo-Sahni-Starobinsky ansatz \citep{shafieloo2009} models the EOS evolution as a ``$\rm{tanh}$'' form that ensures $w=-1$ at early times and $w \rightarrow 0$ at low $z$. This ansatz prevents the crossing of the ``phantom divide'' at $w=-1$, desirable since phantom fluids can not be explained by a minimally coupled scalar field \citep{chevallier2000}. The ansatz we adopt in this work is the Chevallier-Polarski-Linder ansatz \citep{chevallier2000,linder2003}
\begin{equation}
w(a) = w_0 + w_a(1-a),\quad w(z) = w_0 + w_a\left(\frac{z}{1+z}\right).
\end{equation}
This ansatz was adopted by the Dark Energy Task Force \citep{detf-findings-2009}, and has several desirable features. It depends on only two free parameters, it reduces to the linear model at low $z$, and it is well-behaved at high redshift, tending to $w_0+w_a$. Using this EOS,
\begin{equation}
\Omega_{\Lambda}(z) = \Omega_{\Lambda,0}\times {(1+z)}^{3(1+w_0+w_a)}\times e^{-3w_a\left(\frac{z}{1+z}\right)}.
\end{equation}

For different global geometries of the Universe the luminosity distance, $D_L$, is given by
\begin{equation}
D_L(z|\mathcal{C}) = (1+z)\times\mathcal{F}(z|\mathcal{C}),\nonumber
\end{equation}
where
\begin{equation}\label{eq:lum-distance-geometry}
\mathcal{F}(z|\mathcal{C})=
\begin{cases}
 \frac{D_H}{\sqrt{\Omega_{k,0}}}\sinh\left(\sqrt{\Omega_{k,0}}\frac{D_c(z|\mathcal{C})}{D_H}\right), & \Omega_{k,0}>0, \\
 D_c(z|\mathcal{C}), & \Omega_{k,0}=0, \\
 \frac{D_H}{\sqrt{|\Omega_{k,0}|}}\sin\left(\sqrt{|\Omega_{k,0}|}\frac{D_c(z|\mathcal{C})}{D_H}\right), & \Omega_{k,0}<0,
\end{cases}
\end{equation}
in which $D_H$ is the Hubble scale ($c/H_0$) and $\mathcal{C}$=$\{H_0,\Omega_{m,0},\Omega_{\Lambda,0},\Omega_{k,0},w_0,w_a\}$ is the set of cosmological parameters describing the large-scale characteristics of the universe.

The comoving radial distance, $D_c(z)$, is given by
\begin{equation}
D_c(z) = D_H\int_0^z\frac{dz'}{E(z')},
\end{equation}
where $E(z)$ is given by Eq.\ (\ref{eq:hubble-flow}). The redshift derivative of the comoving volume is given generally by
\begin{equation}
\frac{dV_c}{dz}=4\pi D_H\frac{D_L(z)^2}{(1+z)^2E(z)}.
\end{equation}

\section{Making $\&$ Analyzing DNS catalogs}\label{sec:make-analyse-catalogs}
We refer the reader to our previous study \citep{hwth2011} for full details of our calculation, but we summarise the main details here.
\subsection{Distribution of detectable DNS systems}
The two system properties we will use in our analysis are the redshifted chirp mass, $\mathcal{M}_z$, and the luminosity distance, $D_L$. We assume that only systems with an SNR greater than a given threshold will be detected. We can write down the distribution of the number of events per unit time in the observer's frame with $\mathcal{M}$, $z$ and effective $\Theta$ \citep{Finn96,oshaughnessy2010-binary-compact},
\begin{equation} \label{eq:data-generate}
\frac{d^4N}{dtd{\Theta}dzd\mathcal{M}}={\frac{dV_c}{dz}}{\frac{\dot{n}(z)}{(1+z)}}{\mathcal{P}}({\mathcal{M}}){\mathcal{P}}_{\Theta}(\Theta).
\end{equation}
The $1/(1+z)$ factor accounts for the redshifting of the merger rate \citep{oshaughnessy2010-binary-compact}.

Converting this to a distribution in $\mathcal{M}_z$, $D_L$ and $\rho$, and integrating over $\rho$ to find the distribution of \textit{detectable} systems (i.e.,\ systems above SNR threshold) gives
\begin{align} \label{eq:rprob}
{\frac{d^3N}{dtd{D_L}d\mathcal{M}_z}}{\bigg\vert}_{{\rho}>{\rho}_0}=&\frac{dz}{dD_L}{\frac{dV_c}{dz}}{\frac{{\dot{n}}(z)}{(1+z)^2}}\times{\mathcal{P}}\left({\frac{{\mathcal{M}}_{z}}{1+z}}{\bigg\vert}{D_L}\right)\nonumber\\
&\times C_{\Theta}\left[{\frac{\rho_0}{8}}{\frac{D_L}{r_{0}}}\left({\frac{1.2M_{\odot}}{{\mathcal{M}}_{z}}}\right)^{5/6}\right],
\end{align}
where the form of $(dz/dD_L)$ will depend on the curvature of the Universe [see Eq.\ (\ref{eq:lum-distance-geometry})].

To calculate the number of detected systems (given a set of model parameters, $\overrightarrow\mu$) we integrate over this distribution, which is equivalent to integrating the distribution over redshift and chirp mass, i.e.\ $N_{\mu}=T\times{\int_0^{\infty}}{\int_0^{\infty}}\left(\frac{d^3N}{dtdzd\mathcal{M}}\right){\rm{d}}z{\rm{d}}{\mathcal{M}}$, where $T$ is the duration of the observation run.\footnote{We found that, for the purposes of the calculation, assuming the NS mass distribution was a $\delta$-function, centred at the mean given by the trial parameters, allowed at least a tenfold speedup in the calculation. See Appendix \ref{sec:faster-detection-rate} for further details.}

\subsection{Creating mock catalogs of DNS binary inspiraling systems} \label{sec:ref-model}
The model parameter space we investigate is the $7$D space of $[w_0,w_a,{\mu}_{\rm{NS}},{\sigma}_{\rm{NS}},{\alpha},{\beta}_1,{\beta}_2]$. To generate a catalog of events, we choose a set of reference parameters, motivated by previous analysis in the literature. For our reference cosmology, we adopt $H_0=70.4$ kms$^{-1}$Mpc$^{-1}$, ${\Omega}_{m,0}=0.2726$, ${\Omega}_{\Lambda,0}=0.728$, $w_0=-1.0$ and $w_a=0.0$ \citep{wmap_plus}. The reference parameters of the neutron-star mass distribution are ${\mu}_{\rm{NS}}=1.35M_{\odot}$ and ${\sigma}_{\rm{NS}}=0.06M_{\odot}$ \citep{kiziltan2010}. We have previously discussed the delay-time distribution and SFR density in Sec.\ \ref{sec:merger-rate-density}. We adopt a power-law merger-delay distribution with reference power-law index $\alpha=-1.0$, and we take the SFR density to be given by the SF2 ansatz \citep{porciani2001}, with $\beta_1=3.4$ and $\beta_2=3.4$.

These reference parameters are used to calculate an expected number of events\footnote{The observation time, $T$, is assumed to be $1$ yr, and the mass efficiency is assumed to be $\lambda=10^{-5}M_{\odot}^{-1}$ (as mentioned earlier).}, and the number of observed events, $N_o$, is drawn from a Poisson distribution (assuming each binary system is independent of all others) with that mean. Monte-Carlo acceptance/rejection sampling is used to draw random redshifts and chirp masses from the distribution in Eq.\ (\ref{eq:data-generate}) for each event. The $D_L$ and ${\mathcal{M}}_z$ are then computed from the sampled  ${\mathcal{M}}$ and $z$.

For these reference parameters, which give a local merger-rate density of $\sim 2\times10^{-7}$ Mpc$^{-3}$yr$^{-1}$, and assuming detected systems must have a network SNR greater than $8$, we find that the expected number of detections in $1$ yr is $\sim 10^5$.\footnote{This reference local merger-rate density is roughly five times smaller than the realistic value quoted in \citet{abadie-rate2010}, but $20$ times larger than the pessimistic estimate. Whilst we could scale our merger-rate density calculations to match the realistic value of $10^{-6}$ Mpc$^{-3}$yr$^{-1}$, our modified likelihood statistic makes our analysis insensitive to such scalings. A rescaling to the realistic local merger-rate density of Ref.\ \citep{abadie-rate2010} would lead to an expected detection rate of approximately half a million sources.}

\subsection{Likelihood statistic}
In the measurement-parameter space of redshifted chirp mass and luminosity distance, the measured number of detections in a given bin will be a Poisson random variate with a model-dependent mean. If we take the continuum limit of this, such that bin sizes are infinitesimally small and contain either $0$ or $1$ events, then we can formulate the likelihood of a catalog of discrete events,
\begin{equation} \label{eq:cont-prob}
{\mathcal{L}}(\overrightarrow{{\overrightarrow{\Lambda}}}|{\overrightarrow{\mu}},\mathcal{H})={e^{-N_{\mu}}{\displaystyle\prod_{i=1}^{N_o}}}r({\overrightarrow{{\lambda}_i}}|{\overrightarrow{\mu}}),
\end{equation}
{\noindent}where ${\overrightarrow{{\overrightarrow{\Lambda}}}}={\{}{\overrightarrow{{\lambda}_1}},{\overrightarrow{{\lambda}_2}},{\ldots},{\overrightarrow{{\lambda}_{N_o}}}{\}}$ is the vector of measured system properties, with ${\overrightarrow{{\lambda}_i}}=({\mathcal{M}}_{z}, D_L)_{i}$ for system $i$. $N_o$ is the \textit{actual} number of detected systems, while $N_{\mu}$ is the number of DNS inspiral detections \textit{predicted} by the model with parameters $\overrightarrow{\mu}$. Finally, $r({\overrightarrow{{\lambda}_i}}|{\overrightarrow{\mu}})$ is the rate of events with properties ${\mathcal{M}}_{z}$ and $D_L$, evaluated for the $i^{\rm{th}}$ detection under model parameters $\overrightarrow{\mu}$, which is given by Eq.\ (\ref{eq:rprob}). The trial cosmological parameters are used to calculate a model-dependent redshift from the cataloged luminosity distance, and, in turn, this redshift is used to infer a model-dependent intrinsic chirp mass from the cataloged value of ${\mathcal{M}}_{z}$. These values of ${\mathcal{M}}_{z}$, $D_L$, as well as the model-dependent values of $z$ and $\mathcal{M}$ are inserted into Eq.\ (\ref{eq:rprob}) to compute the likelihood.

In our previous study, we employed a modified likelihood statistic which marginalized over the local merger-rate density of DNS systems. This approach reflects our current lack of knowledge of this quantity, estimates of which vary over several orders of magnitude. We adopt the same approach in this analysis, to eliminate the dependence on poorly known scaling factors. This includes the mass efficiency parameter, $\lambda$, which is a quantity derived from population synthesis studies and can vary over several orders of magnitude.

The modified likelihood statistic is
\begin{equation} \label{eq:new-stat}
\tilde{\mathcal{L}}(\overrightarrow{{\overrightarrow{\Lambda}}}|{\overrightarrow{\mu}},\mathcal{H})\propto{N_{\mu}}^{-(N_o+1)}\displaystyle\prod_{i=1}^{N_o}r({\overrightarrow{{\lambda}_i}}|{\overrightarrow{\mu}}),
\end{equation}

We note that we have not included a prior on the scaling factors in the above, which is equivalent to using an improper flat prior over the range $[0,\infty]$. This reflects our current lack of knowledge of the mass efficiency.

\subsection{Calculating the posterior probability} \label{sec:posterior-calc}

We use a weakly informative prior on the model parameters, so that it does not prejudice our analysis. As a prior on ${\mu}_{\rm{NS}}$ we take a normal distribution with parameters ${\mu}=1.35M_{\odot}$, ${\sigma}=0.13M_{\odot}$. This is motivated by the posterior predictive density estimate for a neutron star in a DNS binary system given in Ref.\ \cite{kiziltan2010}. 

Given that ET will most likely not be operational until the mid-2020s, we must consider what constraints conventional observational cosmology can put on cosmological parameters. In the recent study by \citet{et-dark-energy}, the authors investigated how the dark-energy EOS could be probed by ET observations of DNS systems to provide distance measurements, complemented by electromagnetic measurements of the associated short gamma-ray burst (sGRB) to provide the redshift. They estimated that a combination of the Planck cosmic microwave background (CMB) prior, JDEM BAO results, and future Type Ia supernova observations could provide cosmological constraints by the ET era of
\begin{align}
\Delta\Omega_{m,0} = 3.46\times 10^{-3},&\quad\Delta\Omega_{k,0} = 5.91\times 10^{-4},\nonumber\\
\Delta H_0 = 0.336,\quad\Delta w_0 = & 0.048,\quad\Delta w_a = 0.184.
\end{align}
Hence, we assume that  $H_0$, $\Omega_{m,0}$ and $\Omega_{k,0}$ are precisely known quantities, with values of $70.4$ kms$^{-1}$Mpc$^{-1}$, $0.2726$ and $-0.0006$, respectively. As a prior on $w_0$ and $w_a$, we adopt the constraint that $w(z)<-1/3$ at all redshifts. Hence we use uniform priors on the EOS parameters with $w_0<-1/3$ \textit{and} $w_0+w_a<-1/3$ and lower bounds set low enough so as not to affect the posterior probability distribution. We also adopt uniform priors for all other parameters under investigation.

\subsection{Bayesian analysis and an adaptive Markov chain Monte Carlo technique} \label{sec:bayes-adaptive-mcmc}
Bayes' theorem states that the \textit{posterior} probability distribution of the parameters $\overrightarrow{{\mu}}$ describing a hypothesis model $\mathcal{H}$, and given data $D$ is given by
\begin{equation} \label{eq:bayes-theorem}
p(\overrightarrow{{\mu}}|D,\mathcal{H}) = \frac{{\mathcal{L}}(D|\overrightarrow{{\mu}},\mathcal{H})\pi(\overrightarrow{{\mu}}|\mathcal{H})}{p(D|\mathcal{H})},
\end{equation}
where ${\mathcal{L}}(D|\overrightarrow{{\mu}},\mathcal{H})$ is the \textit{likelihood} (the probability of measuring the data, given the model $\mathcal{H}$ with parameters $\overrightarrow{{\mu}}$), $\pi(\overrightarrow{{\mu}}|\mathcal{H})$ is the \textit{prior} (any constraints already existing on the model parameters) and finally $p(D|\mathcal{H})$ is the \textit{evidence} (this is important in model selection, but in the subsequent analysis in this paper can be ignored as a normalization constant).

Markov chain Monte Carlo (MCMC) techniques provide an efficient way to explore the model-parameter space. An initial point, $\overrightarrow{x_0}$, is drawn from the \textit{prior} distribution and then at each subsequent iteration, $i$, a new point, $\overrightarrow{y}$, is drawn from a \textit{proposal distribution}, $q({\overrightarrow{y}}|{\overrightarrow{x}})$ and the Metropolis-Hastings ratio evaluated,
\begin{equation}
R=\frac{{\pi}(\overrightarrow{y}){\mathcal{L}}(D|{\overrightarrow{y}},\mathcal{H})q({\overrightarrow{x_i}}|\overrightarrow{y})}{{\pi}(\overrightarrow{x_i}){\mathcal{L}}(D|{\overrightarrow{x_i}},\mathcal{H})q({\overrightarrow{y}}|{\overrightarrow{x_i}})}.
\end{equation}

A random sample is drawn from a uniform distribution, $u\in U[0,1]$, and if $u<R$ the move to the new point is accepted and we set ${\overrightarrow{x_{i+1}}}={\overrightarrow{y}}$. If $u>R$, the move is rejected and we set ${\overrightarrow{x_{i+1}}}={\overrightarrow{x_i}}$. 

The MCMC samples can be used to carry out integrals over the posterior
\begin{equation}
\int f(\overrightarrow{x})p(\overrightarrow{x}|D,\mathcal{H})d\overrightarrow{x}\approx{\frac{1}{N}}\displaystyle\sum_{i=1}^Nf(\overrightarrow{x_i}).
\end{equation}
The $1$D marginalized posterior probability distributions in individual model parameter follows by binning the chain samples in that parameter.

The trick to using this technique efficiently is to choose an appropriate proposal distribution. In our analysis we employ an adaptive MCMC procedure, which utilises an ``in-flight'' estimation of the sampled chain's covariance matrix to construct an updating proposal distribution. This covariance matrix is updated at each iteration, with a certain chain memory \citep{haario1999,haario2001,dunkley2005}. We use several of the procedures outlined in Ref.\ \citep{dunkley2005}.

For the first $100$ points in the chain, simple Gaussian proposal distributions for each individual parameter are used. These first $100$ points are merely used to provide a starting point for the covariance matrix evaluation and so the exact proposal distribution used in this stage is not important. After the first hundred points are sampled, we begin generating points via the adaptive procedure.
For a $D$-dimensional target posterior distribution, we suppose that at the $i^{\rm{th}}$ iteration we have sampled at least $H$ points, where the fixed integer $H$ is the \textit{memory} parameter. We then generate a $D$-dimensional vector of trial parameters, $\overrightarrow{y}$, via a linear mapping of an $H$-dimensional vector of unit-variance Gaussian random scalars, $\overrightarrow{\xi}$,
\begin{equation}
\overrightarrow{y}=\textbf{C}^{1/2}\overrightarrow{\xi},
\end{equation}
where $\textbf{C}^{1/2}$ is the positive-definite square root of the $D\times D$ covariance matrix evaluated using the previous $H$ points. The covariance matrix may be calculated by collecting the previous $H$ points in the chain into an $H\times D$ matrix $\textbf{K}$, with each row representing one sampled point. Then,
\begin{equation}
\textbf{C}=\frac{1}{H-1}\tilde{\textbf{K}}^T\tilde{\textbf{K}},
\end{equation} 
where the centered matrix, $\tilde{\textbf{K}}$, is constructed by centering each column of $\textbf{K}$ around the means of the respective parameters, calculated from the $H$ samples.

We then generate the trial parameter vector $\overrightarrow{y}$ via
\begin{equation}
\overrightarrow{y}\sim\mathcal{N}(\overrightarrow{x},c_d^2\textbf{C})\sim\overrightarrow{x}+\frac{c_d}{\sqrt{H-1}}\tilde{\textbf{K}}^T\overrightarrow{\xi},
\end{equation}
where $c_d$ is a variable which depends only on the dimensionality of the target distribution. This variable is used to optimise the efficiency of the sampling process, and we use the value of $\approx 2.4/\sqrt{D}$ \citep{gelman1995,dunkley2005}. 

With a memory parameter which is less than the total past history of the chain, this is denoted as the Adaptive Proposal (AP) algorithm \citep{haario1999}. Since the proposal distribution is updated constantly and relies on previous chain information, this procedure is not Markovian, and does not have the correct ergodicity properties for an MCMC algorithm \citep{haario1999}. In principle this can bias the reconstruction of the target posterior; however this bias is ignorable in many practical applications, and for well-behaved target posterior distributions \citep{haario1999,haario2001}. If the entire previous chain is used to update the covariance matrix, then this algorithm is known as the Adaptive Metropolis (AM) algorithm \citep{haario2001}. The AM algorithm does not suffer from the biases which can occur in the AP algorithm, and ergodicity is retained.\footnote{In the AM algorithm the covariance of the proposal distribution is actually taken to be $\textbf{C}+\epsilon\textbf{I}_D$, where $\textbf{I}_D$ is the $D$-dimensional identity matrix. Choosing $\epsilon>0$ allows for the correct ergodicity properties of an MCMC algorithm to be retained, and in practice is useful if the covariance of the chain has a tendency to degenerate. However, this parameter can be set very small with respect to the size of the target space, and in practice can be set to zero.} We use the AM algorithm in our work.
  
\section{Results}\label{sec:results}
\subsection{Posterior recovery}
An analysis using the full $10^5$ observations expected in a year of ET data is computationally prohibitive, so we use a working reference sample of $\sim 4500$ detections (corresponding to a shorter observation time or a lower merger rate density) and extrapolate to the expected number of detections, as discussed in Sec.~\ref{sec:ndetect}. For each analysis, we ran $120$ independent adaptive MCMC chains of 5000 points on the same data catalog. We then used the last point from each chain to initialize a follow-up run of another $5000$ iterations. The first $2000$ points from each chain of the follow-up run were discarded as burn-in. This procedure therefore generated $360,000$ points, with an average acceptance rate of $\sim 30\%$. The analysis of the $4500$-event reference catalog took $\sim 3.5$ hrs in total. Our sampled points were analyzed using the \textsc{CosmoloGUI} package \citep{bridle2011}. 

\subsection{Marginalized posterior distributions}
\begin{figure*}
   \subfloat[]{\label{fig:w0_mu}\incgraph{0}{0.33}{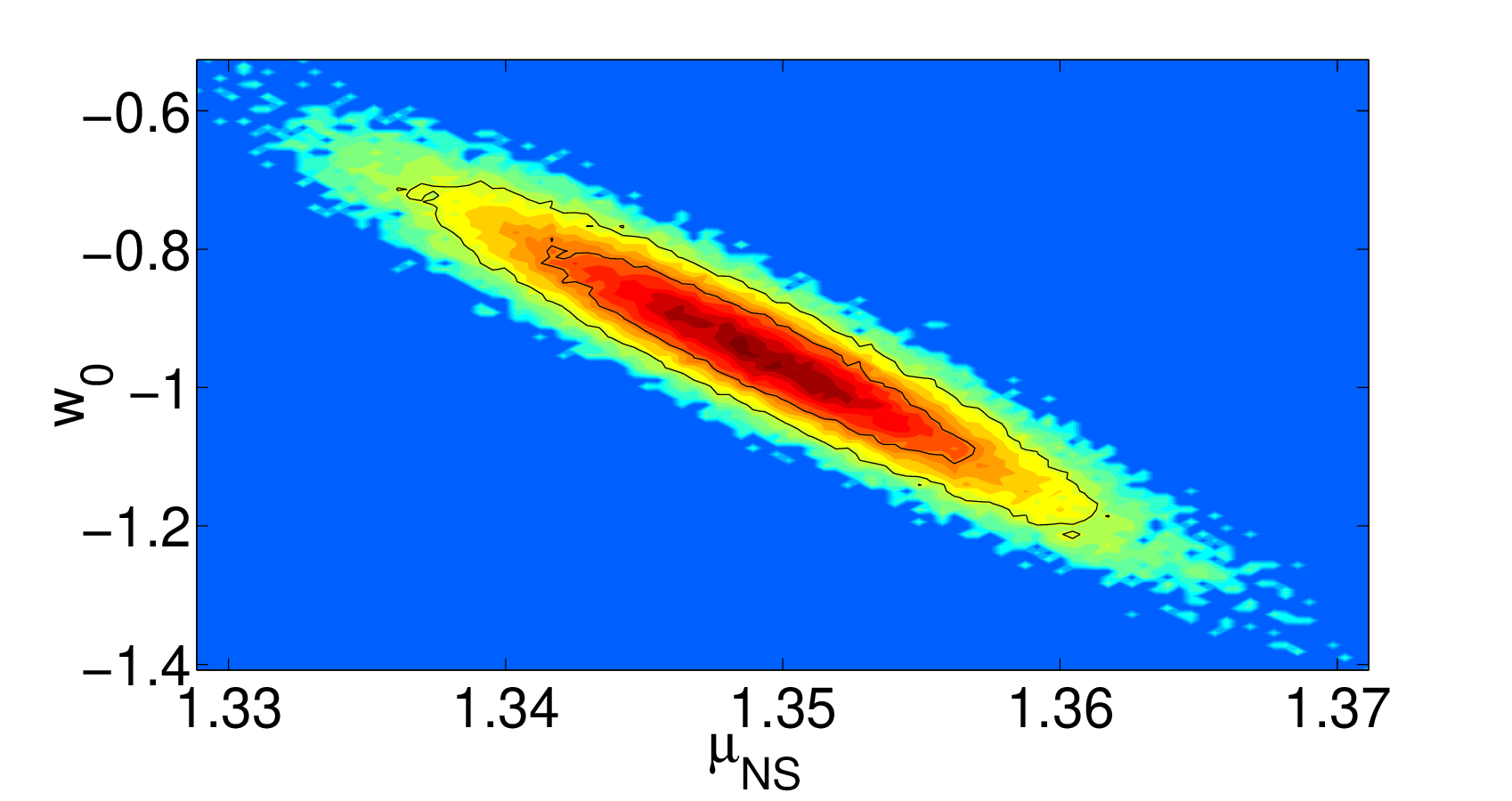}} \hspace{-7mm}
   \subfloat[]{\label{fig:wa_mu}\incgraph{0}{0.33}{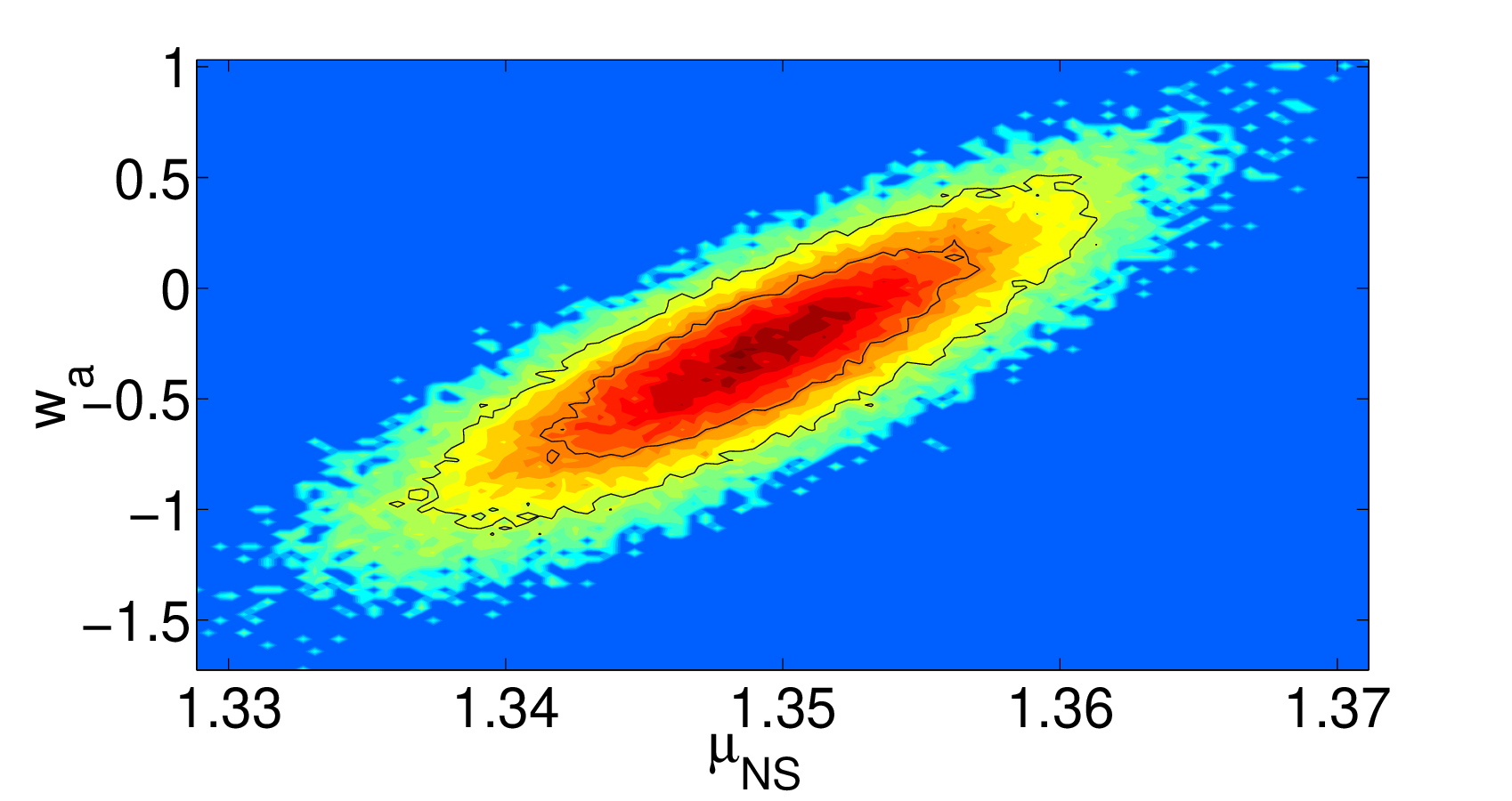}} \hspace{-7mm}
   \subfloat[]{\label{fig:w0_wa}\incgraph{0}{0.33}{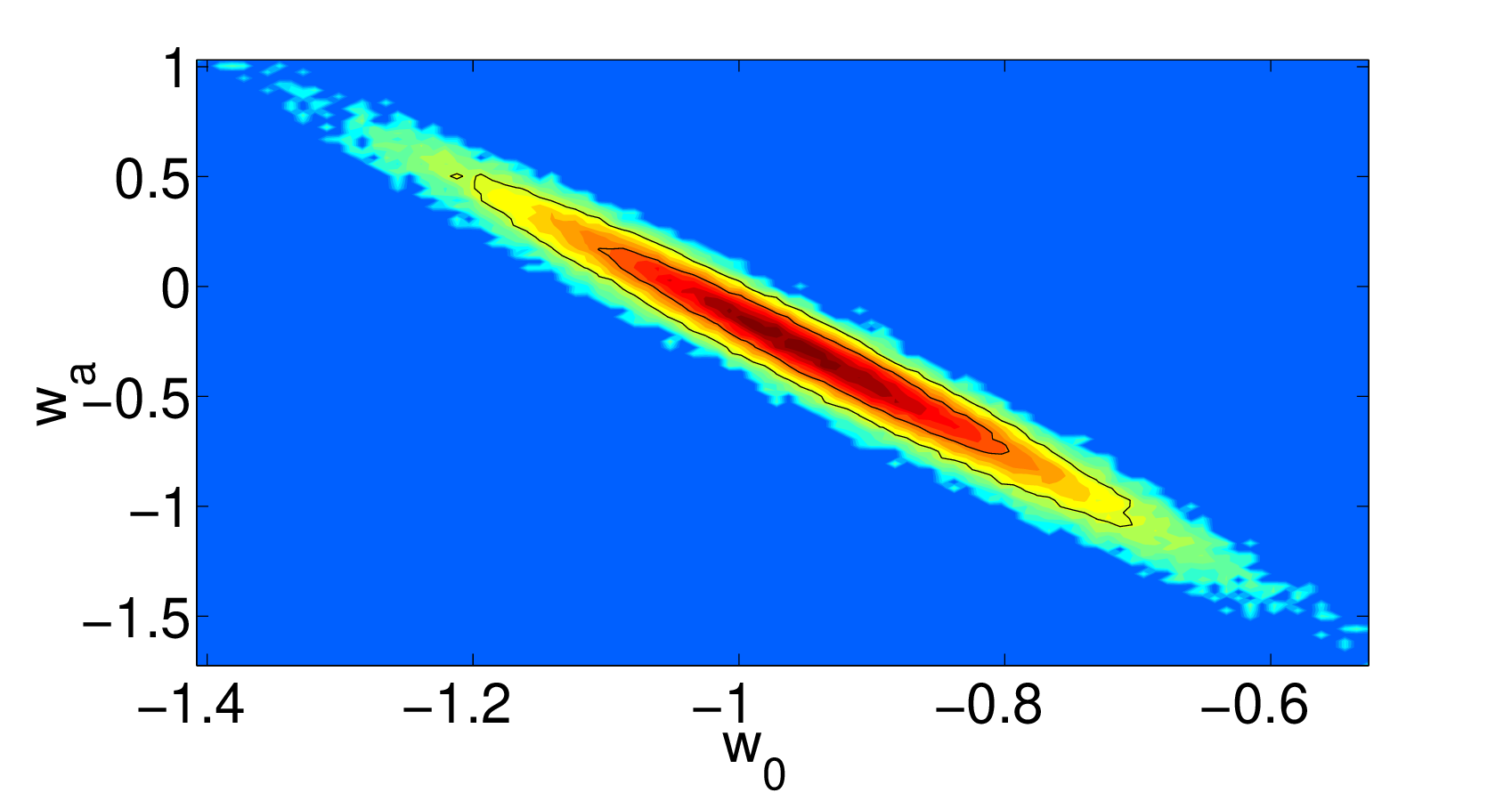}}\\
   \subfloat[]{\label{fig:beta1_beta2}\incgraph{0}{0.33}{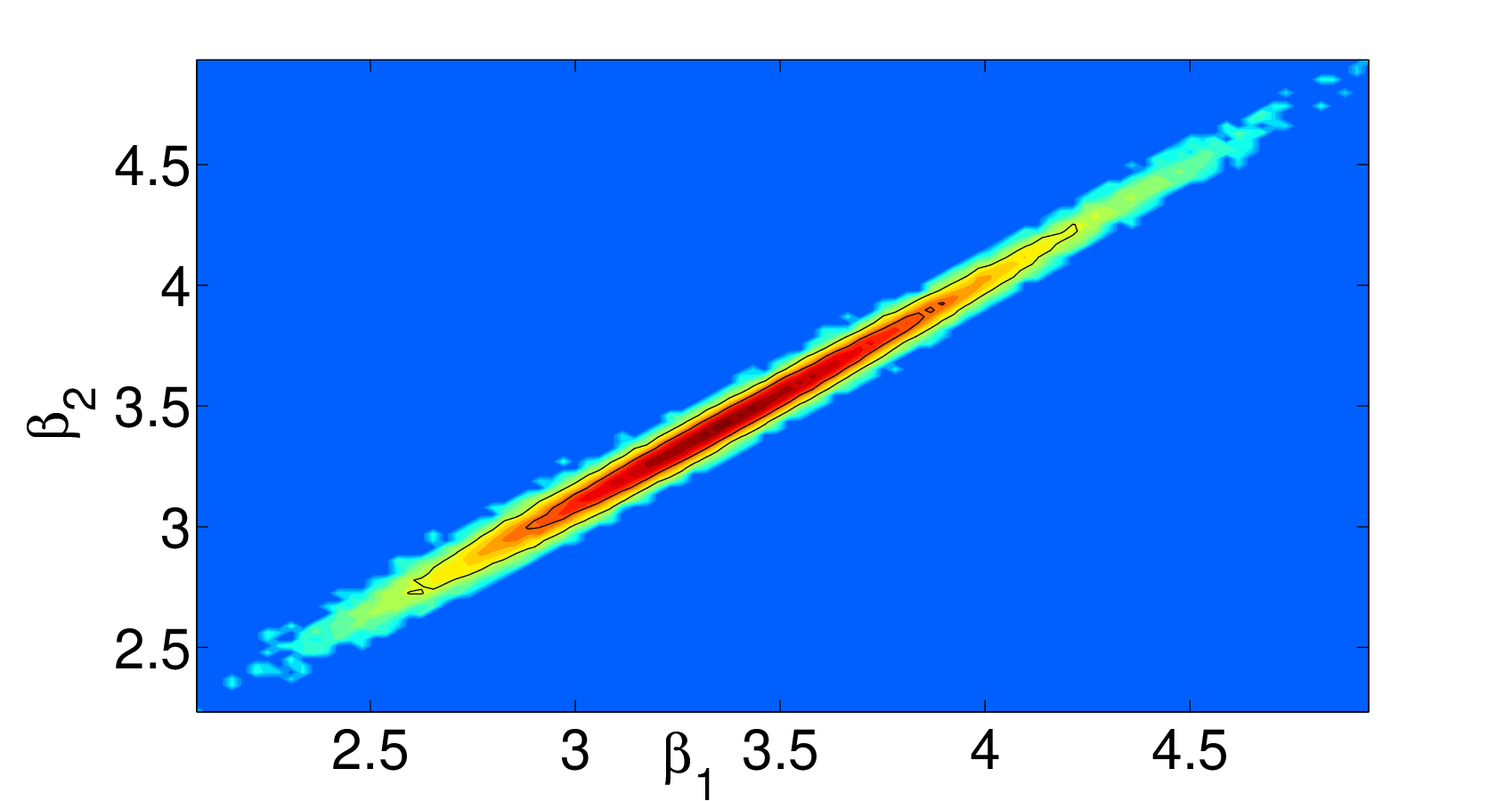}} \hspace{-7mm}
   \subfloat[]{\label{fig:alpha_beta1}\incgraph{0}{0.33}{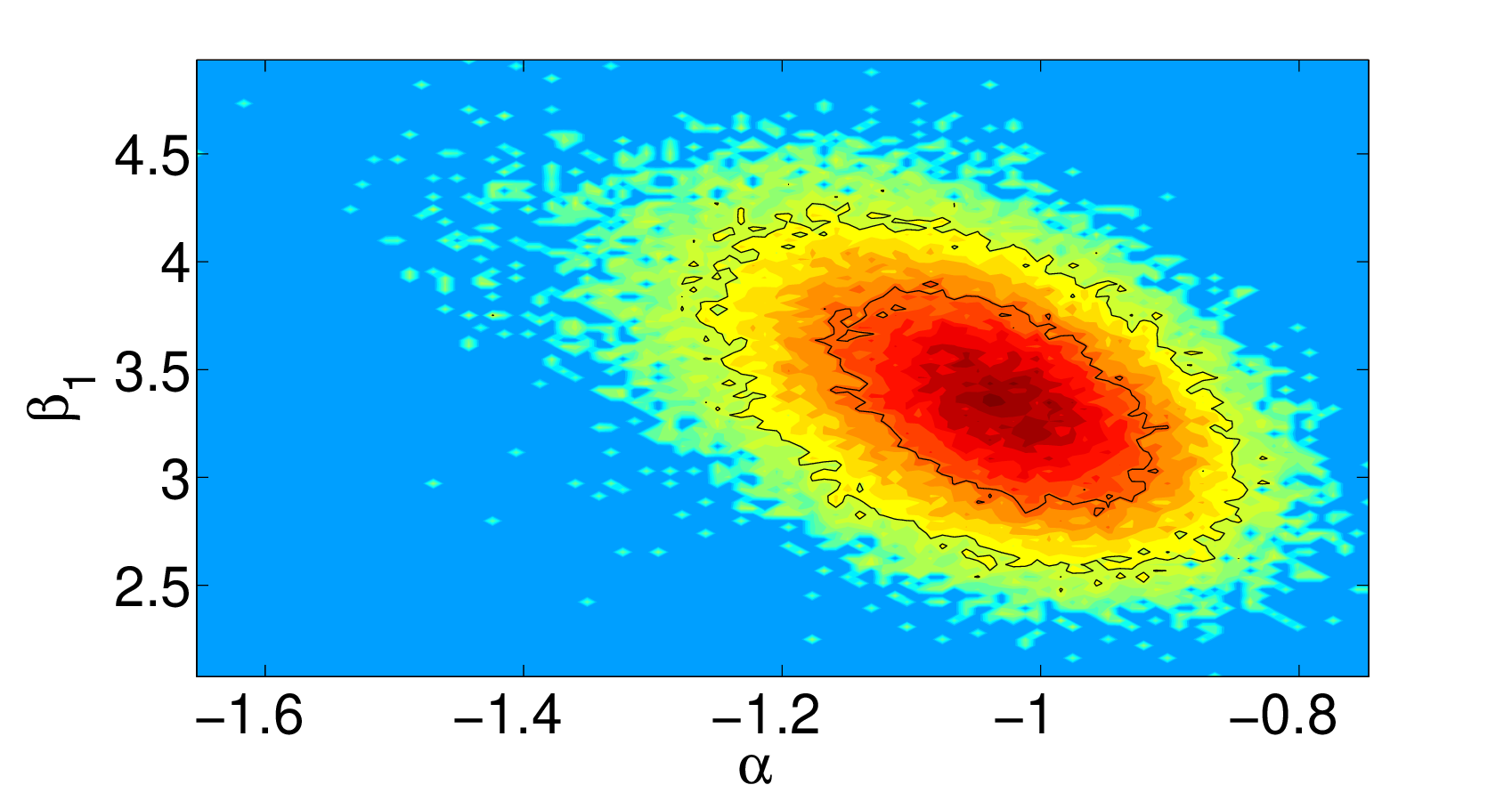}} \hspace{-7mm}
   \subfloat[]{\label{fig:w0_wa_mu}\incgraph{0}{0.33}{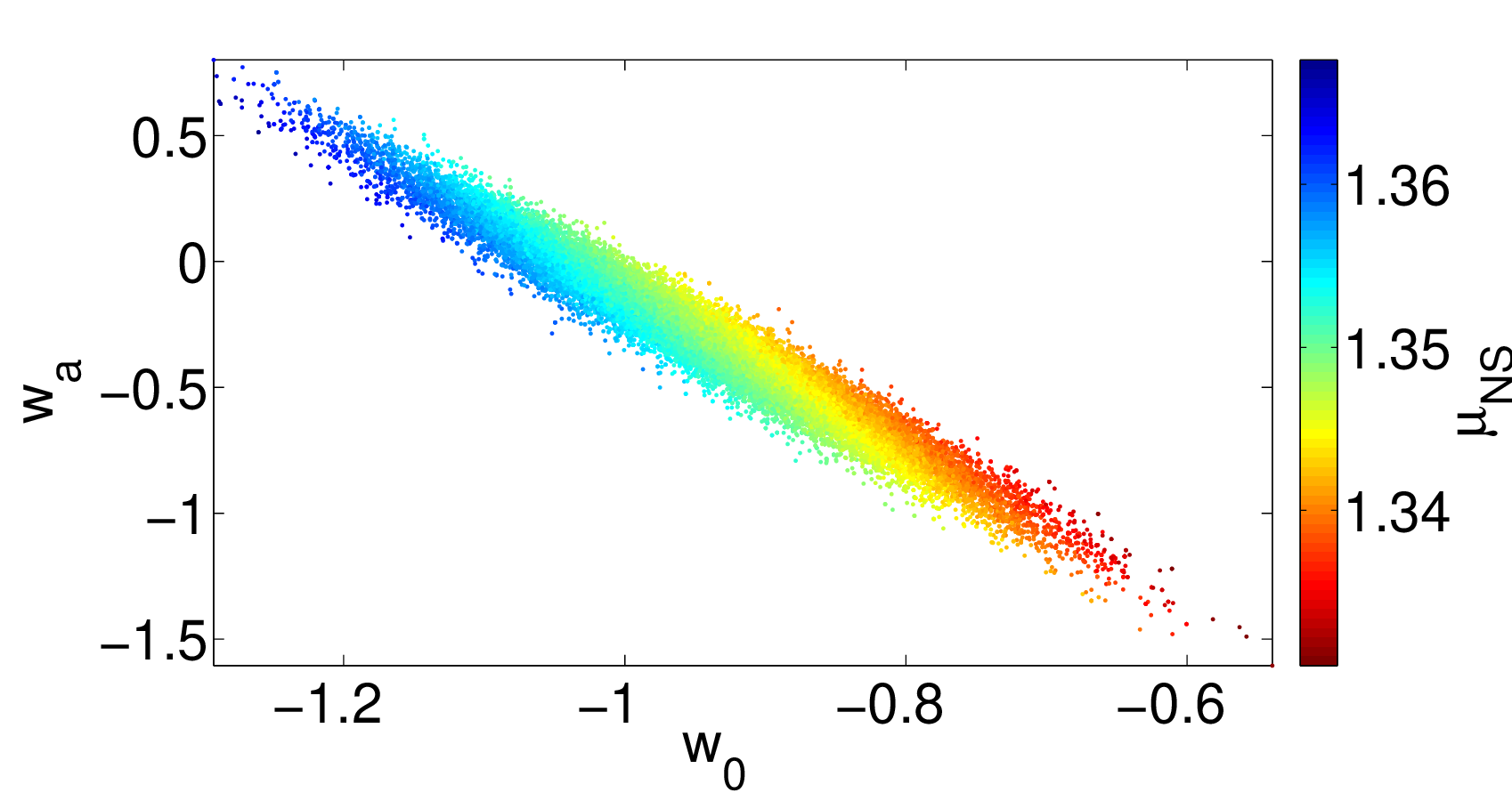}} 
   \caption{\label{fig:2d-correlations}Marginalized $2$D posterior distributions for the reference catalog of $4500$ detections. Only those $2$D distributions showing correlations between parameters are shown. The reference parameters are $\mu_{\rm{NS}}=1.35 M_{\odot}$, $\sigma_{\rm{NS}}=0.06 M_{\odot}$, $w_0=-1$, $w_a=0$, $\alpha=-1$ and $\beta_1=\beta_2=3.4$.}
 \end{figure*}

In Fig.\ \ref{fig:2d-correlations} we show the recovered marginalized $2$D posterior distributions (with $68\%$ and $95\%$ confidence contours) for the reference catalog. In Fig.\ \ref{fig:2d-correlations}\subref{fig:w0_wa} we observe a correlation between the recovered dark-energy parameters. This is easily explained by the fact that a given cataloged luminosity distance may be consistent\footnote{Here, by ``consistent'' we mean within $\pm 1\%$ of the reference value.} with a set of $w_0$ and $w_a$ combinations, which will depend on the redshift of the source. Since the majority of detected systems will be centred around $z\sim 1$, the $w_0-w_a$ correlation will be dominated by these sources.\footnote{The correlation between the two dark-energy EOS parameters can be reduced by rebinning the MCMC samples using the Wang parametrization \citep{wang-parametrise-2008}. This simply involves a tranformation from the $(w_0,w_a)$ parametrisation to $(w_0,w_{0.5})$, where $w_{0.5}=w_0+(w_a/3)$.} In Fig.\ \ref{fig:2d-correlations}\subref{fig:w0_mu} a negative correlation is observed between the recovered values for $w_0$ and $\mu_{\rm{NS}}$. For a given cataloged luminosity distance and fixed $w_a$, a low value of $w_0$ will imply a low redshift in that model. When this redshift is used to compute $\mathcal{M}$ from $\mathcal{M}_z$, we obtain a large value of the chirp mass, which is consistent with a chirp-mass distribution (and hence a NS-mass distribution) centered at larger values. Figure \ref{fig:2d-correlations}\subref{fig:wa_mu} merely shows the combined information of Figures \ref{fig:2d-correlations}\subref{fig:w0_mu} and \ref{fig:2d-correlations}\subref{fig:w0_wa} (where the recovered $w_a$ values are negatively correlated with the $w_0$ values); therefore the correlation observed in Fig.\ \ref{fig:2d-correlations}\subref{fig:wa_mu} is positive. 

A strong positive correlation is observed between the SFR-density SF2 ansatz parameters, $\beta_1$ and $\beta_2$, as seen in Fig.\ \ref{fig:2d-correlations}\subref{fig:beta1_beta2}, while Fig.\ \ref{fig:2d-correlations}\subref{fig:alpha_beta1} shows a weak negative correlation between $\alpha$ and $\beta_1$. These correlations correspond to keeping the merger-rate density approximately constant. We calculated which combinations of $\alpha$, $\beta_1$ and $\beta_2$ were consistent with a given merger-rate density, at a variety of redshifts. We found that there was a strong positive correlation in these points between $\beta_1$ and $\beta_2$, but the correlation between $\alpha$ and $\beta_1$ changed sign as the redshift increased. The greatest change occurred as the redshift was increased from $0$ to $1$, where the correlation then reversed; however at $z=4$ the magnitude of the correlation was still not as large as it was at $z=0.1$. This leads us to believe that although the $D_L$ distribution of detected sources is peaked around $\sim 6$ Gpc, with a long tail to $\sim 45$ Gpc, the lower distance sources dominate the $\alpha-\beta_1$ correlation, giving an overall negative correlation.

\begin{figure*}
   \subfloat[]{\incgraph{0}{0.33}{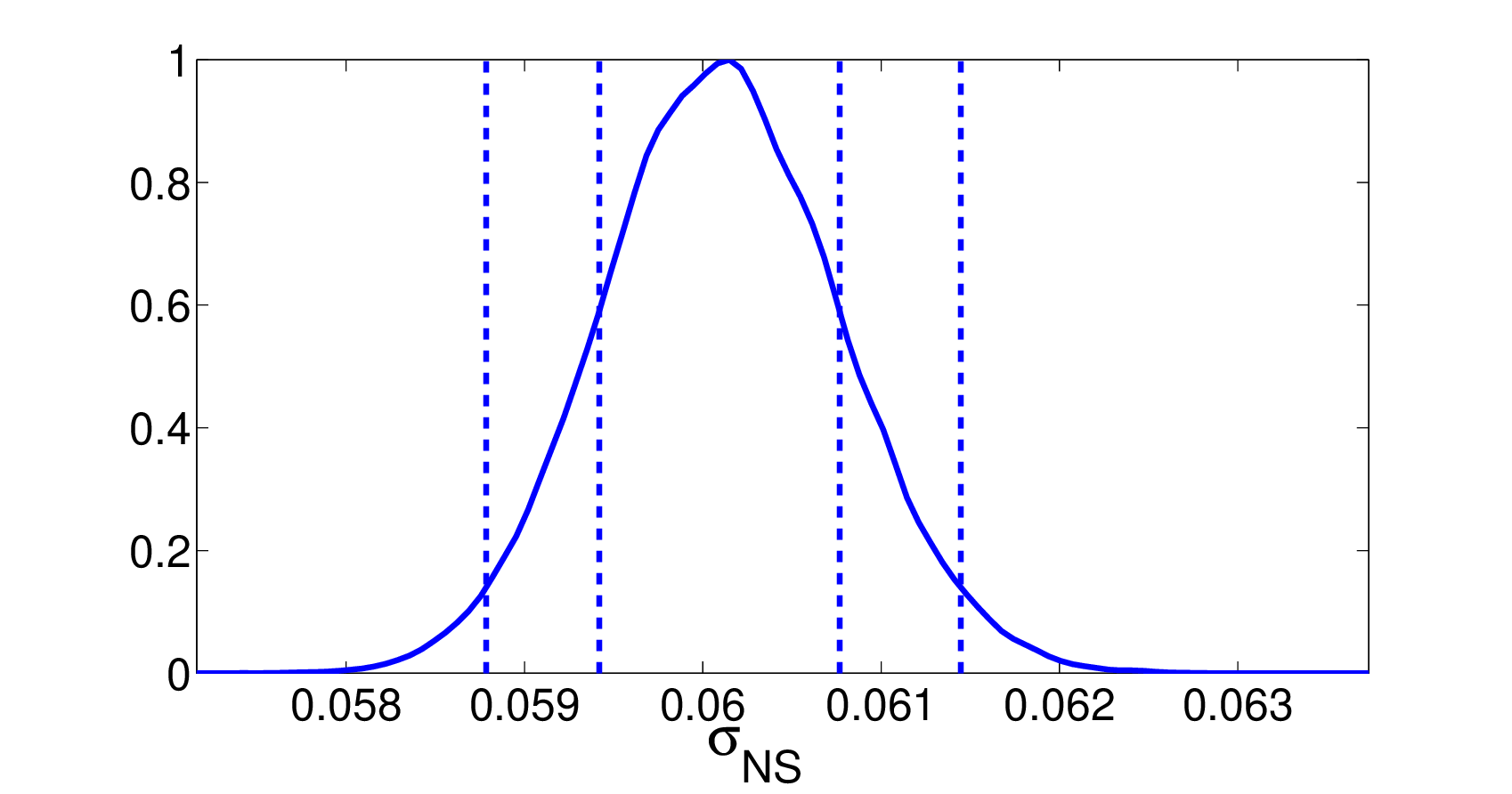}} \hspace{-7mm}
   \subfloat[]{\incgraph{0}{0.33}{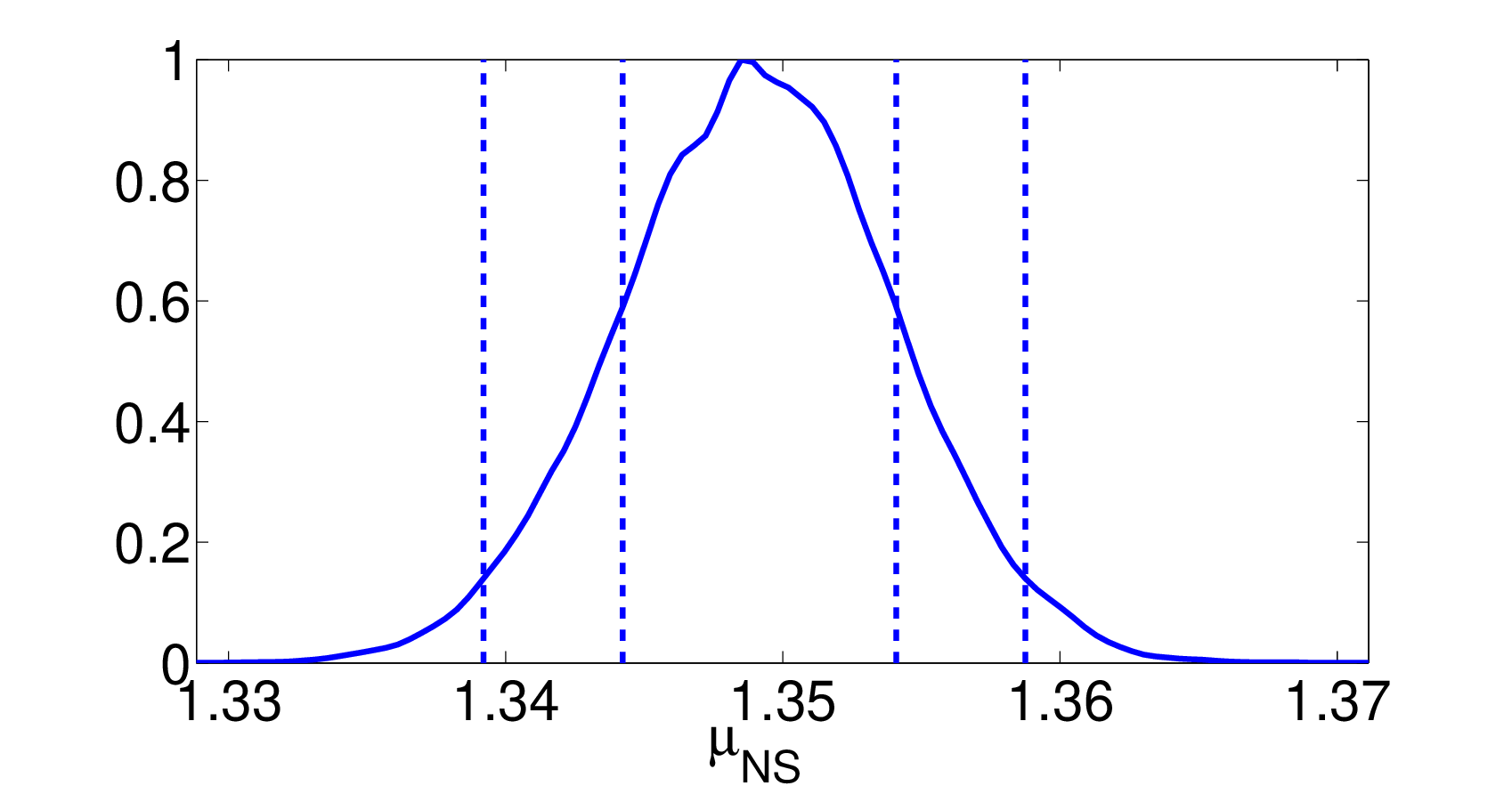}} \hspace{-7mm}
   \subfloat[]{\incgraph{0}{0.33}{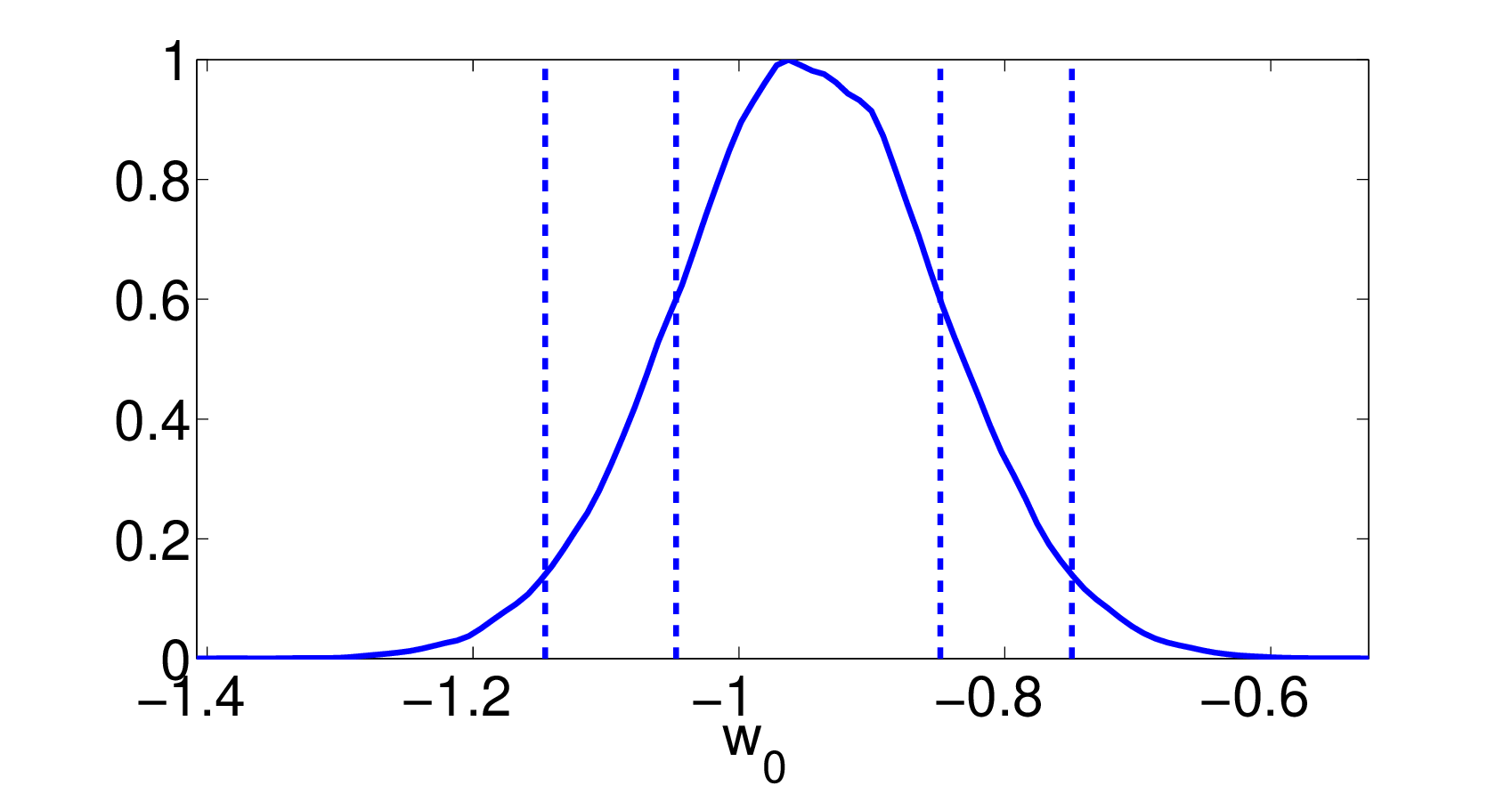}}\\
   \subfloat[]{\incgraph{0}{0.33}{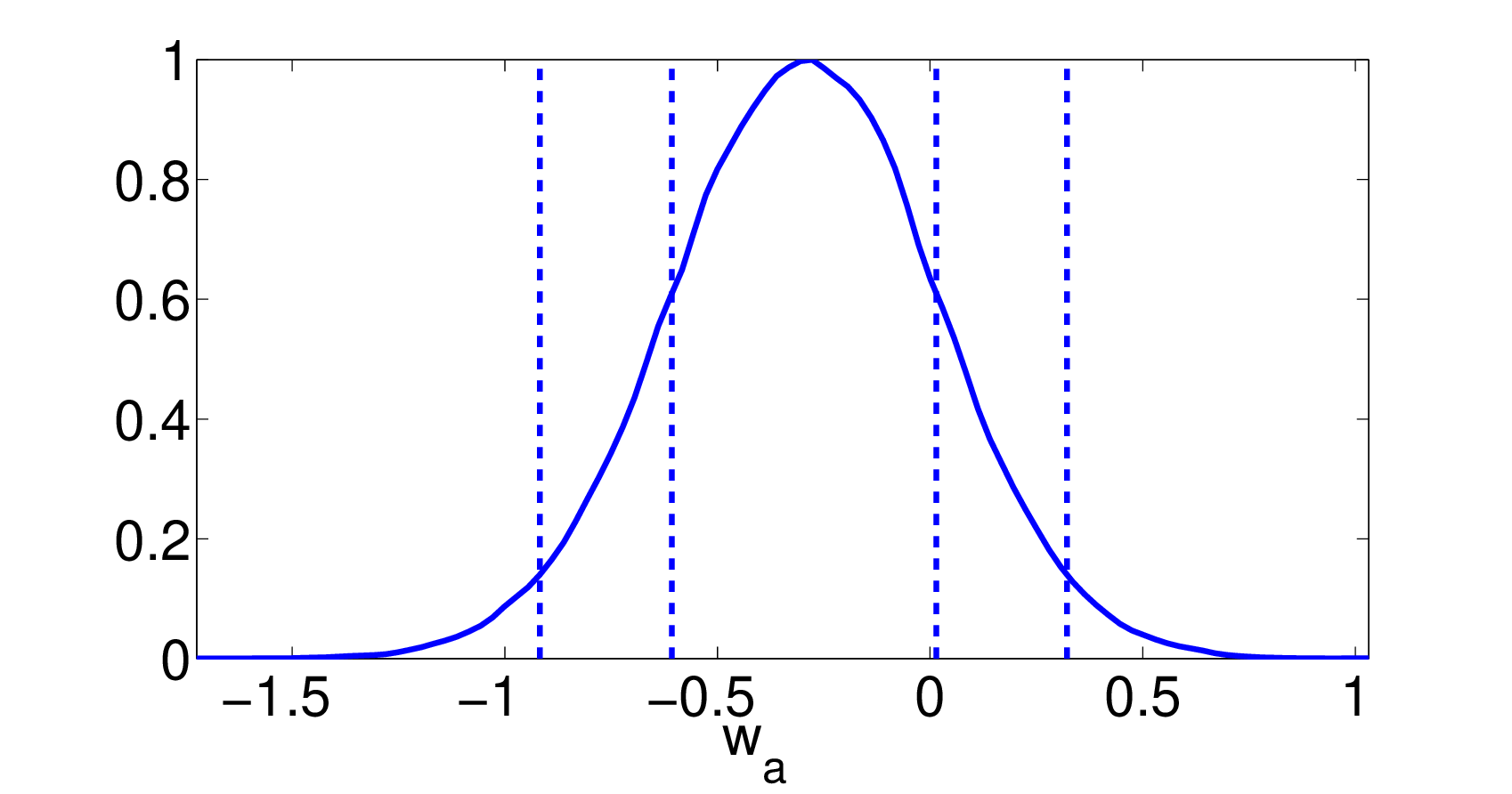}} \hspace{-7mm} 
   \subfloat[]{\incgraph{0}{0.33}{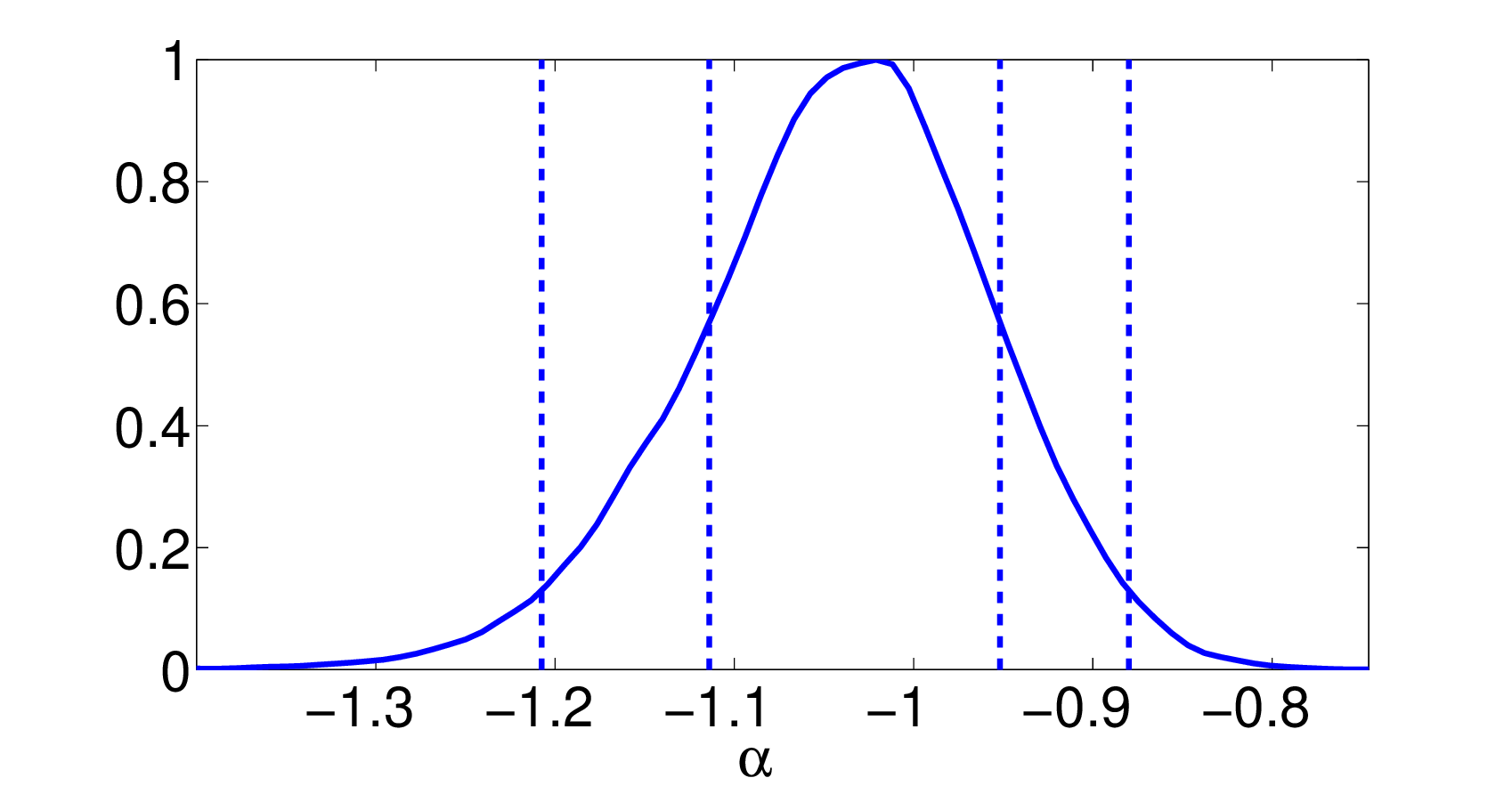}} \hspace{-7mm}
   \subfloat[]{\incgraph{0}{0.33}{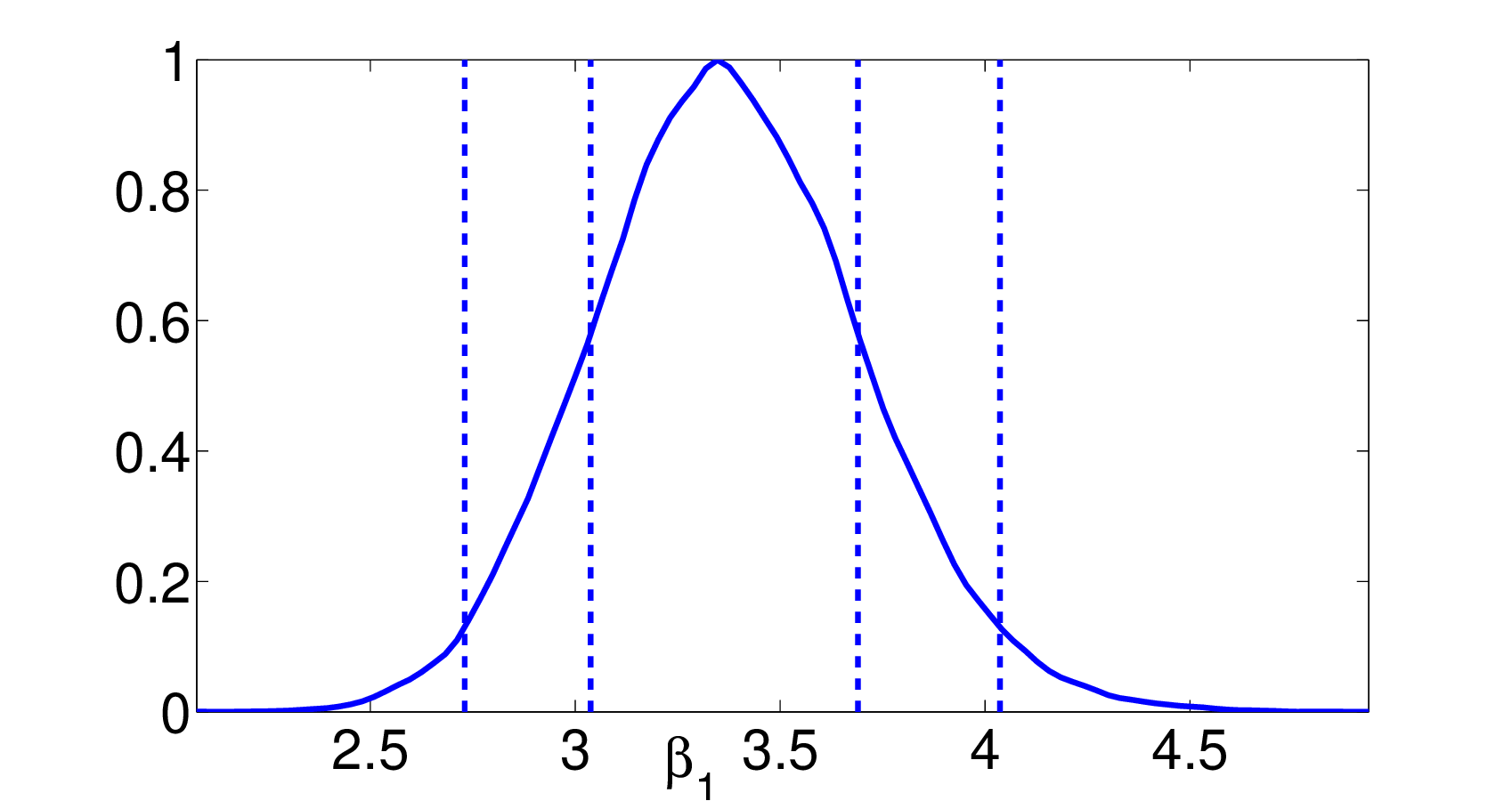}} 
   \caption{\label{fig:1d-posteriors}Marginalized $1$D posterior distributions for the reference catalog of $4500$ detections. Dotted lines indicate the boundaries of the $95\%$ and $68\%$ confidence intervals.  The reference parameters are $\mu_{\rm{NS}}=1.35 M_{\odot}$, $\sigma_{\rm{NS}}=0.06 M_{\odot}$, $w_0=-1$, $w_a=0$, $\alpha=-1$ and $\beta_1=3.4$.}
 \end{figure*}

In Fig.\ \ref{fig:1d-posteriors}, we show the marginalized $1$D posterior distributions for the model parameters. The dotted lines in the plots indicate the $68\%$ and $95\%$ confidence regions of the marginalized distributions.\footnote{While these results were computed using the fast merger-rate approximation, we also analysed a catalog using the full merger-rate density. The $95\%$ confidence intervals of the marginalized posterior distributions were consistent with our approximate analysis, justifying the use of the approximation to compute the rest of our results. No correlations between the merger-rate density parameters and the dark-energy EOS parameters were found, which supports our earlier statement that the dependence of the merger-rate density on the underlying cosmological parameters is weak within the applied priors.}

\subsection{Precision scaling with number of detections}\label{sec:ndetect}
\begin{figure}
    \incgraph{270}{0.5}{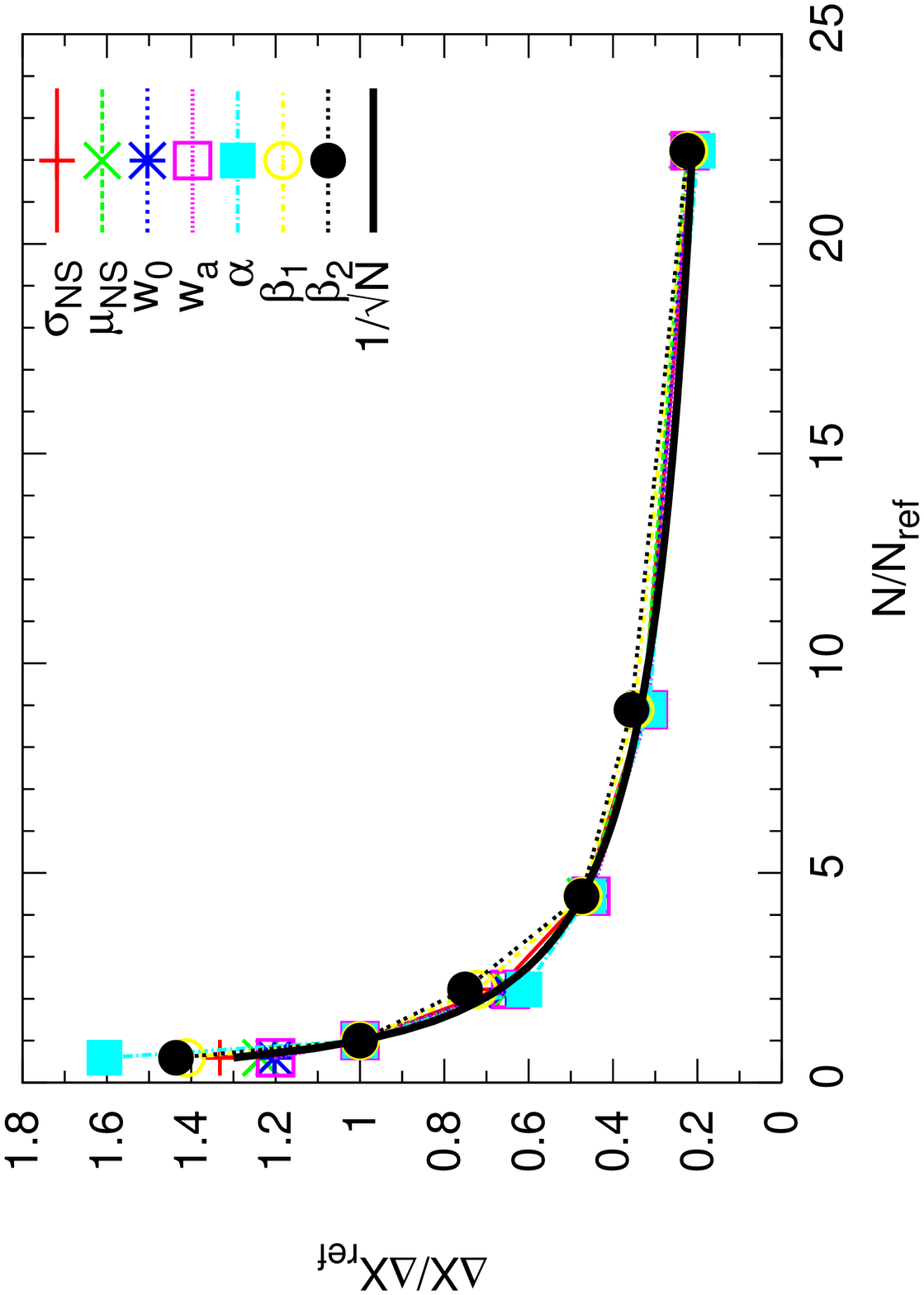}
   \caption{\label{fig:num-vary} $95\%$ confidence intervals of the $1$D marginalized distributions relative to those of the $4500$-event reference catalog, shown as a function of the number of cataloged events. The same intrinsic parameters of the underlying distributions are used to create the mock catalogs. The expected $\sim 1/\sqrt{N}$ relationship is overlaid on the plot.}
\end{figure}

We performed similar analyses on catalogs containing various numbers of detections, culminating in a run with $10^5$ detections. We can characterize the precision with which we can measure the various model parameters by the $95\%$ confidence intervals. Recording these intervals for all parameters for varying catalog sizes, and dividing by the reference sample intervals gave the results shown in Fig.\ \ref{fig:num-vary}. This clearly shows that the precisions scale as $1/\sqrt{N_o}$ as we would expect. Parameter measurement accuracies for the $10^5$-event catalog are shown in Table \ref{tab:all-accuracy-data}. We see that the measurement precisions of the dark-energy EOS parameters are the same order of magnitude as those forecast for CMB$+$BAO$+$SNIa \citep{et-dark-energy}, as discussed in Sec.\ \ref{sec:posterior-calc}.

\begin{table}
\caption{$95\%$ confidence intervals obtained from a catalog of $10^5$ detections, with reference parameters used to generate the data. $\Delta X$ gives the width of the $95\%$ confidence interval.\label{tab:all-accuracy-data}}
\begin{ruledtabular}
\begin{tabular}{c c c c c}
Parameter & Reference value & $95\%$ conf.\ interval & ${\Delta}X$\\
\hline
$\sigma_{\rm{NS}} / M_{\odot}$ & 0.06 & [0.059688 , 0.060254] & 0.000566\\ 
$\mu_{\rm{NS}} / M_{\odot}$ & 1.35 & [1.347408 , 1.351789] & 0.00438\\
$w_0$ & -1.0 & [-1.036403 , -0.949623] & 0.0869\\
$w_a$ & 0.0 & [-0.195630 , 0.073602] & 0.269\\
$\alpha$ & -1.0 & [-1.026691 , -0.961659] & 0.0650\\
$\beta_1$ & 3.4 & [3.318136 , 3.605810] & 0.288\\
$\beta_2$ & 3.4 & [3.310287 , 3.582895] & 0.273\\
\end{tabular}
\end{ruledtabular}
\end{table}

\subsection{Including and accounting for errors}\label{sec:error-section}
\begin{table*}\scriptsize
\caption{$95\%$ confidence intervals derived from the reference sample ($4500$ detections), both in the case where distances are measured precisely, and when distance errors are included and accounted for (using the error averaging technique described in the text, with various numbers of points sampled from the distance posterior PDF). $\Delta X$ gives the width of the $95\%$ confidence interval.\label{tab:err-accuracy-data}}
\begin{ruledtabular}
\begin{tabular}{c | c c c c c c c}
Parameter & No errors & \multicolumn{2}{c}{Errors (50 pts)} & \multicolumn{2}{c}{Errors (100 pts)} & \multicolumn{2}{c}{Errors (400 pts)}\\
& $95\%$ conf.\ interval & $95\%$ conf.\ interval & $\Delta X/\Delta X_{\rm{ref}}$ & $95\%$ conf.\ interval & $\Delta X/\Delta X_{\rm{ref}}$ & $95\%$ conf.\ interval & $\Delta X/\Delta X_{\rm{ref}}$\\
\hline
$\sigma_{\rm{NS}} / M_{\odot}$ & [0.058785 , 0.061447] & [0.066378 , 0.071911] & 2.07851 & [0.063815 , 0.069409] & 2.10143 & [0.059309 , 0.064849] & 2.08114\\
$\mu_{\rm{NS}} / M_{\odot}$ & [1.339198 , 1.358745] & [1.329060 , 1.354066] & 1.27928 & [1.335499 , 1.361690] & 1.3399 & [1.335782 , 1.359339] & 1.20515\\
$w_0$ &  [-1.145894 , -0.749671] & [-0.880092 , -0.338642] & 1.36653 & [-1.146052 , -0.537588] & 1.53566 & [-1.116809 , -0.546566] & 1.4392\\
$w_a$ & [-0.917590 , 0.321901] & [-2.345082 , -0.452114] & 1.52722 & [-1.651230 , 0.463072] & 1.70578 & [-1.605299 , 0.377340] & 1.59956\\
$\alpha$ & [-1.207554 , -0.879888] & [-1.215388 , -0.863579] & 1.07368 & [-1.208005 , -0.874856] & 1.01673 & [-1.198613 , -0.890136] & 0.941437\\
$\beta_1$ & [2.730152 , 4.036099] & [2.851895 , 4.260085] & 1.07829 & [2.729217 , 4.038989] & 1.00293 & [2.771780 , 4.069150] & 0.99343\\
$\beta_2$ & [2.842474 , 4.059874] & [2.954584 , 4.274131] & 1.08391 &  [2.863813 , 4.080100] & 0.999088 & [2.887584 , 4.100009] & 0.995918\\
\end{tabular}
\end{ruledtabular}
\end{table*}
Distance measurements from a third-generation GW-interferometer network will not be error-free. While a network consisting of a single ET plus one other right-angle interferometer can place constraints on a source's sky location and luminosity distance, the precisions of these properties are improved to almost the $3$-ET network level by the inclusion of a second additional right-angle interferometer \citep{eliu-ET}. The redshifted chirp mass is expected to be very well constrained ($\lesssim 0.5\%$ error \citep{regimbau2012}), and so we ignore measurement errors in this parameter. We assume the error in the luminosity distance arising from instrumental noise scales as $\sim 1/\rho$, and include the effects of weak lensing as a further source of error. The weak-lensing error on luminosity distance measurements at $z\sim 1$ is approximately $5\%$, and we linearly extrapolate this to all other redshifts \citep{holz-linder-2005,kocsis2006,et-dark-energy,et-cosmography}. While several techniques have been proposed to reduce this weak-lensing error [e.g.\ Refs.\ \citep{hirata-2010,hilbert-gair-king-2011} and references therein], we assume no correction has been done, corresponding to a worst-case scenario. 

Errors on the distance-redshift relation from binary-system peculiar velocities are much smaller than instrumental and weak-lensing errors at all but the lowest redshifts, becoming comparable with these at $z\sim 0.1$ where the error is $\lesssim 1\%$, and decreasing sharply at higher redshifts [Ref.\ \citep{nishizawa2012} and references therein]. The lowest redshift in our reference catalog is $\sim 0.05$, where the peculiar velocity errors will dominate, but only lead to an error of $\lesssim 2\%$. The sensitivity of the luminosity distance to the dark-energy EOS parameters is very weak in this redshift regime; hence peculiar velocities are unlikely to introduce significant parameter bias/inaccuracy, and we ignore them here. 

We also ignore the effect of detector calibration errors, which, unlike statistical measurement uncertainties, would not be mitigated by boosting the detection rate. Such systematic biases have recently been studied in the case of advanced-era detectors \citep{calibration-errors-2011}, and found to induce a systematic shift in the estimated system parameters which is a small fraction of the statistical measurement errors. We ignore waveform-modelling errors in our analysis, since current post-Newtonian models will only break down close to the onset of the merger-phase, and for the neutron-star binaries considered in this analysis this is at frequencies where the instrumental noise is high and which therefore do not contribute much to the overall signal-to-noise of the system. Furthermore, luminosity-distance determination comes primarily from the network triangulation which will not be significantly affected by modelling uncertainties, and so the distance error will be dominated by instrumental-noise and weak-lensing, as discussed earlier. Similarly, the error in the distribution of possible source-redshifts arising from the measured redshifted chirp mass will be dominated by the intrinsic width of the NS mass distribution rather than the small error in the redshifted chirp mass coming from instrumental noise and modelling uncertainties.

We repeat the run of the working reference sample, offsetting the cataloged luminosity distance by an amount drawn from a Gaussian distribution, with mean at the true distance, and standard deviation,
\begin{equation}\label{eq:error-combine}
\sigma = D_L\times\sqrt{(1/\rho)^2 + (0.05z)^2}.
\end{equation}
The data collected for each event will actually be in the form of posterior probability density functions (PDFs) for the parameters, where previously we have assumed these are $\delta$-functions at the true values. If the offset luminosity distances are assumed to be the true values with a $\delta$-function posterior PDF, the chain does not move away from it's starting point. Hence we must take these errors into account in the likelihood calculation stage.

We can account for these errors in the analysis, by modifying the likelihood in Eq.\ (\ref{eq:cont-prob}) \citep{gair2010} to

\begin{align} \label{eq:error-integral}
{\mathcal{L}}(\overrightarrow{{\overrightarrow{\Lambda}}}|{\overrightarrow{\mu}},\mathcal{H})=&{e^{-N_{\mu}}}{\int}{\int}{\ldots}{\int}\left[p\left(\overrightarrow{n}={\overrightarrow{s}}-{\displaystyle\sum_{i}}{\overrightarrow{h_i}}({\overrightarrow{\lambda_i}})\right)\right. \nonumber\\
&\left.\times{\displaystyle\prod_{i=1}^{N_o}}r({\overrightarrow{{\lambda_i}}}|{\overrightarrow{\mu}})\right]\text{d}^{k}{\overrightarrow{\lambda_1}}\text{d}^{k}{\overrightarrow{\lambda_2}}{\ldots}\text{d}^{k}{\overrightarrow{\lambda_{N_o}}},
\end{align}
{\noindent}where, in our case, the number of parameters $k=2$, and ${\overrightarrow{s}}$ is the detector output, which is a combination of $N_o$ signals, ${\overrightarrow{h_i}}$, and noise, $\overrightarrow{n}$. Equation (\ref{eq:error-integral}) is an integral over all possible values of the source parameters that are consistent with the data. The first term inside the square bracket is the computed posterior PDF for the detected population of sources. Concern has been raised that the high event rate of ET detections may lead to a confusion background. However, the \textit{noise-power-spectral-density-weighted} signals are short enough that there is not expected to be significant overlap \citep{et-cosmography,regimbau2012}. Hence, these detections should be uncorrelated, with independent parameter estimates, and so this first term reduces to the product of the posterior PDFs for each detection.

If the posterior PDF for a given source has been obtained via MCMC techniques, then the integral in Eq.~(\ref{eq:error-integral}) may be computed by summing over the chain samples. Thus, errors may be accounted for by making the following replacement in Eq.\ (\ref{eq:cont-prob}):
\begin{equation}
r({\overrightarrow{{\lambda}_i}}|{\overrightarrow{\mu}})\longrightarrow{\frac{1}{{\mathcal{N}}_i}}{\displaystyle\sum_{j=1}^{{\mathcal{N}}_i}}r({\overrightarrow{{\lambda}_i}}^{(j)}|{\overrightarrow{\mu}}),
\end{equation}
{\noindent}where ${\mathcal{N}}_i$ is the number of points in the chain for the $i^{\rm{th}}$ source's PDF, and ${\lambda}_i^{(j)}$ is the $j^{\rm{th}}$ element of the discrete chain representing this PDF. This technique does not assume a specific form for the PDF, and can be used in the case of multimodal distributions.

We represent the $D_L$ posterior PDF for each source by a chain of $50$ points, drawn from a normal distribution with standard deviation as in Eq.\ (\ref{eq:error-combine}), and a mean equal to the value in the data catalog, which in this analysis, as discussed earlier, includes an error to offset it from the true value. Results are shown in Table \ref{tab:err-accuracy-data}. We see that a significant bias in the reconstructed model parameters still exists. We suspected that this bias arose from using only $50$ points to evaluate the distance posterior PDFs. We therefore repeated the analysis with an increasing number of points sampled from the distance posterior PDF.\footnote{The posterior distributions obtained via this analysis should be considered estimates of the true distributions, since the long likelihood computation time required by this error-analysis means that we did not collect as many samples as when errors are ignored. We performed burn-in runs, and then follow-up runs to estimate the posterior distributions as well as feasible.} With $100$ points, all biases are corrected expect for that in $\sigma_{\rm{NS}}$, and the ratio of the $95\%$ confidence interval widths to the reference widths is not significantly different from the $50$ point case. This suggests that a larger number of points in the error averaging technique will be necessary to correct all biases, but this is not neccesary to estimate parameter measurement accuracies in the presence of distance errors. Using $400$ points sampled from the distance posterior PDF all biases in the parameter posterior distributions appear to be corrected, in the sense that the reference parameters then lay within the $95\%$ confidence intervals of the $1$D marginalized posterior distributions.

Overall, we find that the result of properly accounting for instrumental and weak-lensing errors is that parameter measurement precisions are, at worst, approximately halved. Thus instrumental and weak-lensing induced errors should not affect our general conclusions about the science capabilities of a third-generation GW-interferometer network. We carry out the remainder of this study using catalogs which are generated and analyzed without including errors.

\subsection{Precision scaling with intrinsic parameters}
\begin{table*}
\caption{The expected detection rates for different choices of intrinsic parameters of the underlying distribution are compared to the reference expected detection rate. One parameter is varied at a time, with all other parameters kept fixed at their reference values. The variation of the expected detection rate with the intrinsic value of $\sigma_{\rm{NS}}$ is not shown, since this parameter is not used in the rate calculation (see Appendix \ref{sec:faster-detection-rate}).\label{tab:intrinsic-param-num-vary}}
\begin{ruledtabular}
\begin{tabular}{c c | c c | c c | c c | c c | c c | c c}
$\mu_{\rm{NS}} / M_{\odot}$ & $N/N_{\rm{ref}}$ & $w_0$ & $N/N_{\rm{ref}}$ & $w_a$ & $N/N_{\rm{ref}}$ & $\alpha$ & $N/N_{\rm{ref}}$ & $\beta_1$ & $N/N_{\rm{ref}}$ & $\beta_2$ & $N/N_{\rm{ref}}$ & $\beta_1=\beta_2$ & $N/N_{\rm{ref}}$\\
\hline
1.31 & 0.952 & -1.50 & 1.08 & -0.50 & 1.02 & -1.10 & 1.04 & 2.90 & 0.405 & 3.40 & 1.00 & 3.00 & 0.929\\
1.33 & 0.976 & -1.25 & 1.05 & -0.25 & 1.01 & -1.00 & 1.00 & 3.00 & 0.475 & 3.60 & 0.696 & 3.20 & 0.966\\
1.35 & 1.00 & -1.00 & 1.00 & 0.00 & 1.00 & -0.90 & 0.958 & 3.10 & 0.561 & 3.70 & 0.595 & 3.40 & 1.00\\
1.37 & 1.02 & -0.75 & 0.935 & 0.25 & 0.986 & -0.60 & 0.810 & 3.20 & 0.671 & 3.80 & 0.514 & 3.60 & 1.03\\
1.39 & 1.05 & -0.50 & 0.844 & 0.50 & 0.970 & -0.50 & 0.757 & 3.40 & 1.00 & 4.00 & 0.394 & 3.80 & 1.06\\
\end{tabular}
\end{ruledtabular}
\end{table*}
\begin{figure*}
   \subfloat[]{\incgraph{270}{0.31}{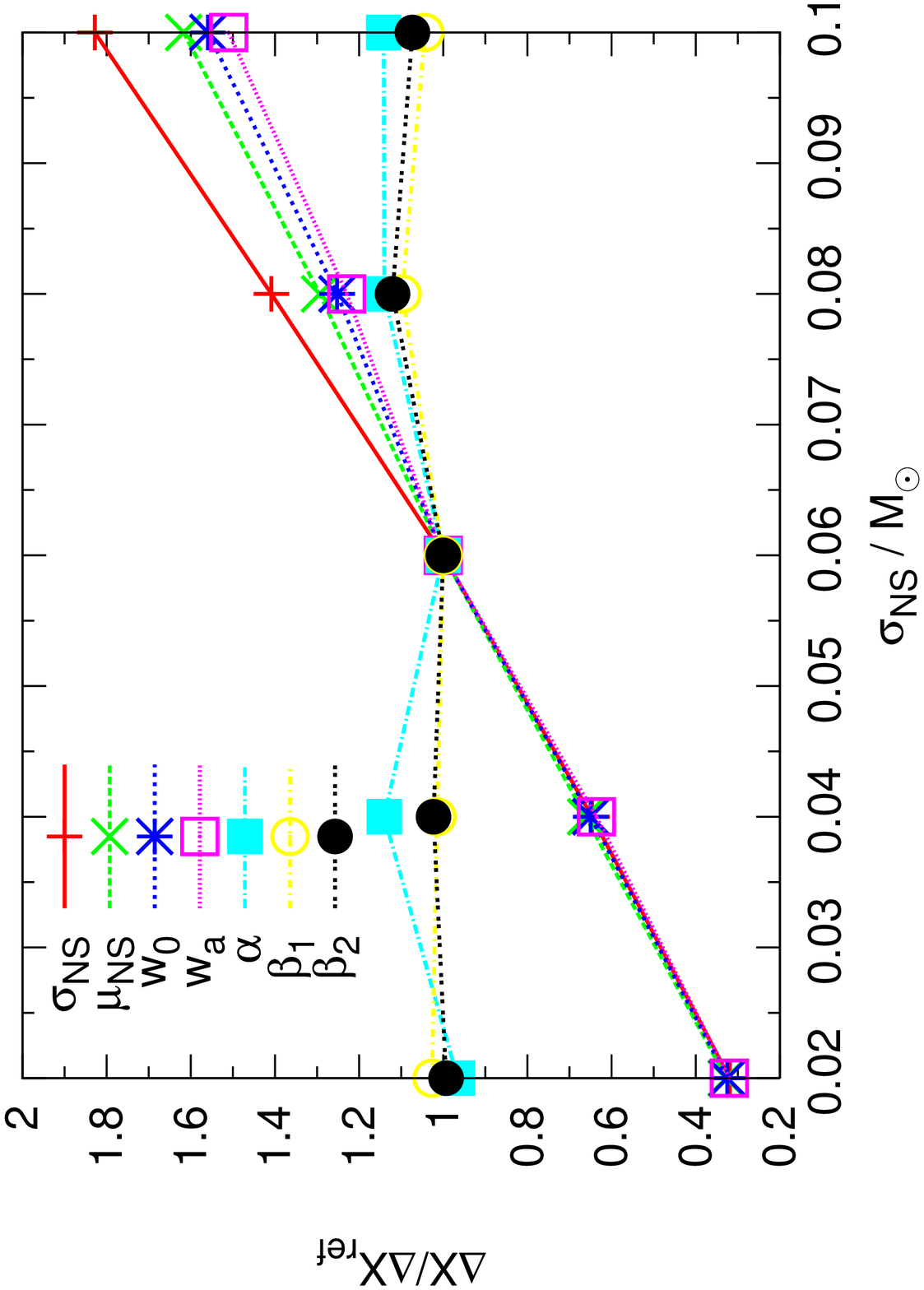}} %\hspace{-5mm}
   \subfloat[]{\incgraph{270}{0.31}{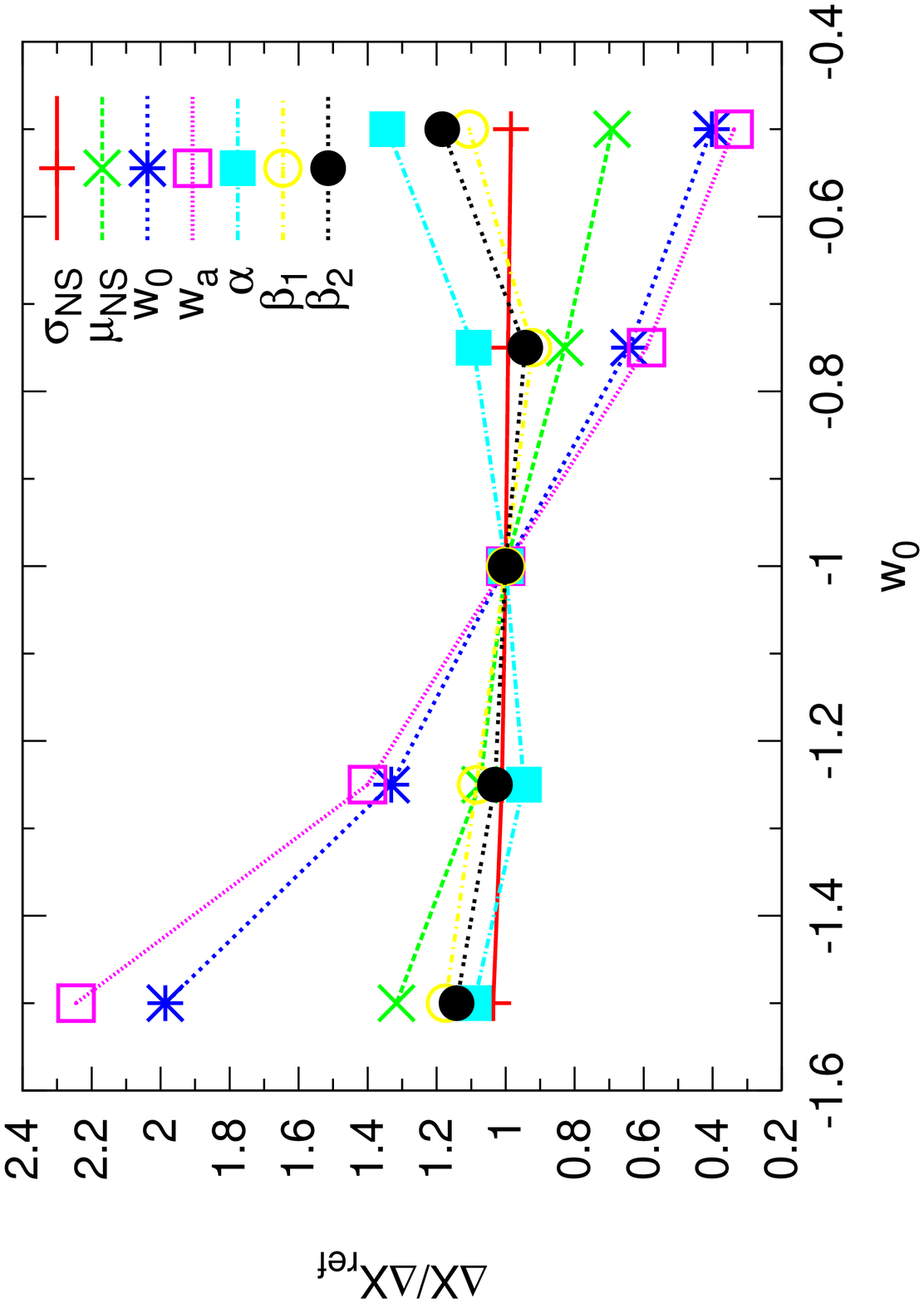}}  %\hspace{-5mm}
   \subfloat[]{\incgraph{270}{0.31}{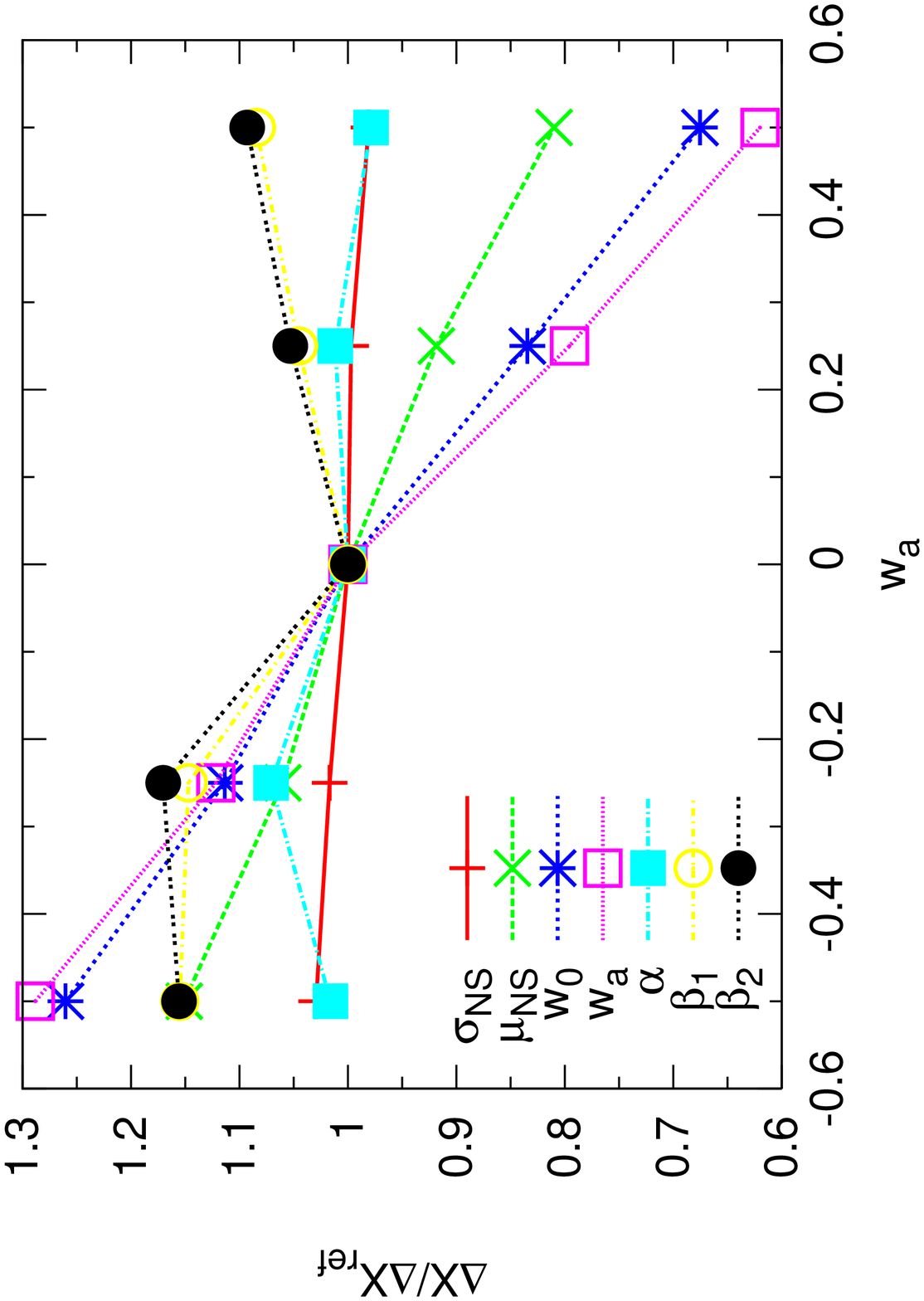}}\\
   \subfloat[]{\incgraph{270}{0.31}{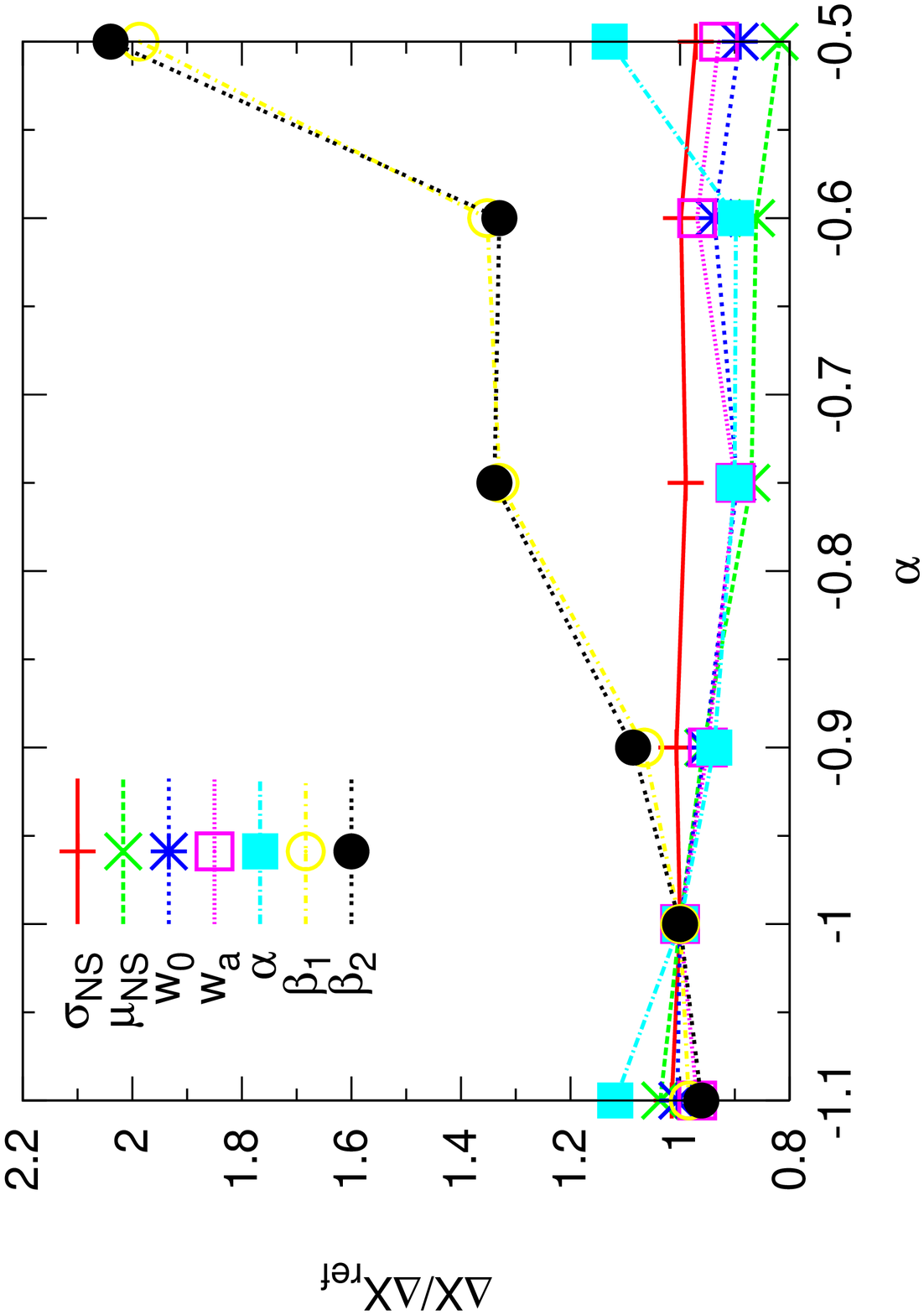}} %\hspace{-5mm}
   \subfloat[]{\incgraph{270}{0.31}{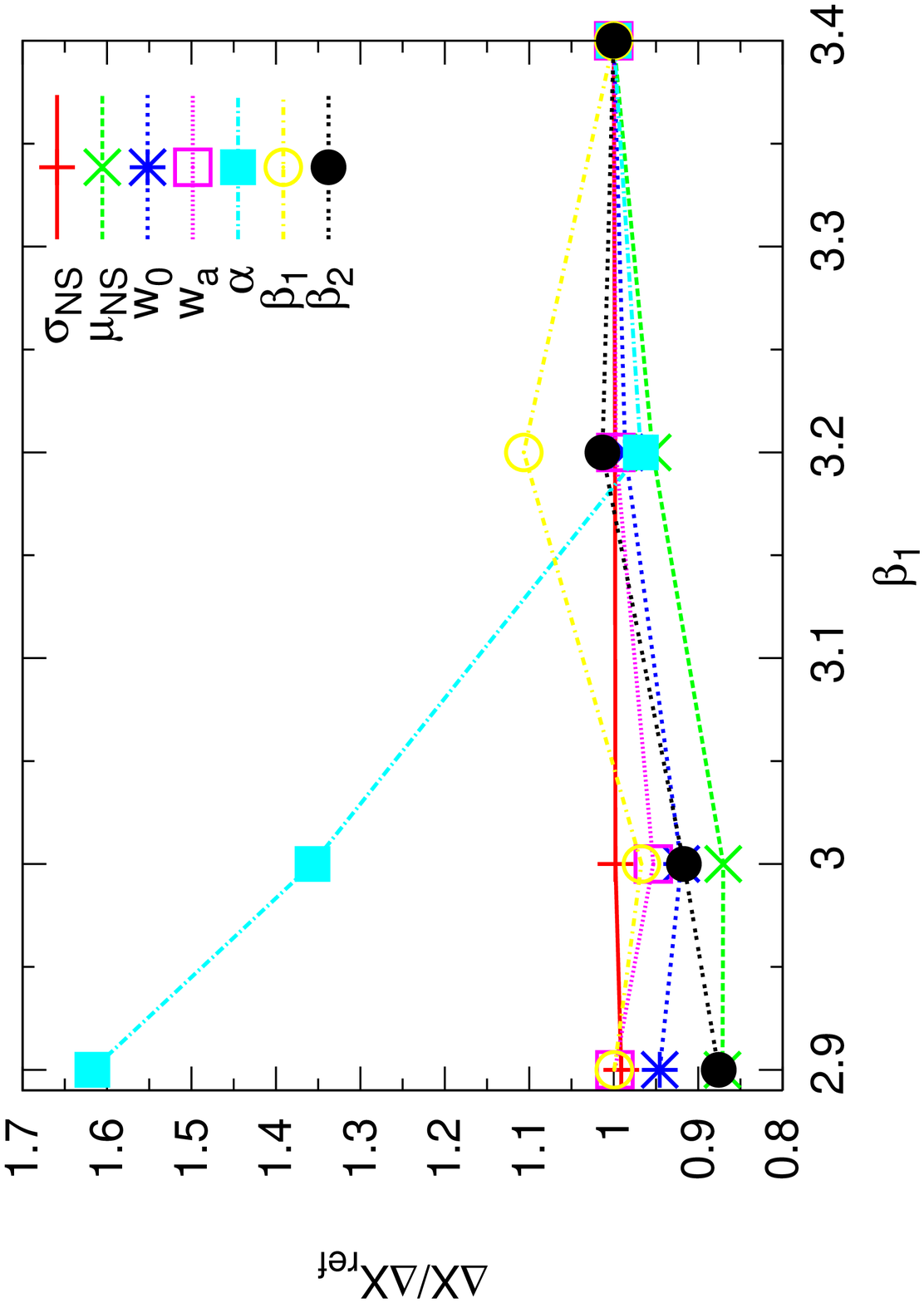}} %\hspace{-5mm}
   \subfloat[]{\incgraph{270}{0.31}{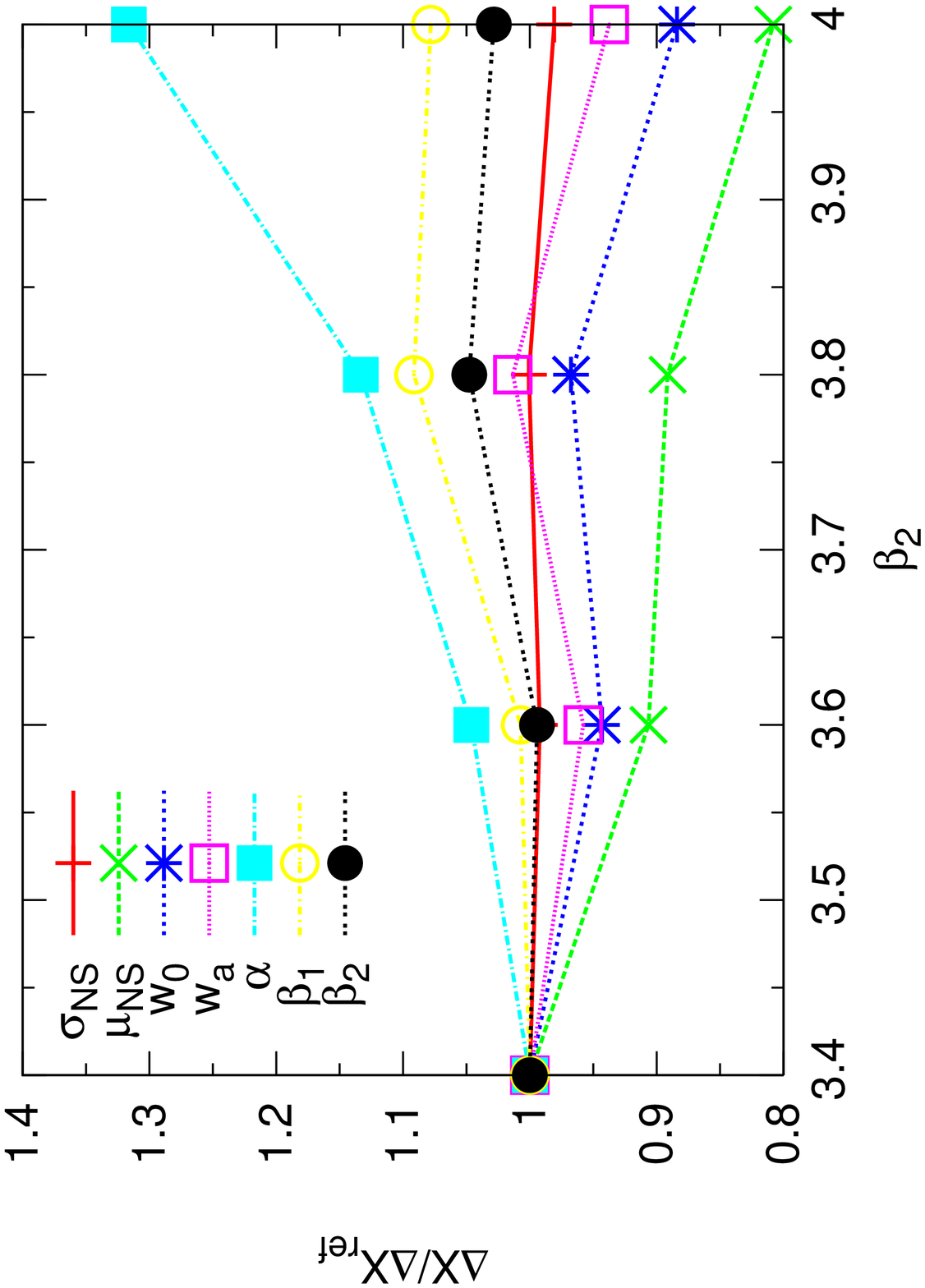}}
   \caption{The variation of measurement precision with different choices of the intrinsic parameters of the underlying distributions. One parameter is varied at a time, and in the interest of testing how the precision of parameter recovery is affected by the underlying \textit{distribution} of events, all catalogs are generated with the same number of events ($4500$ to match the reference catalog). Each point in each panel represents the average $95\%$ confidence interval width of $3$ realizations of the catalog.\label{fig:intrinsic-param-vary}}
 \end{figure*}
We now investigate how the ability of ET to constrain the parameters of the underlying distributions is affected by the values of the intrinsic parameters themselves. This is similar to the kind of analysis performed in our previous study with second-generation interferometers \citep{hwth2011}. We perform analyses of data catalogs generated with different intrinsic parameter combinations; multiple runs are performed on each parameter combination. We vary one parameter at a time, fixing all others to the reference values. 

Varying the intrinsic parameters with fixed SNR threshold will alter the expected detection rate. This is illustrated in Table \ref{tab:intrinsic-param-num-vary}, where the model with reference parameters has an expected detection rate of $\sim 10^5$ yr$^{-1}$. The expected $\sim1/\sqrt{N}$ relationship is well established, as shown in Fig.\ \ref{fig:num-vary}. Hence, we remove this number effect by generating catalogs with the same number of events ($4500$ each in order to compare against the reference catalog). Therefore we are testing how the cosmological, astrophysical and intrinsic-mass \textit{distributions} of coalescing DNS binaries impact the precision of parameter recovery.

The results of these analyses are shown in Fig.\ \ref{fig:intrinsic-param-vary}. We see that as $\sigma_{\rm{NS}}$ is increased the measurement precision of both the NS mass distribution and dark-energy EOS parameters decreases. We found a similar trend in Ref.\ \citep{hwth2011}. This makes sense, since if we have an intrinsically narrow NS mass distribution, then we have a good idea of what the intrinsic masses of the systems are and the range of candidate redshifts produced from a measured redshifted chirp mass will be narrow, improving the precision with which we can recover cosmological parameters.

A variation in the intrinsic $\mu_{\rm{NS}}$ (not shown) produces accuracies comparable to the reference accuracies. Hence, the impact of the intrinsic value of the NS mass-distribution mean on parameter accuracies is predominantly through the change to the expected detection rates i.e.,\ a larger mean, on average, will lead to larger chirp masses, so that detections can be made from a larger volume (see Eq.\ (\ref{eq:snr})).

Increasing the value of the EOS parameter $w_0$ increases the precision with which we can recover $w_0$, $w_a$ and $\mu_{\rm{NS}}$. As $w_0$ is increased, while the intrinsic $w_a$ is fixed at zero, the recovered posteriors for these parameters are squeezed by the prior restrictions, $w_0<-1/3$ and $(w_0+w_a)<-1/3$. A larger intrinsic $w_0$ increases the horizon distance of detections, which permits greater sensitivity to the dark-energy EOS parameters. Furthermore, a narrowed range of cosmological parameters means that the range of candidate redshifts is also narrowed, such that the precision of the recovered NS mass-distribution mean (deduced from the redshifted chirp mass) improves. We also notice these effects when the intrinsic $w_a$ is increased with the intrinsic $w_0$ fixed at the reference value. However, the effect is less pronounced in this case, since $w_a$ is a first-order correction to $w_0$.

As the power-law index, $\alpha$, is increased the average delay between the formation of the massive progenitor system and the merging of the final compact-system increases. This means that more systems formed at higher redshifts survive to merge at lower redshifts, and hence the merger-rate density is boosted to higher values at lower redshifts. In addition, as $\alpha$ increases the merger-rate density tracks the underlying SFR-density to a lesser extent, so it becomes more difficult to extract the details of the SFR-density. Hence the $\beta_{1,2}$ distributions widen to reflect this reduced sensitivity to the underlying SFR-density.

When we keep the intrinsic values of $\beta_1$ and $\beta_2$ equal (not shown), we find that varying these by $\pm 0.4$ around the reference value has a negligible impact on the measurement precision of the parameters. A higher common $\beta_{1,2}$ value leads to a larger expected detection rate, but this is a small effect.

Lowering the intrinsic value of $\beta_1$, with $\beta_2$ fixed, shifts the distribution of events to lower distances, and changes the shape of the underlying merger-rate density. This distribution is consistent with a wider range of $\alpha$ values than the reference distribution, since the sensitivity of the merger-rate density to $\alpha$ is reduced at lower redshifts. This causes the marginalized $\alpha$-posterior distribution to widen. The same is true when the intrinsic value of $\beta_2$ is increased.

\subsection{Varying the SNR threshold}\label{sec:threshold-change}
\begin{figure}
    \incgraph{270}{0.5}{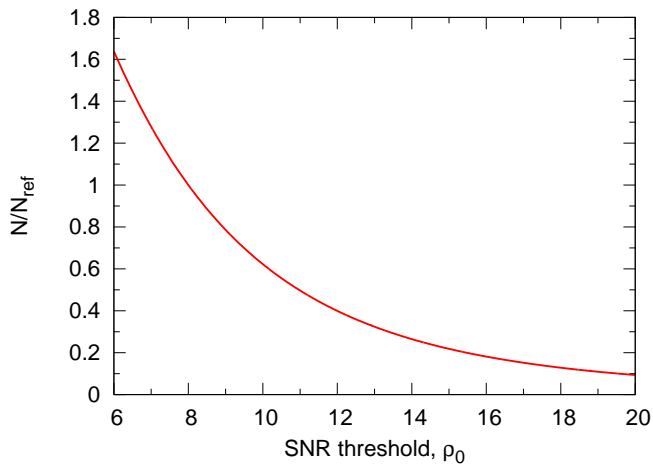}
   \caption{\label{fig:threshold_num}We show the variation of the expected detection rate as the SNR threshold, $\rho_0$, is raised. This can also be interpreted as \textit{lowering} the characteristic distance reach of the network. Since $\rho$ scales as $1/D_L$ (see Eq.\ (\ref{eq:snr})), and the difference between the luminosity distance and radial comoving distance becomes smaller at lower redshift, one would expect that at high enough values of $\rho_0$ the comoving detection volume (and hence detection rate) would scale as $1/\rho_0^3$. This is approximately valid for $\rho_0\gtrsim 15$.}
\end{figure}
\begin{figure*}
   \subfloat[]{\incgraph{270}{0.45}{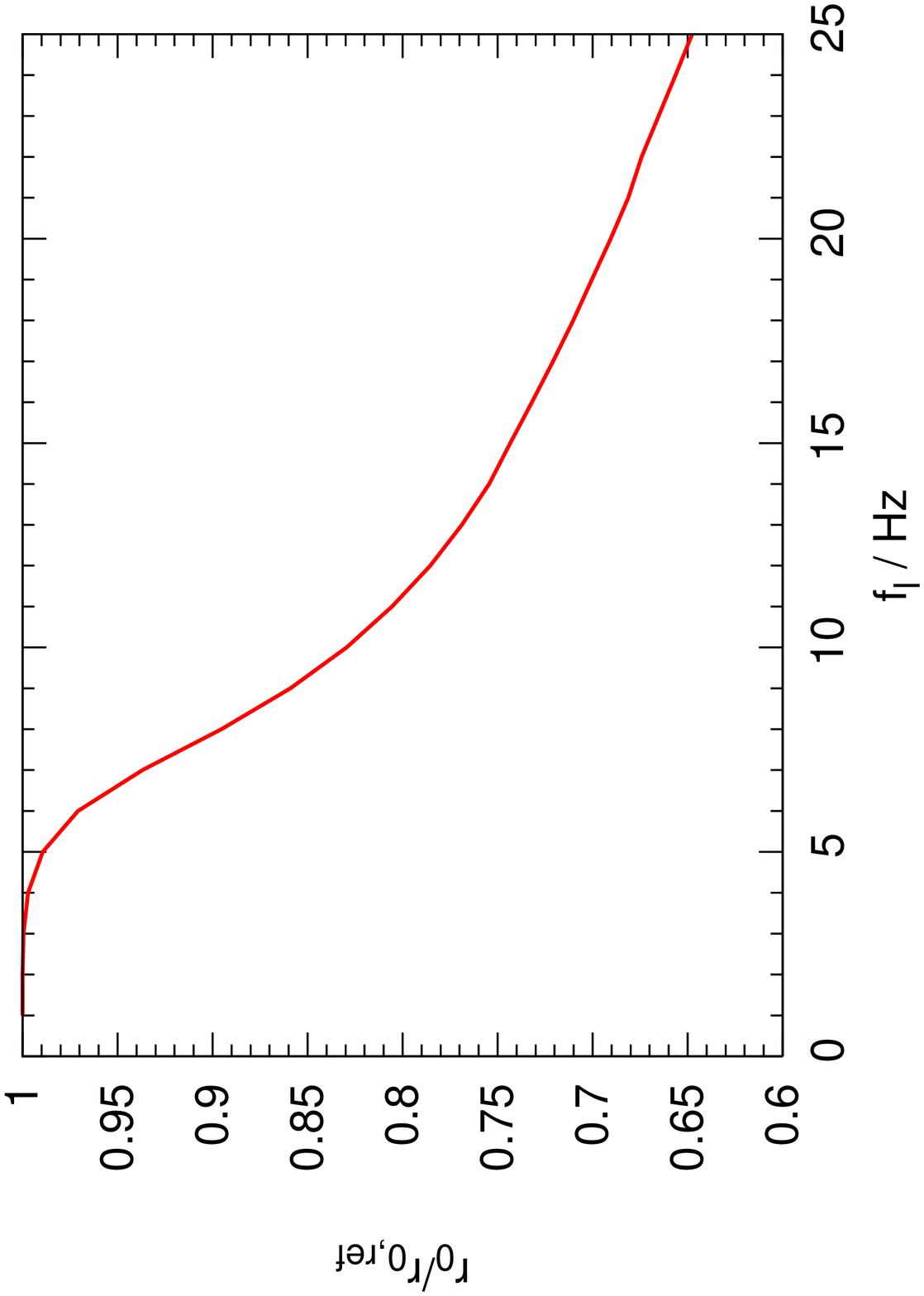}} %\hspace{-5mm}
   \subfloat[]{\incgraph{270}{0.45}{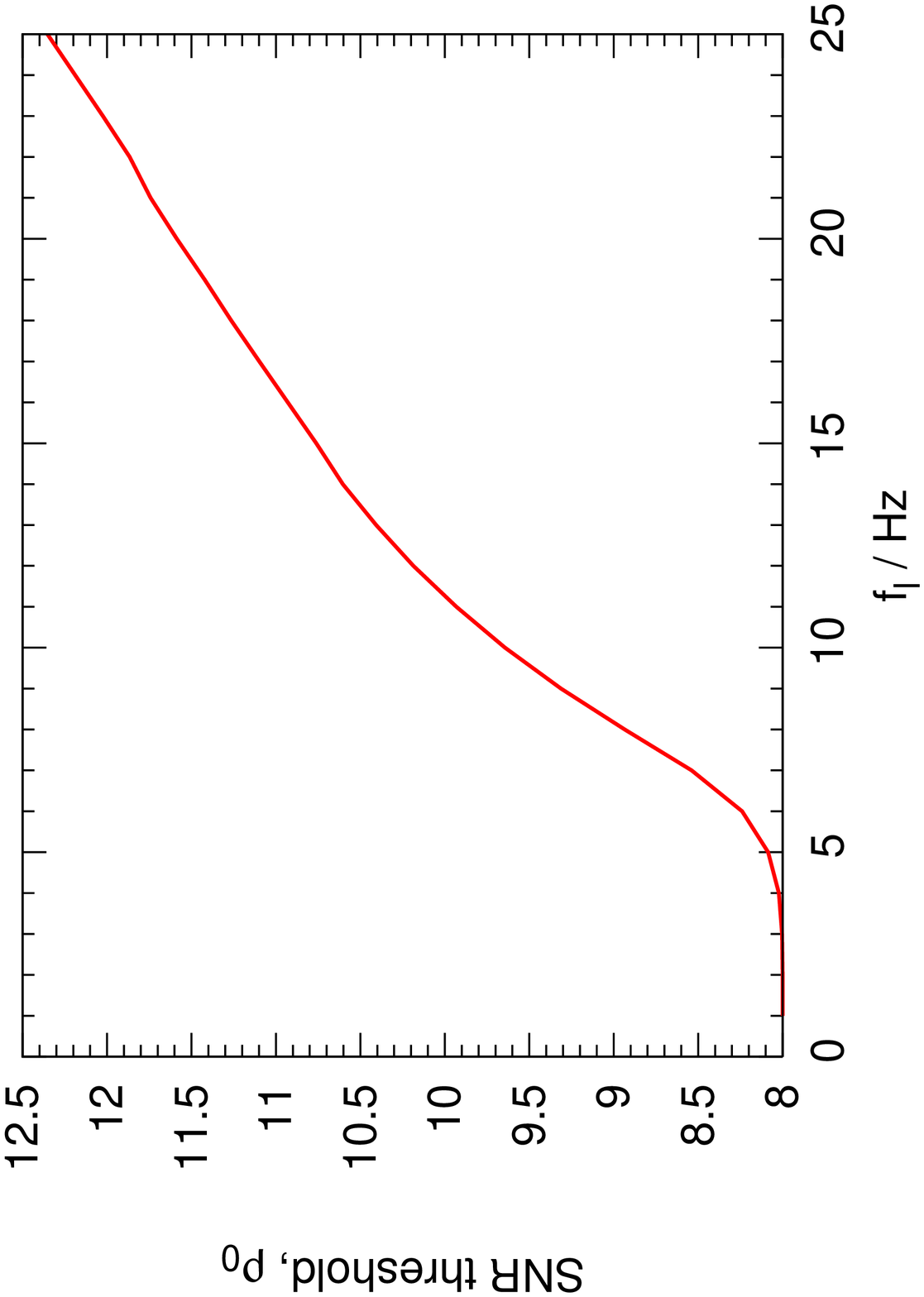}} 
   \caption{The reduction of the characteristic distance reach associated with raising the low-frequency cutoff, $f_l$, of $3^{\rm{rd}}$-generation detectors. This can also be interpreted, via Eq.\ (\ref{eq:rprob}), as raising the network SNR threshold. The figures were produced using the ET-D noise curve \citep{ET-site}.\label{fig:reach-snr-freq-cutoff}}
 \end{figure*}
\begin{figure*}
    \subfloat[\label{fig:merger-threshold}]{\incgraph{270}{0.45}{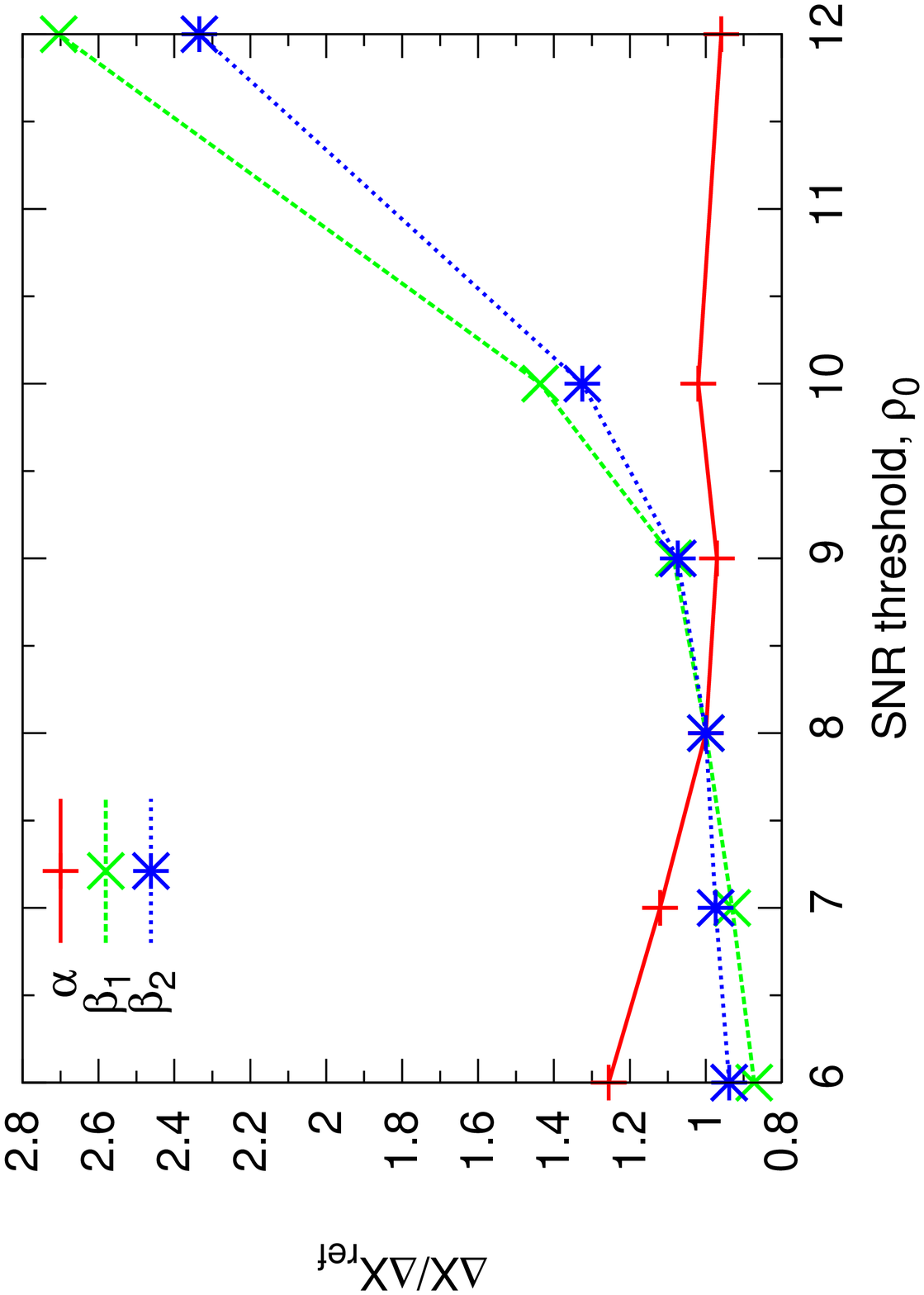}} %\hspace{-5mm}
    \subfloat[\label{fig:cosmo-threshold}]{\incgraph{270}{0.45}{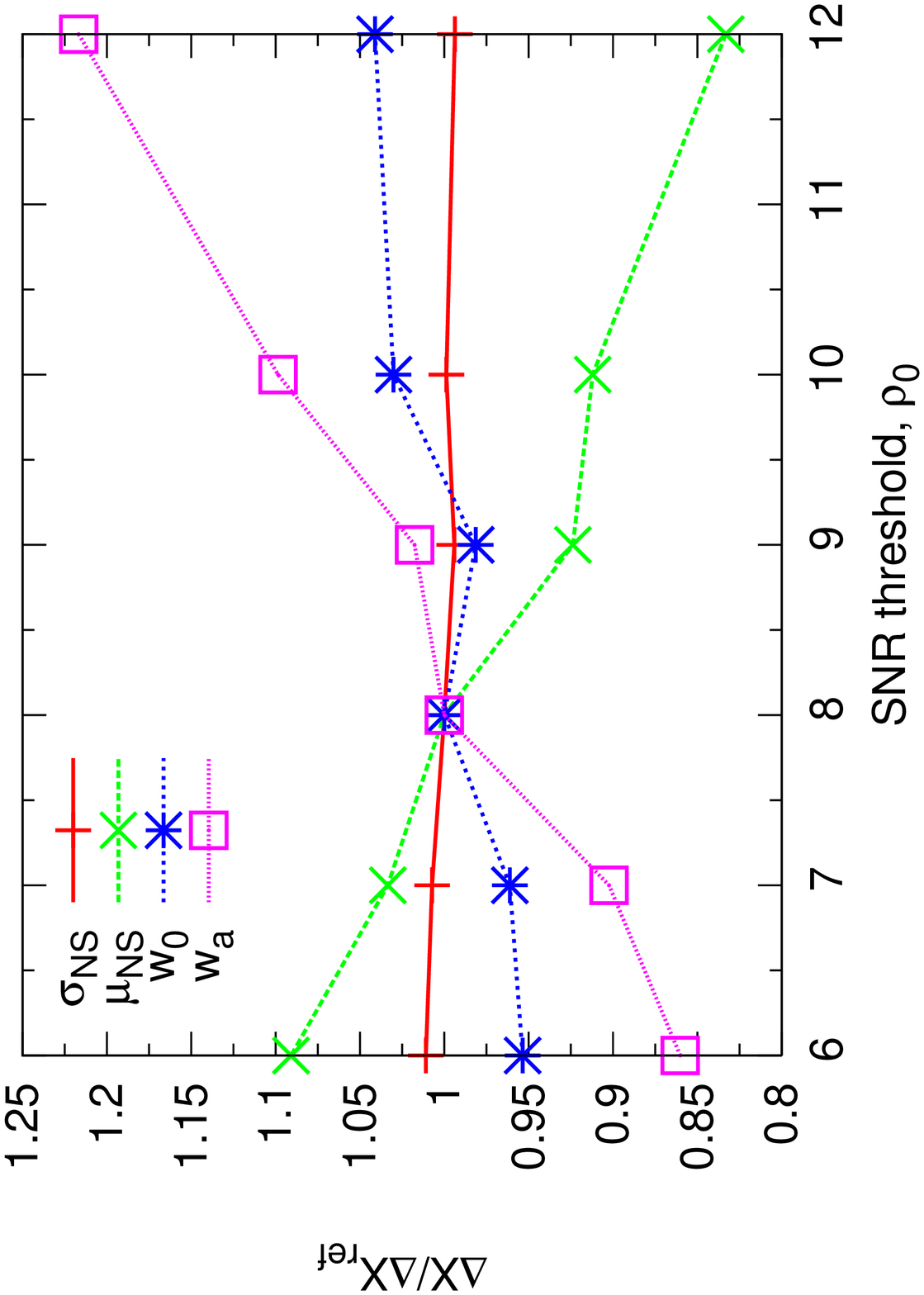}} 
   \caption{The variation of the parameter measurement precisions with the network SNR threshold. The left panel shows precisions, characterized by the width of the $68\%$ confidence intervals, for the merger-rate density parameters, while the right panel shows precisions, characterized by the $95\%$ confidence intervals used elsewhere, for all other parameters. We use the narrower confidence intervals for the merger-rate parameters to mitigate the effct of poor sampling in the low-$\alpha$ region which was observed in some AM-MCMC chains in this analysis. All catalogs contain the same numbers of events at each threshold value, which, as in the previous subsection, is $4500$ to match the reference catalog.\label{fig:snr-threshold-vary}}
\end{figure*}
\begin{figure}
    \incgraph{270}{0.5}{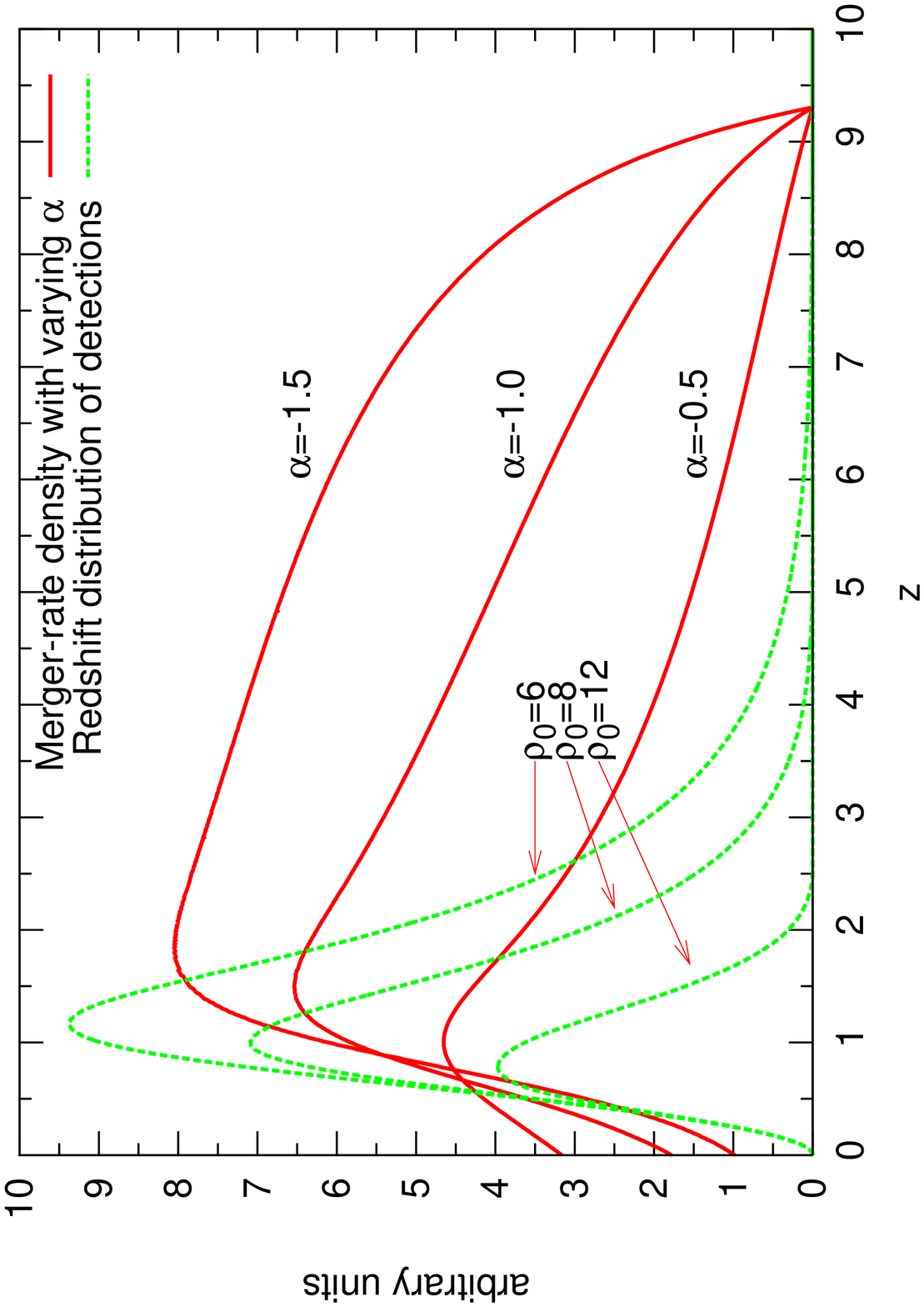}
   \caption{\label{fig:distribution-merger-compare}We show the redshift distribution of the DNS merger-rate density for various choices of the power-law index of the delay-time distribution, $\alpha$ (all other parameters are fixed at their reference values). The merger-rate density is relatively featureless beyond $z \sim 2.5$, making it difficult for our analysis (which is insensitive to linear scalings of the merger-rate density) to discriminate between values of $\alpha$ in this range. Overlaid on this figure we show the redshift distribution of detections for various SNR thresholds.}
\end{figure}
We also generated catalogs for different SNR thresholds defining the detectability of merging systems. Multiple catalogs were analyzed for each SNR threshold, but once again the number of events was fixed at $4500$ to match the reference catalog (see Fig.\ \ref{fig:threshold_num}). An increase in the SNR threshold is equivalent to the characteristic distance reach of the detectors decreasing. Hence we would expect the sensitivity of the data to varying dark-energy EOS parameters, which have a greater influence at larger redshifts, to be reduced. A reduced characteristic reach would also result from a larger low-frequency cutoff, $f_l$, in the detector's noise power spectral density. In the recent mock ET data challenge, \citet{regimbau2012} found that confusion between two or more signals rarely affected the analysis performance when $f_l=25$ Hz. Standard algorithms currently employed for LSC-Virgo analyses cannot handle templates longer than a few minutes; however multiband filter methods are being developed which will allow $f_l$ to be pushed below $25$ Hz. In Fig.\ \ref{fig:reach-snr-freq-cutoff} one can see that with $f_l=25$ Hz, the effective SNR threshold is raised from the reference value of $8$ (with $f_l=1$ Hz) to $\sim 12.4$.

From Fig.\ \ref{fig:snr-threshold-vary}\subref{fig:merger-threshold}, we see that as the SNR threshold is increased, with the number of cataloged events fixed, the accuracies of $\beta_1$ and $\beta_2$ degrade sharply. At higher SNR thresholds (or, equivalently, at lower distance reaches) the sensitivity of the merger-rate density to varying $\beta_{1,2}$ is reduced; hence the wider posterior distributions. The measurement precision of $\alpha$ increases slightly as the SNR threshold is increased from $6-12$. One might expect $\alpha$ to show the same trend as $\beta_1$ and $\beta_2$, since an increasing SNR threshold pushes the events to lower redshifts where the sensitivity of the merger-rate density to $\alpha$ is reduced. However, we see in Fig.\ \ref{fig:distribution-merger-compare} that the merger-rate density, for various choices of $\alpha$ (but all other parameters fixed), is relatively featureless beyond $\sim 2.5$. The distribution of the merger-rate density in the redshift window of $\sim 2.5-7$ could be approximately linearly scaled to satisfy a large range of $\alpha$.\footnote{The same argument did not apply when $\beta_{1,2}$ was varied, since this not only shifted the distribution to lower redshifts but altered the \textit{shape} of the merger-rate density in a way that could not be equated with a linear scaling in any redshift window.} Therefore, given that our likelihood statistic is insensitive to linear scalings of the merger-rate density [see Eq.\ (\ref{eq:new-stat})], the significant number of high-redshift detections in a $\rho_0=6$ catalog will widen the $\alpha$ posterior distribution, while most $\alpha$-information is found in the redshift window $\sim 1-2$, where the merger-rate density has more features.

In Fig.\ \ref{fig:snr-threshold-vary}\subref{fig:cosmo-threshold}, we see that the measurement accuracy of $w_0$ and $w_a$ is slightly reduced for higher SNR thresholds; this is a small effect, and is expected with a catalog shifted to lower redshifts, where distances are less sensitive to varying dark-energy EOS parameters. The accuracy of $w_0$ only varies by $\sim\pm 5\%$, since we remain sensitive to detections at tens of Gpcs even with an SNR threshold of $12$. However, $w_a$ shows a stronger variation since it is a higher order correction to the EOS parameter, and distances become sensitive to this parameter at higher redshifts than they do to $w_0$. We also see that the measurement precision of $\mu_{\rm{NS}}$ and $\sigma_{\rm{NS}}$ is increased slightly as we move to larger SNR thresholds. This small effect is probably due to the fact that a lower redshift range in the data catalog will mean that the redshifted chirp mass is closer to the intrinsic chirp mass.
\begin{figure}
   \incgraph{270}{0.5}{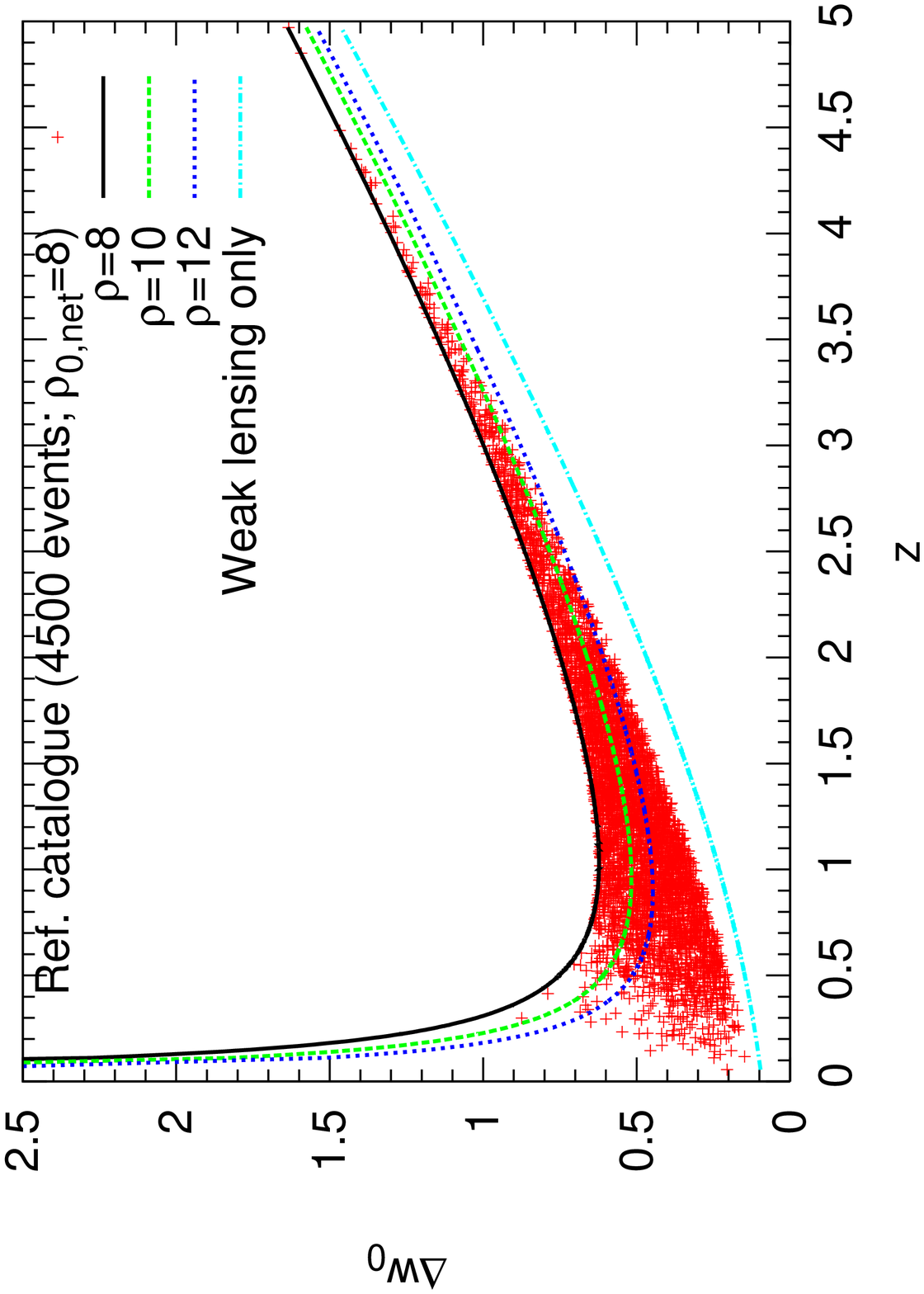}
   \caption{\label{fig:w0-sens-redshift}The redshift dependence of the sensitivity of the luminosity distance to $w_0$. This parameter has a very weak intrinsic impact on $D_L$ at low redshifts, whilst distance errors from instrumental noise and weak-lensing dominate at higher redshifts. This results in a redshift ``sweet-spot'', where these effects are minimised for lines of constant SNR. We also plot the individual $\Delta w_0$ values calculated for the reference catalog events.}
 \end{figure}

Although this suggests that a greater distance reach will improve the precision of cosmological parameter recovery, we have so far ignored distance errors. In fact, instrumental and weak-lensing errors impart an interesting redshift evolution to the $w_0$-sensitivity, which we approximate as \citep{wambsganss-1996}
\begin{equation}
\Delta w(z)\sim\bigg\vert\frac{\partial w_0}{\partial D_L}\bigg\vert\times D_L\times\sqrt{(1/\rho)^2 + (0.05z)^2}.
\end{equation}
In Fig.\ \ref{fig:w0-sens-redshift} we see that the sensitivity of the luminosity distance to the cosmological parameter $w_0$ is greatest at $z\sim 1$, since $w_0$ has a very weak intrinsic impact on $D_L$ at low redshifts and distance errors dominate at higher redshifts. Increasing a detector's distance reach will raise the fraction of high-redshift cataloged events. We calculate the effective measurement precision of $w_0$ from our reference catalog by adding the $\Delta w_0$ values from each event in quadrature i.e.\ $1/{\Delta w_{0,\rm{eff}}}=\sqrt{\sum (1/{\Delta w_{0,i}})^2}$. This is repeated for various lower and higher SNR threshold values. We perform these calculations for catalogs containing the same number of events ($4500$ to match the reference catalog), and for catalogs with the number of events scaled by the ratio of the expected detection rate for each SNR threshold to the reference threshold (which in this analysis is $8$). The results are shown in Table \ref{tab:w0-sweet-spot}, where we see that for catalogs with the same number of events, lowering the SNR threshold actually worsens the precision of $w_0$ recovery since the distribution of events is weighted to higher redshifts, where distance errors degrade the precision. Increasing the SNR threshold reduces the number of events at high redshift and hence mitigates the degradation of precision due to distance errors (see Fig.\ \ref{fig:w0-sens-redshift}). However, this effect slows down with increasing SNR threshold. For catalogs with numbers of events scaled to match the expected detection rate for each SNR threshold, we see that the increased number of events associated with a lower SNR threshold is enough to compensate for degradation of precision from higher redshift events. However this loss of precision means that lowering the SNR threshold does not lead to the $1/\sqrt{N}$, or $1/\rho_0^{3/2}$ improvement in parameter measurement precision which one would naively expect.
\begin{table}\scriptsize
\caption{The events from catalogs with different SNR thresholds are used to compute an effective $w_0$ precision, by adding the $\Delta w_0$ values of each event in quadrature. This analysis is performed for catalogs with the same number of events, and for catalogs with the numbers of events scaled to match the expected detection rate for each SNR threshold.\label{tab:w0-sweet-spot}}
\begin{ruledtabular}
\begin{tabular}{c c c c}
\multirow{2}{*}{$\rho_{0,\rm{net}}$} & \multirow{2}{*}{$f=N/N_{\rm{ref}}$} & \multicolumn{2}{c}{$\Delta w_{0,\rm{eff}} / 10^{-3}$}\\
& & ($N_o=N_{\rm{ref}}$) & ($N_o=f\times N_{\rm{ref}}$)\\
\hline
6 & 1.64 & 8.08 & 6.33\\
8 & 1.00 & 6.82 & 6.82\\
12 & 0.399 & 5.33 & 8.35\\
20 & 0.0936 & 4.00 & 13.1\\  
30 & 0.0271 & 3.38 & 20.6
\end{tabular}
\end{ruledtabular}
\end{table}

Finally we address the issue of having assumed that Earth motion does not modulate the antenna patterns of the detectors. The time spent ``in-band'' by an inspiraling-event scales as \citep{regimbau2012}
\begin{equation}
\tau\sim 5.4\left(\frac{\mathcal{M}_z}{1.22M_{\odot}}\right)^{-5/3}f_l^{-8/3}\quad\text{days}.
\end{equation}
Hence, a detector with a low-frequency cutoff of $1$ Hz (as we have assumed) could have events in band for as long as $\sim 5$ days. In this case, a correct treatment of the antenna pattern modulation would be needed. However, if we increase $f_l$ to $\sim 8$ Hz, then the maximum time spent in-band is less than $30$ minutes, and ignoring the antenna pattern modulation is reasonable. In Fig.\ \ref{fig:reach-snr-freq-cutoff} we see that a low-frequency cutoff of $8$ Hz is equivalent to raising the SNR threshold to $\sim 9$, and from Fig.\ \ref{fig:snr-threshold-vary} we see that the precision of parameter recovery is within $\sim\pm10\%$ of the reference precisions for an SNR threshold of $9$. Therefore our approximate treatment of the network antenna patterns would seem reasonable.

\section{Conclusions}\label{sec:conclusions}

We have built on our previous work \citep{hwth2011} which explored the capabilities of an advanced (i.e.\ second-generation) GW-interferometer network to constrain aspects of the NS mass distribution in DNS systems, as well as the Hubble constant. The technique we employed used only information obtained via analysis of the GWs detected in such a network. In this paper we extended the analysis to a possible third-generation network, consisting of the proposed Einstein Telescope, and complemented by third-generation right-angled interferometers at LIGO Livingston and LIGO-India. The target for the Einstein Telescope is a broadband factor of $10$ sensitivity increase with respect to advanced detectors, but to also extend the low-frequency sensitivity of ground-based GW interferometers below $10$ Hz. The current design for a single ET consists of three overlapping interferometers, arranged in an equilateral configuration with arm-opening angles of $60^{\circ}$ \citep{triple-michelson2009,sathy-et-potential2011,3rd-gen-science-reach2010}. Each interferometer will consist of a cryogenically-cooled, underground low-frequency detector, and a high laser-power, high-frequency detector in a ``xylophone'' configuration -- these two detectors work in tandem to suppress noise over the entire band \citep{ET-xylophone-2010,sensitivity-studies2011}. Current projections for funding and construction of ET place ``first-light'' sometime in the mid-2020's.

The sources of interest in this paper are inspiraling double NS systems, which could be observed at rates of $\sim 40$ yr$^{-1}$ by advanced detectors \citep{abadie-rate2010} and rates of $\mathcal{O}(10^5-10^6)$ yr$^{-1}$ may be achieved by a third-generation network \citep{3rd-gen-science-reach2010,regimbau2012,et-cosmography}. These sources are commonly referred to as \textit{self-calibrating standard sirens}, since their distance from us is directly encoded in the emitted GWs. Combined with a method of redshift determination, these sources can be used to probe the distance-redshift relation and hence extract constraints on background cosmological parameters which are independent of the cosmic distance ladder \citep{Schutz86,hughes-holz-2003,et-cosmography,et-dark-energy}.

Our method of cosmography using only GWs relies on the narrowness of the distributions of masses of NSs in these DNS systems. Recent analysis indicates that this mass distribution is indeed narrow, with a Gaussian mean of $\sim 1.35M_{\odot}$ and standard deviation of $0.06M_{\odot}$, which may be a product of a distinct evolutionary path \cite{kiziltan2010,valentim2011,ozel2012}. Using a measurement of a source's redshifted chirp mass, we can therefore obtain a narrow candidate redshift distribution. A narrower intrinsic NS mass distribution will obviously mean the precision of redshift determination increases. We can combine these with GW-interferometer network determinations of the luminosity distance to constrain cosmological parameters. 

We used a Bayesian theoretical framework to assess the capability of a third-generation network to measure cosmological and astrophysical parameters. We performed $7$-dimensional adaptive MCMC analysis on the catalogs of detections, using reference parameters $H_0=70.4$ kms$^{-1}$Mpc$^{-1}$, $\Omega_{m,0}=0.2726$, $\Omega_{k,0}=-0.0006$, $w_0=-1$, $w_a=0$, $\mu_{\rm{NS}}=1.35M_{\odot}$ and $\sigma_{\rm{NS}}=0.06M_{\odot}$. Keeping $H_0$, $\Omega_{m,0}$ and $\Omega_{\Lambda,0}$ fixed, we found that the measurement precisions of the dark-energy EOS parameters possible with a $10^5$-event catalog were of the same order of magnitude as forecasted constraints from future CMB$+$BAO$+$SNIa measurements \citep{et-dark-energy}. Furthermore the power-law index of the merger delay-time distribution, $\alpha$, and the parameters of the underlying star-formation-rate (SFR) density were constrained to within $\sim 10\%$. Accounting for measurement errors degraded precisions by a factor of $\lesssim 2$, while increasing the network SNR threshold required for detection from $8$ to $9$ (which is equivalent to considering only the $\sim30$ minute section of inspiral above $8$ Hz) changed the precisions by only $\sim10\%$.

We also investigated how the precision of parameter recovery scaled with the values of the intrinsic parameters themselves, keeping the number of detected events fixed to factor out pure number-of-event effects. Varying the intrinsic $\sigma_{\rm{NS}}$ showed a linear scaling of parameter precision, with narrower intrinsic NS mass distributions favouring tighter parameter constraints. The precisions of the merger-rate density parameters did not appear to be affected in this case. Increasing the intrinsic $w_0$ and $w_a$ had the effect of increasing their measurement precision, as well as that of $\mu_{\rm{NS}}$. This was probably due to the fact that larger $w_0$ and $w_a$ give detections out to greater distances, where the sensitivity to these parameters is higher. Tighter cosmological constraints implies narrower candidate redshift distributions from the cataloged distances, which improves $\mu_{\rm{NS}}$ precision. Increasing the intrinsic value of $\alpha$ meant that the merger-rate density tracked the underlying SFR density to a lesser extent and hence worsened the precision of SFR-density parameter recovery. As we changed the shape of the underlying SFR density to favor closer detections, the measurement precision of $\alpha$ worsened, since the sensitivity of the merger-rate density to $\alpha$ is lower at lower redshifts.

Finally, we varied the criterion for a network detection, which we denoted by a threshold value of the network signal-to-noise ratio. This could also be interpreted as varying the characteristic distance reach of the network, which, in turn, could be caused by varying the detector's low-frequency cutoff. Varying the SNR threshold between $6-12$ caused a slight decrease in $w_0$ and $w_a$ precision, as catalogs with lower distance events will be less sensitive to these cosmological parameters. However, catalogs with, on average, closer events will provide better NS mass-distribution parameter precision, since the redshifted chirp mass will be less offset with respect to the intrinsic chirp mass. Increasing the SNR threshold, and hence decreasing the characteristic distance reach of the network, caused a significant decrease in SFR-density parameter precision, since the merger-rate density is less sensitive to the SFR-density parameters at lower redshift. 

While the sensitivity of distances to the dark-energy EOS will obviously be intrinsically weak at low redshifts, distance-measurement errors begin to dominate at higher redshifts. We found that for a fixed number of events in a catalog, lowering the SNR threshold actually worsened the precision of $w_0$ recovery since events are weighted to higher redshifts, where distance errors degrade the measurement precision. The larger expected detection rate associated with lower SNR thresholds is enough to reverse this effect, but means that lowering the SNR threshold (or increasing the network's distance reach) does not lead to the great improvement in parameter measurement precision which one would naively expect.

We have not considered association of GW detections with an EM counterpart, either through precise sGRB \citep{et-cosmography,et-dark-energy} or host-galaxy association. The latter technique may only be possible with $\sim 0.01\%$ of detectable GW events \citep{nishizawa-phase-shift-2012}. However, the technique we have used has been shown previously \citep{hwth2011} to be well complemented by precision redshift information. In particular, we found that if redshift information (measured to much greater precision than the luminosity distance) is available for $\sim 10\%$ of a GW-event catalog, then measurement precisions of parameters were more than doubled.

This paper completes our proof-of-principle study of this GW-only cosmographic technique. We have shown the significant potential for a third-generation network including the Einstein Telescope to place interesting constraints on the NS mass distributions in DNS systems, the dark-energy EOS, the average delay between the formation of the DNS-system progenitors and the final merger and the underlying SFR density in the Universe. Over the following decade tighter constraints will be derived for the NS mass distribution, delay-time distribution of DNS systems, and the SFR density, which can be readily incorporated within this technique. We intend to test this technique in the upcoming ET mock data challenge, as well as study the ability of this technique to discriminate between NS mass distributions from different metallicity progenitors, different delay-time distributions resulting from different formation paths, and possibly multimodal NS mass distributions. Unshackling GW cosmography from its reliance on EM counterparts will be an important step in establishing DNS systems as physical distance indicators, and contribute to GW-analysis becoming a precision astrophysical tool. 

\begin{acknowledgments}
S.R.T is supported by the STFC. J.R.G is supported by the Royal Society. We thank Ilya Mandel for useful discussions and proofreading. This work was performed using the Darwin Supercomputer of the University of Cambridge High Performance Computing Service (http://www.hpc.cam.ac.uk/), provided by Dell Inc. using Strategic Research Infrastructure Funding from the Higher Education Funding Council for England.

\end{acknowledgments}
\appendix
\section{DNS Merger-rate density}
Here, we provide a more detailed discussion of the astrophysics of DNS mergers and justification for the ansatz we employ for the merger-rate density.
\subsection{Merger-delay distribution, $dP_m/dt$} \label{appsec:merger-delay} 
Assuming the number, $N$, of DNS binaries born with separation $a$ follows $dN/da\propto a^{\gamma}$ \citep{totani1997}, the merger-delay distribution is
\begin{equation}
\frac{dP_m}{dt}\propto\frac{dN}{d\tau_{\rm gr}}=\frac{dN}{da}\frac{da}{d\tau_{\rm gr}}\propto t^{\gamma/4}t^{-3/4} = t^\alpha.
\end{equation}
An early suggestion by Piran was to consider newly formed DNS binaries as having the same orbital separation distribution as normal-abundance main-sequence stars \citep{piran1992,abt1983}. For normal-abundance main-sequence binary systems, the distribution of periods has been found to be flat in $\ln{(P)}$ (where $P$ is the binary period) \citep{abt1983}, or lognormal \citep{duquennoy1991,raghavan2010}. If we follow Ref.\ \citep{piran1992}, and ignoring the progenitor ellipticity distribution, then the initial DNS orbital separation distribution will be flat in $\ln{(a)}$ (where $a$ is the semi-major axis of the binary orbit). Therefore $\gamma=-1$ and $dP_m/dt\propto t^{-1}$.\footnote{The caveats here are that ellipticity can have a significant effect on inspiral timescales, and it is not obvious that DNS systems should have the same orbital separation distribution as main-sequence binaries, since the two supernovae the systems survive would likely modify it. Furthermore the distribution functions for progenitor evolutionary timescales and merger timescales are not independent. The evolutionary timescale depends on the mass of the progenitor system components and the gravitational inspiral timescale depends on the chirp mass of the system. Strictly speaking the joint probability consideration should be considered \citep{schneider2001}; however we ignore this subtlety here.}

The catalog of DNS systems in Ref.\ \citep{champion2004} is used by Refs.\ \citep{guetta2005,guetta2006} to estimate the merger-delay distribution of observed systems; it approximately follows $(1/t)$, but there appears to be an excess of systems below an inspiral time of $100$ Myr. Selection effects having to do with the difficulty of detecting  binary pulsars in close orbit, due to the large and rapidly varying Doppler shift, may significantly affect the reconstructed merger-delay distribution below a few hundred Myr. The authors comment that with such a small sample of systems ($\sim 6$) it is difficult to make predictive conclusions about this distribution, but that it is the best one can presently do with the observations. 

Population synthesis calculations in Refs.\ \citep{schneider2001,oshaughnessy2008} appear to show the cumulative merger-delay distribution being approximately linear in $\ln{(t)}$ (in which case the PDF varies as $\sim t^{-1}$), while the studies in Refs.\ \citep{belczynski2006,belczynski-twin-2007,dominik2012} show that a $(1/t)$ PDF is an appropriate approximation over several orders of magnitude of the delay-time. Furthermore, the population synthesis calculations of Ref.\ \citep{belczynski2010}, in their study of the formation rates of short- and long-GRBs, indicates an approximate $(1/t)$ delay-time distribution for NS-NS and NS-black-hole systems. 

Population synthesis calculations have also proposed previously unconsidered DNS-formation channels; specifically Ref.\ \citep{belczynski2000-dce} suggests a formation channel via a double common-envelope phase between two low-mass helium stars, such that the subsequently formed NSs would not have had time to accrete matter and be recycled. These DNS systems would be under-represented in Galactic pulsar surveys, since they would be observable as radio pulsars for a much shorter time scale than recycled pulsar systems. Hence DNS coalescence rate calculations based on the observed Galactic pulsar sample need to take into account any observational biases. Another of these new formation channels involves a stage of hypercritical common-envelope accretion from a low-mass helium giant to the firstborn NS, resulting in a population of tight, short-lived binaries (with merger-timescales $\lesssim 1.0$ Myr) which may contribute significantly to the total number of coalescences \citep{belczynski2002-mergersites}. 

In the 2004 study of merging DNS systems as the source of sGRBs by \citet{ando2004}, the merger-delay distribution is modelled as a power-law ($\sim t^{\alpha}$), and the calculated GRB rate densities are found to be relatively insensitive to the lower-cutoff time necessitated by such a parametrization, but considerably sensitive to $\alpha$. The characteristic upper inspiral timescale is also of interest; several known DNS systems are calculated to have inspiral times exceeding $10$ Gyr (Refs.\ \citep{champion2004,oslowski2011} and references therein). If these are representatives of a class of DNS systems resulting from a different evolutionary path than the lower time scale systems, then this evolutionary path need not be considered for detectable GW sources. 

With the above considerations in mind, in the present study we adopt a power-law merger-delay distribution for DNS systems, with a reference index of $-1$. This is supported by the existing (albeit sparse) observational data on the gravitational inspiral times of Galactic DNS systems, and the prevalence of power-law delay distributions found in population synthesis studies. For normalization purposes, we adopt a lower delay time of $50$ Myr, since the massive progenitor system (containing components with masses between $\sim 8-20 M_{\odot}$ for NS-NS system formation) may require an \textit{evolutionary} timescale of $\gtrsim 50$ Myr.\footnote{This evolutionary time scale is an approximate main-sequence lifetime for a $8 M_{\odot}$ star, burning $\sim 10\%$ of its core hydrogen, and obtained via the simple scaling relationship, $\tau_{\rm{evol}}\sim 10^{4}(M/M_{\odot})^{-2.5}$ Myr.} Taking this as a lower delay time avoids considerations of the (possibly significant) DNS-formation channel with the extra mass-transfer episode (which creates a peak in the delay-time distribution around $\sim 20$ Myr, and corresponds to a population of tight, short-lived DNS systems \citep{belczynski2002-mergersites}). We model only DNS systems formed via the classical formation channel \citep{bhatta-1991,belczynski2002-comprehensive,belczynski2006} for which the $\sim t^{-1}$ delay-time distribution is an appropriate approximation over several orders of magnitude. The power-law index will have a greater impact on merger-rate density calculations than the lower cutoff time. We assume an upper inspiral timescale equal to the cosmology-dependent age of the Universe. For the present study, the power-law index in this merger-delay distribution is labeled $\alpha$.

\subsection{Star-formation rate density, $d\rho_*/dt$}\label{appsec:sfrdens}
The determination of the low-redshift SFR density has been achieved via a wide variety of techniques, utilizing light at different wavelengths. However, these measurements become more difficult at higher redshifts since many of the techniques successfully employed in the low-$z$ Universe rely on light at wavelengths that cannot be detected beyond $z\sim 4$. This leaves us with only a handful of available techniques to probe the high-redshift star-formation history.\footnote{See \citep{bouwens2009} for references of low-$z$ techniques for probing the SFR density, as well as the non-UV techniques possible at $z\sim 2-4$.}  

The estimations of star-formation rates at high-redshift are obtained from measurements of UV luminosity functions (LFs), which tells us how many galaxies emit light in the UV-band in a given epoch. Excepting galaxies with the largest SFRs (which likely suffer from significant dust extinction) UV light has been shown to be a good tracer of the SFR (Refs.\ \citep{bouwens2007,bouwens2009} and references therein). Dust-extinction of UV light can be investigated, and hence corrected for, via the measurement of the UV-continuum slope, which has been shown to be well-correlated with dust extinction in the local Universe (Ref.\ \citep{bouwens2009} and references therein). A systematic study of the high-redshift SFR density was undertaken in Ref.\ \citep{bouwens2009} using \textit{Hubble Space Telescope} data. Correcting their UV luminosity density calculations for dust extinction, and converting this to an estimate of the SFR density, yielded significant evolution of the SFR density between $0< z \lesssim 6$. The SFR density is shown to rise out to $z\sim2-4$, followed by a decrease out to $z\sim 6$. This decrease is shown to continue out to $z\sim 8.5$ \citep{gonzalez2010}.

Given that only a handful of techniques exist to probe the high-redshift star-formation history, we will have to wait until further studies are carried out, or new techniques are developed, to complement the analyses in Refs.\ \citep{bouwens2009,gonzalez2010}. In our present study, we are only interested in a sensible model of the redshift evolution of the SFR density, which we can parametrize for a Bayesian inference analysis. Several of the studies (Refs.\ \citep{guetta2005,guetta2006,ando2004}) mentioned in Sec.\ \ref{sec:merger-delay}, as well as several other studies which attempt to fit GRB densities to delayed SFR-density models (e.g.,\ Ref.\ \citep{virgili2011}), employed the SF2 model of \citet{porciani2001}. Of the three models considered in the aforementioned paper, the SF2 model attempts to factor in the uncertainties in the incompleteness of data sets and the amount of dust extinction at early epochs. As such, the SFR density remains roughly constant at $z\gtrsim 2$. Its form is,
\begin{align}\label{eq:SF2}
\frac{d\rho_*}{dt}(z) \approx &\quad0.16\times\left(\frac{\exp{(3.4z)}}{\exp{(3.4z)}+22}\right)\nonumber\\
&\times\frac{E(z)}{(1+z)^{3/2}}\quad\text{$M_{\odot}$Mpc$^{-3}$yr$^{-1}$},
\end{align}
where
\begin{equation}
E(z)=\sqrt{{\Omega}_{m,0}(1+z)^3+{\Omega}_{k,0}(1+z)^2+{\Omega}_{\Lambda}(z)}.
\end{equation}
Obviously, if studies in the following decade confirm the SFR-density trends found in Refs.\ \citep{bouwens2009,gonzalez2010} we would not attempt to fit any ET data with the SF2 model. This model would need to be updated with a more realistic parametrisation. But for now, we adopt the SF2 model as a useful ansatz. For the present study, we parametrize the SF2 ansatz by making the factors of $3.4$ in the numerator and denominator of Eq.\ (\ref{eq:SF2}) variables, labeled $\beta_1$ and $\beta_2$ respectively.

\section{A faster calculation of the expected detection rate}\label{sec:faster-detection-rate}
The expected detection rate of inspiraling NS-NS binaries is given by
\begin{align} \label{eq:model-num}
N_D=T\times&{\int_0^{\infty}}{\int_0^{\infty}}{\frac{4{\pi}{D_c(z)}^2D_H}{E(z)}}{\frac{{\dot{n}(z)}}{(1+z)}}\times{\mathcal{P}}({\mathcal{M}})\nonumber\\
&\times C_{\Theta}\left[{\frac{\rho_0}{8}}{\frac{D_L(z)}{r_{0}}}\left({\frac{1.2M_{\odot}}{(1+z)\mathcal{M}}}\right)^{5/6}\right]dzd{\mathcal{M}}.
\end{align}

In our previous study \citep{hwth2011} we found that a simple parametrization of the expected detection rate provided a good approximation to the slower multi-dimensional integration necessitated by Eq.\ (\ref{eq:model-num}). However, we now want to extend our model-parameter space to a larger number of dimensions, for which the simple ansatz method becomes cumbersome. We found in our previous analysis that the standard deviation of the NS mass distribution had very little impact on the expected detection rate. Changing ${\sigma}_{\rm NS}$ from $0.02M_{\odot}$ to $0.12M_{\odot}$ led to a change in the expected detection rate of $\lesssim O(1\%)$.

Although the precision with which we are able to constrain ${\sigma}_{\rm NS}$ scales with the number of detections as $1/\sqrt{N}$, it is the {\it distribution} of detectable systems rather than their number that provides information on ${\sigma}_{\rm NS}$. If we approximate the Gaussian chirp mass distribution by a $\delta$-function centered on the mean of the chirp mass distribution, we can replace Eq.\ (\ref{eq:model-num}) by a 1D integral
\begin{align} \label{eq:model-num-1d}
N_D=T\times&{\int_0^{\infty}}{\frac{4{\pi}{D_c(z)}^2D_H}{E(z)}}{\frac{{\dot{n}(z)}}{(1+z)}}\times\nonumber\\
&\times C_{\Theta}\left[{\frac{\rho_0}{8}}{\frac{D_L(z)}{r_{0}}}\left({\frac{1.2M_{\odot}}{(1+z)\mu_{\mathcal{M}}}}\right)^{5/6}\right]dz.
\end{align}

This integration can be solved at least an order of magnitude faster by standard routines and gives results consistent with the full $2$D integration procedure. We also checked this faster method against the ansatz parametrization method, finding that the method used in our previous analysis was sufficiently accurate.

\bibliography{et_bib_resubmit}

\begin{thebibliography}{96}
\expandafter\ifx\csname natexlab\endcsname\relax\def\natexlab#1{#1}\fi
\expandafter\ifx\csname bibnamefont\endcsname\relax
  \def\bibnamefont#1{#1}\fi
\expandafter\ifx\csname bibfnamefont\endcsname\relax
  \def\bibfnamefont#1{#1}\fi
\expandafter\ifx\csname citenamefont\endcsname\relax
  \def\citenamefont#1{#1}\fi
\expandafter\ifx\csname url\endcsname\relax
  \def\url#1{\texttt{#1}}\fi
\expandafter\ifx\csname urlprefix\endcsname\relax\def\urlprefix{URL }\fi
\providecommand{\bibinfo}[2]{#2}
\providecommand{\eprint}[2][]{\url{#2}}

\bibitem[{\citenamefont{{Abbott} et~al.}(2009)\citenamefont{{Abbott}, {Abbott},
  {Adhikari}, {Ajith}, {Allen}, {Allen}, {Amin}, {Anderson}, {Anderson},
  {Arain} et~al.}}]{ligo2009}
\bibinfo{author}{\bibfnamefont{B.~P.} \bibnamefont{{Abbott}}},
  \bibinfo{author}{\bibfnamefont{R.}~\bibnamefont{{Abbott}}},
  \bibinfo{author}{\bibfnamefont{R.}~\bibnamefont{{Adhikari}}},
  \bibinfo{author}{\bibfnamefont{P.}~\bibnamefont{{Ajith}}},
  \bibinfo{author}{\bibfnamefont{B.}~\bibnamefont{{Allen}}},
  \bibinfo{author}{\bibfnamefont{G.}~\bibnamefont{{Allen}}},
  \bibinfo{author}{\bibfnamefont{R.~S.} \bibnamefont{{Amin}}},
  \bibinfo{author}{\bibfnamefont{S.~B.} \bibnamefont{{Anderson}}},
  \bibinfo{author}{\bibfnamefont{W.~G.} \bibnamefont{{Anderson}}},
  \bibinfo{author}{\bibfnamefont{M.~A.} \bibnamefont{{Arain}}},
  \bibnamefont{et~al.}, \bibinfo{journal}{Reports on Progress in Physics}
  \textbf{\bibinfo{volume}{72}}, \bibinfo{pages}{076901}
  (\bibinfo{year}{2009}), \eprint{0711.3041}.

\bibitem[{\citenamefont{{Grote} and {the LIGO Scientific
  Collaboration}}(2008)}]{geo2008}
\bibinfo{author}{\bibfnamefont{H.}~\bibnamefont{{Grote}}} \bibnamefont{and}
  \bibinfo{author}{\bibnamefont{{the LIGO Scientific Collaboration}}},
  \bibinfo{journal}{Classical and Quantum Gravity}
  \textbf{\bibinfo{volume}{25}}, \bibinfo{pages}{114043}
  (\bibinfo{year}{2008}).

\bibitem[{\citenamefont{{Acernese} et~al.}(2006)\citenamefont{{Acernese},
  {Amico}, {Alshourbagy}, {Antonucci}, {Aoudia}, {Avino}, {Babusci},
  {Ballardin}, and et~al.}}]{virgo2006}
\bibinfo{author}{\bibfnamefont{F.}~\bibnamefont{{Acernese}}},
  \bibinfo{author}{\bibfnamefont{P.}~\bibnamefont{{Amico}}},
  \bibinfo{author}{\bibfnamefont{M.}~\bibnamefont{{Alshourbagy}}},
  \bibinfo{author}{\bibfnamefont{F.}~\bibnamefont{{Antonucci}}},
  \bibinfo{author}{\bibfnamefont{S.}~\bibnamefont{{Aoudia}}},
  \bibinfo{author}{\bibfnamefont{S.}~\bibnamefont{{Avino}}},
  \bibinfo{author}{\bibfnamefont{D.}~\bibnamefont{{Babusci}}},
  \bibinfo{author}{\bibfnamefont{G.}~\bibnamefont{{Ballardin}}},
  \bibnamefont{and} \bibinfo{author}{\bibnamefont{et~al.}},
  \bibinfo{journal}{Classical and Quantum Gravity}
  \textbf{\bibinfo{volume}{23}}, \bibinfo{pages}{S635} (\bibinfo{year}{2006}).

\bibitem[{\citenamefont{{Takahashi} and {the TAMA
  Collaboration}}(2004)}]{tama2004}
\bibinfo{author}{\bibfnamefont{R.}~\bibnamefont{{Takahashi}}} \bibnamefont{and}
  \bibinfo{author}{\bibnamefont{{the TAMA Collaboration}}},
  \bibinfo{journal}{Classical and Quantum Gravity}
  \textbf{\bibinfo{volume}{21}}, \bibinfo{pages}{S403} (\bibinfo{year}{2004}).

\bibitem[{\citenamefont{{Kuroda} and {LCGT Collaboration}}(2010)}]{lcgt2010}
\bibinfo{author}{\bibfnamefont{K.}~\bibnamefont{{Kuroda}}} \bibnamefont{and}
  \bibinfo{author}{\bibnamefont{{LCGT Collaboration}}},
  \bibinfo{journal}{Classical and Quantum Gravity}
  \textbf{\bibinfo{volume}{27}}, \bibinfo{pages}{084004}
  (\bibinfo{year}{2010}).

\bibitem[{\citenamefont{{Somiya}}(2012)}]{kagra2012}
\bibinfo{author}{\bibfnamefont{K.}~\bibnamefont{{Somiya}}},
  \bibinfo{journal}{Classical and Quantum Gravity}
  \textbf{\bibinfo{volume}{29}}, \bibinfo{pages}{124007}
  (\bibinfo{year}{2012}), \eprint{1111.7185}.

\bibitem[{\citenamefont{{Mandel} and
  {O'Shaughnessy}}(2010{\natexlab{a}})}]{mandel-oshaughnessy2010}
\bibinfo{author}{\bibfnamefont{I.}~\bibnamefont{{Mandel}}} \bibnamefont{and}
  \bibinfo{author}{\bibfnamefont{R.}~\bibnamefont{{O'Shaughnessy}}},
  \bibinfo{journal}{Classical and Quantum Gravity}
  \textbf{\bibinfo{volume}{27}}, \bibinfo{pages}{114007}
  (\bibinfo{year}{2010}{\natexlab{a}}), \eprint{0912.1074}.

\bibitem[{\citenamefont{{Abadie}
  et~al.}(2010{\natexlab{a}})\citenamefont{{Abadie}, {Abbott}, {Abbott},
  {Abernathy}, {Accadia}, {Acernese}, {Adams}, {Adhikari}, {Ajith}, {Allen}
  et~al.}}]{ligo-s5-virgo-vsr1}
\bibinfo{author}{\bibfnamefont{J.}~\bibnamefont{{Abadie}}},
  \bibinfo{author}{\bibfnamefont{B.~P.} \bibnamefont{{Abbott}}},
  \bibinfo{author}{\bibfnamefont{R.}~\bibnamefont{{Abbott}}},
  \bibinfo{author}{\bibfnamefont{M.}~\bibnamefont{{Abernathy}}},
  \bibinfo{author}{\bibfnamefont{T.}~\bibnamefont{{Accadia}}},
  \bibinfo{author}{\bibfnamefont{F.}~\bibnamefont{{Acernese}}},
  \bibinfo{author}{\bibfnamefont{C.}~\bibnamefont{{Adams}}},
  \bibinfo{author}{\bibfnamefont{R.}~\bibnamefont{{Adhikari}}},
  \bibinfo{author}{\bibfnamefont{P.}~\bibnamefont{{Ajith}}},
  \bibinfo{author}{\bibfnamefont{B.}~\bibnamefont{{Allen}}},
  \bibnamefont{et~al.}, \bibinfo{journal}{\prd} \textbf{\bibinfo{volume}{82}},
  \bibinfo{pages}{102001} (\bibinfo{year}{2010}{\natexlab{a}}).

\bibitem[{\citenamefont{Abadie et~al.}(2012)\citenamefont{Abadie, Abbott,
  Abbott, Abbott, Abernathy, Accadia, Acernese, Adams, Adhikari, Affeldt
  et~al.}}]{ligo-s6}
\bibinfo{author}{\bibfnamefont{J.}~\bibnamefont{Abadie}},
  \bibinfo{author}{\bibfnamefont{B.~P.} \bibnamefont{Abbott}},
  \bibinfo{author}{\bibfnamefont{R.}~\bibnamefont{Abbott}},
  \bibinfo{author}{\bibfnamefont{T.~D.} \bibnamefont{Abbott}},
  \bibinfo{author}{\bibfnamefont{M.}~\bibnamefont{Abernathy}},
  \bibinfo{author}{\bibfnamefont{T.}~\bibnamefont{Accadia}},
  \bibinfo{author}{\bibfnamefont{F.}~\bibnamefont{Acernese}},
  \bibinfo{author}{\bibfnamefont{C.}~\bibnamefont{Adams}},
  \bibinfo{author}{\bibfnamefont{R.}~\bibnamefont{Adhikari}},
  \bibinfo{author}{\bibfnamefont{C.}~\bibnamefont{Affeldt}},
  \bibnamefont{et~al.} (\bibinfo{collaboration}{LIGO Scientific Collaboration
  and Virgo Collaboration}), \bibinfo{journal}{Phys. Rev. D}
  \textbf{\bibinfo{volume}{85}}, \bibinfo{pages}{082002}
  (\bibinfo{year}{2012}),
  \urlprefix\url{http://link.aps.org/doi/10.1103/PhysRevD.85.082002}.

\bibitem[{\citenamefont{{Harry} and {the LIGO Scientific
  Collaboration}}(2010)}]{AdvLIGO}
\bibinfo{author}{\bibfnamefont{G.~M.} \bibnamefont{{Harry}}} \bibnamefont{and}
  \bibinfo{author}{\bibnamefont{{the LIGO Scientific Collaboration}}},
  \bibinfo{journal}{Classical and Quantum Gravity}
  \textbf{\bibinfo{volume}{27}}, \bibinfo{pages}{084006}
  (\bibinfo{year}{2010}).

\bibitem[{\citenamefont{{Abadie}
  et~al.}(2010{\natexlab{b}})\citenamefont{{Abadie}, {Abbott}, {Abbott},
  {Abernathy}, {Accadia}, {Acernese}, {Adams}, {Adhikari}, {Ajith}, {Allen}
  et~al.}}]{abadie-rate2010}
\bibinfo{author}{\bibfnamefont{J.}~\bibnamefont{{Abadie}}},
  \bibinfo{author}{\bibfnamefont{B.~P.} \bibnamefont{{Abbott}}},
  \bibinfo{author}{\bibfnamefont{R.}~\bibnamefont{{Abbott}}},
  \bibinfo{author}{\bibfnamefont{M.}~\bibnamefont{{Abernathy}}},
  \bibinfo{author}{\bibfnamefont{T.}~\bibnamefont{{Accadia}}},
  \bibinfo{author}{\bibfnamefont{F.}~\bibnamefont{{Acernese}}},
  \bibinfo{author}{\bibfnamefont{C.}~\bibnamefont{{Adams}}},
  \bibinfo{author}{\bibfnamefont{R.}~\bibnamefont{{Adhikari}}},
  \bibinfo{author}{\bibfnamefont{P.}~\bibnamefont{{Ajith}}},
  \bibinfo{author}{\bibfnamefont{B.}~\bibnamefont{{Allen}}},
  \bibnamefont{et~al.}, \bibinfo{journal}{Classical and Quantum Gravity}
  \textbf{\bibinfo{volume}{27}}, \bibinfo{pages}{173001}
  (\bibinfo{year}{2010}{\natexlab{b}}), \eprint{1003.2480}.

\bibitem[{Adv(2009)}]{AdvVirgo}
\bibinfo{type}{Virgo Technical Report} \bibinfo{number}{VIR-0027A-09},
  \bibinfo{institution}{Virgo} (\bibinfo{year}{2009}).

\bibitem[{ind()}]{indigo}
\emph{\bibinfo{title}{{Indian Initiative in Gravitational-wave Observations,
  IndIGO}}},
  \bibinfo{note}{http://www.gw-indigo.org/tiki-index.php?page=Welcome}.

\bibitem[{\citenamefont{{Schutz}}(1986)}]{Schutz86}
\bibinfo{author}{\bibfnamefont{B.~F.} \bibnamefont{{Schutz}}},
  \bibinfo{journal}{\nat} \textbf{\bibinfo{volume}{323}}, \bibinfo{pages}{310}
  (\bibinfo{year}{1986}).

\bibitem[{\citenamefont{{Hughes} and {Holz}}(2003)}]{hughes-holz-2003}
\bibinfo{author}{\bibfnamefont{S.~A.} \bibnamefont{{Hughes}}} \bibnamefont{and}
  \bibinfo{author}{\bibfnamefont{D.~E.} \bibnamefont{{Holz}}},
  \bibinfo{journal}{Classical and Quantum Gravity}
  \textbf{\bibinfo{volume}{20}}, \bibinfo{pages}{S65} (\bibinfo{year}{2003}),
  \eprint{arXiv:astro-ph/0212218}.

\bibitem[{\citenamefont{{Sathyaprakash}
  et~al.}(2010)\citenamefont{{Sathyaprakash}, {Schutz}, and {Van Den
  Broeck}}}]{et-cosmography}
\bibinfo{author}{\bibfnamefont{B.~S.} \bibnamefont{{Sathyaprakash}}},
  \bibinfo{author}{\bibfnamefont{B.~F.} \bibnamefont{{Schutz}}},
  \bibnamefont{and} \bibinfo{author}{\bibfnamefont{C.}~\bibnamefont{{Van Den
  Broeck}}}, \bibinfo{journal}{Classical and Quantum Gravity}
  \textbf{\bibinfo{volume}{27}}, \bibinfo{pages}{215006}
  (\bibinfo{year}{2010}), \eprint{0906.4151}.

\bibitem[{\citenamefont{{Zhao} et~al.}(2011)\citenamefont{{Zhao}, {van den
  Broeck}, {Baskaran}, and {Li}}}]{et-dark-energy}
\bibinfo{author}{\bibfnamefont{W.}~\bibnamefont{{Zhao}}},
  \bibinfo{author}{\bibfnamefont{C.}~\bibnamefont{{van den Broeck}}},
  \bibinfo{author}{\bibfnamefont{D.}~\bibnamefont{{Baskaran}}},
  \bibnamefont{and} \bibinfo{author}{\bibfnamefont{T.~G.~F.}
  \bibnamefont{{Li}}}, \bibinfo{journal}{\prd} \textbf{\bibinfo{volume}{83}},
  \bibinfo{eid}{023005} (\bibinfo{year}{2011}), \eprint{1009.0206}.

\bibitem[{\citenamefont{{Holz} and {Hughes}}(2005)}]{HolzHughes:2005}
\bibinfo{author}{\bibfnamefont{D.~E.} \bibnamefont{{Holz}}} \bibnamefont{and}
  \bibinfo{author}{\bibfnamefont{S.~A.} \bibnamefont{{Hughes}}},
  \bibinfo{journal}{\apj} \textbf{\bibinfo{volume}{629}}, \bibinfo{pages}{15}
  (\bibinfo{year}{2005}), \eprint{arXiv:astro-ph/0504616}.

\bibitem[{\citenamefont{{Nissanke} et~al.}(2010)\citenamefont{{Nissanke},
  {Holz}, {Hughes}, {Dalal}, and {Sievers}}}]{nissanke2010}
\bibinfo{author}{\bibfnamefont{S.}~\bibnamefont{{Nissanke}}},
  \bibinfo{author}{\bibfnamefont{D.~E.} \bibnamefont{{Holz}}},
  \bibinfo{author}{\bibfnamefont{S.~A.} \bibnamefont{{Hughes}}},
  \bibinfo{author}{\bibfnamefont{N.}~\bibnamefont{{Dalal}}}, \bibnamefont{and}
  \bibinfo{author}{\bibfnamefont{J.~L.} \bibnamefont{{Sievers}}},
  \bibinfo{journal}{\apj} \textbf{\bibinfo{volume}{725}}, \bibinfo{pages}{496}
  (\bibinfo{year}{2010}), \eprint{0904.1017}.

\bibitem[{\citenamefont{{MacLeod} and {Hogan}}(2008)}]{macleod-hogan}
\bibinfo{author}{\bibfnamefont{C.~L.} \bibnamefont{{MacLeod}}}
  \bibnamefont{and} \bibinfo{author}{\bibfnamefont{C.~J.}
  \bibnamefont{{Hogan}}}, \bibinfo{journal}{\prd}
  \textbf{\bibinfo{volume}{77}}, \bibinfo{pages}{043512}
  (\bibinfo{year}{2008}), \eprint{0712.0618}.

\bibitem[{\citenamefont{{Del Pozzo}}(2011)}]{delpozzo2011}
\bibinfo{author}{\bibfnamefont{W.}~\bibnamefont{{Del Pozzo}}},
  \bibinfo{journal}{ArXiv e-prints}  (\bibinfo{year}{2011}),
  \eprint{1108.1317}.

\bibitem[{\citenamefont{{Taylor} et~al.}(2012)\citenamefont{{Taylor}, {Gair},
  and {Mandel}}}]{hwth2011}
\bibinfo{author}{\bibfnamefont{S.~R.} \bibnamefont{{Taylor}}},
  \bibinfo{author}{\bibfnamefont{J.~R.} \bibnamefont{{Gair}}},
  \bibnamefont{and} \bibinfo{author}{\bibfnamefont{I.}~\bibnamefont{{Mandel}}},
  \bibinfo{journal}{\prd} \textbf{\bibinfo{volume}{85}}, \bibinfo{eid}{023535}
  (\bibinfo{year}{2012}), \eprint{1108.5161}.

\bibitem[{\citenamefont{{Markovi{\'c}}}(1993)}]{markovic1993}
\bibinfo{author}{\bibfnamefont{D.}~\bibnamefont{{Markovi{\'c}}}},
  \bibinfo{journal}{\prd} \textbf{\bibinfo{volume}{48}}, \bibinfo{pages}{4738}
  (\bibinfo{year}{1993}).

\bibitem[{\citenamefont{{Chernoff} and {Finn}}(1993)}]{chernoff-finn-1993}
\bibinfo{author}{\bibfnamefont{D.~F.} \bibnamefont{{Chernoff}}}
  \bibnamefont{and} \bibinfo{author}{\bibfnamefont{L.~S.}
  \bibnamefont{{Finn}}}, \bibinfo{journal}{\apjl}
  \textbf{\bibinfo{volume}{411}}, \bibinfo{pages}{L5} (\bibinfo{year}{1993}),
  \eprint{arXiv:gr-qc/9304020}.

\bibitem[{\citenamefont{{Finn}}(1996)}]{Finn96}
\bibinfo{author}{\bibfnamefont{L.~S.} \bibnamefont{{Finn}}},
  \bibinfo{journal}{\prd} \textbf{\bibinfo{volume}{53}}, \bibinfo{pages}{2878}
  (\bibinfo{year}{1996}), \eprint{arXiv:gr-qc/9601048}.

\bibitem[{\citenamefont{{Freise} et~al.}(2009)\citenamefont{{Freise},
  {Chelkowski}, {Hild}, {Del Pozzo}, {Perreca}, and
  {Vecchio}}}]{triple-michelson2009}
\bibinfo{author}{\bibfnamefont{A.}~\bibnamefont{{Freise}}},
  \bibinfo{author}{\bibfnamefont{S.}~\bibnamefont{{Chelkowski}}},
  \bibinfo{author}{\bibfnamefont{S.}~\bibnamefont{{Hild}}},
  \bibinfo{author}{\bibfnamefont{W.}~\bibnamefont{{Del Pozzo}}},
  \bibinfo{author}{\bibfnamefont{A.}~\bibnamefont{{Perreca}}},
  \bibnamefont{and}
  \bibinfo{author}{\bibfnamefont{A.}~\bibnamefont{{Vecchio}}},
  \bibinfo{journal}{Classical and Quantum Gravity}
  \textbf{\bibinfo{volume}{26}}, \bibinfo{pages}{085012}
  (\bibinfo{year}{2009}), \eprint{0804.1036}.

\bibitem[{\citenamefont{{Sathyaprakash}
  et~al.}(2012)\citenamefont{{Sathyaprakash}, {Abernathy}, {Acernese}, {Ajith},
  {Allen}, {Amaro-Seoane}, {Andersson}, {Aoudia}, {Arun}, {Astone}
  et~al.}}]{sathy-et-potential2011}
\bibinfo{author}{\bibfnamefont{B.}~\bibnamefont{{Sathyaprakash}}},
  \bibinfo{author}{\bibfnamefont{M.}~\bibnamefont{{Abernathy}}},
  \bibinfo{author}{\bibfnamefont{F.}~\bibnamefont{{Acernese}}},
  \bibinfo{author}{\bibfnamefont{P.}~\bibnamefont{{Ajith}}},
  \bibinfo{author}{\bibfnamefont{B.}~\bibnamefont{{Allen}}},
  \bibinfo{author}{\bibfnamefont{P.}~\bibnamefont{{Amaro-Seoane}}},
  \bibinfo{author}{\bibfnamefont{N.}~\bibnamefont{{Andersson}}},
  \bibinfo{author}{\bibfnamefont{S.}~\bibnamefont{{Aoudia}}},
  \bibinfo{author}{\bibfnamefont{K.}~\bibnamefont{{Arun}}},
  \bibinfo{author}{\bibfnamefont{P.}~\bibnamefont{{Astone}}},
  \bibnamefont{et~al.}, \bibinfo{journal}{Classical and Quantum Gravity}
  \textbf{\bibinfo{volume}{29}}, \bibinfo{pages}{124013}
  (\bibinfo{year}{2012}), \eprint{1206.0331}.

\bibitem[{\citenamefont{{Punturo} et~al.}(2010)\citenamefont{{Punturo},
  {Abernathy}, {Acernese}, {Allen}, {Andersson}, {Arun}, {Barone}, {Barr},
  {Barsuglia}, {Beker} et~al.}}]{3rd-gen-science-reach2010}
\bibinfo{author}{\bibfnamefont{M.}~\bibnamefont{{Punturo}}},
  \bibinfo{author}{\bibfnamefont{M.}~\bibnamefont{{Abernathy}}},
  \bibinfo{author}{\bibfnamefont{F.}~\bibnamefont{{Acernese}}},
  \bibinfo{author}{\bibfnamefont{B.}~\bibnamefont{{Allen}}},
  \bibinfo{author}{\bibfnamefont{N.}~\bibnamefont{{Andersson}}},
  \bibinfo{author}{\bibfnamefont{K.}~\bibnamefont{{Arun}}},
  \bibinfo{author}{\bibfnamefont{F.}~\bibnamefont{{Barone}}},
  \bibinfo{author}{\bibfnamefont{B.}~\bibnamefont{{Barr}}},
  \bibinfo{author}{\bibfnamefont{M.}~\bibnamefont{{Barsuglia}}},
  \bibinfo{author}{\bibfnamefont{M.}~\bibnamefont{{Beker}}},
  \bibnamefont{et~al.}, \bibinfo{journal}{Classical and Quantum Gravity}
  \textbf{\bibinfo{volume}{27}}, \bibinfo{pages}{084007}
  (\bibinfo{year}{2010}).

\bibitem[{\citenamefont{{Hild} et~al.}(2011)\citenamefont{{Hild}, {Abernathy},
  {Acernese}, {Amaro-Seoane}, {Andersson}, {Arun}, {Barone}, {Barr},
  {Barsuglia}, {Beker} et~al.}}]{sensitivity-studies2011}
\bibinfo{author}{\bibfnamefont{S.}~\bibnamefont{{Hild}}},
  \bibinfo{author}{\bibfnamefont{M.}~\bibnamefont{{Abernathy}}},
  \bibinfo{author}{\bibfnamefont{F.}~\bibnamefont{{Acernese}}},
  \bibinfo{author}{\bibfnamefont{P.}~\bibnamefont{{Amaro-Seoane}}},
  \bibinfo{author}{\bibfnamefont{N.}~\bibnamefont{{Andersson}}},
  \bibinfo{author}{\bibfnamefont{K.}~\bibnamefont{{Arun}}},
  \bibinfo{author}{\bibfnamefont{F.}~\bibnamefont{{Barone}}},
  \bibinfo{author}{\bibfnamefont{B.}~\bibnamefont{{Barr}}},
  \bibinfo{author}{\bibfnamefont{M.}~\bibnamefont{{Barsuglia}}},
  \bibinfo{author}{\bibfnamefont{M.}~\bibnamefont{{Beker}}},
  \bibnamefont{et~al.}, \bibinfo{journal}{Classical and Quantum Gravity}
  \textbf{\bibinfo{volume}{28}}, \bibinfo{pages}{094013}
  (\bibinfo{year}{2011}), \eprint{1012.0908}.

\bibitem[{\citenamefont{{Messenger} and {Read}}(2012)}]{messenger-read-2011}
\bibinfo{author}{\bibfnamefont{C.}~\bibnamefont{{Messenger}}} \bibnamefont{and}
  \bibinfo{author}{\bibfnamefont{J.}~\bibnamefont{{Read}}},
  \bibinfo{journal}{Physical Review Letters} \textbf{\bibinfo{volume}{108}},
  \bibinfo{eid}{091101} (\bibinfo{year}{2012}), \eprint{1107.5725}.

\bibitem[{\citenamefont{{Nishizawa}
  et~al.}(2012{\natexlab{a}})\citenamefont{{Nishizawa}, {Yagi}, {Taruya}, and
  {Tanaka}}}]{nishizawa-phase-shift-2012}
\bibinfo{author}{\bibfnamefont{A.}~\bibnamefont{{Nishizawa}}},
  \bibinfo{author}{\bibfnamefont{K.}~\bibnamefont{{Yagi}}},
  \bibinfo{author}{\bibfnamefont{A.}~\bibnamefont{{Taruya}}}, \bibnamefont{and}
  \bibinfo{author}{\bibfnamefont{T.}~\bibnamefont{{Tanaka}}},
  \bibinfo{journal}{ArXiv e-prints}  (\bibinfo{year}{2012}{\natexlab{a}}),
  \eprint{1204.2877}.

\bibitem[{adl(2010)}]{adligo-noise}
\emph{\bibinfo{title}{{Advanced LIGO anticipated sensitivity curves}}}
  (\bibinfo{year}{2010}), \bibinfo{note}{https://dcc.ligo.org/cgi-bin/
  DocDB/ShowDocument?docid=2974.}

\bibitem[{\citenamefont{{Hild} et~al.}(2008)\citenamefont{{Hild}, {Chelkowski},
  and {Freise}}}]{conventional-ET-2008}
\bibinfo{author}{\bibfnamefont{S.}~\bibnamefont{{Hild}}},
  \bibinfo{author}{\bibfnamefont{S.}~\bibnamefont{{Chelkowski}}},
  \bibnamefont{and} \bibinfo{author}{\bibfnamefont{A.}~\bibnamefont{{Freise}}},
  \bibinfo{journal}{ArXiv e-prints}  (\bibinfo{year}{2008}),
  \eprint{0810.0604}.

\bibitem[{\citenamefont{{Hild} et~al.}(2010)\citenamefont{{Hild}, {Chelkowski},
  {Freise}, {Franc}, {Morgado}, {Flaminio}, and {DeSalvo}}}]{ET-xylophone-2010}
\bibinfo{author}{\bibfnamefont{S.}~\bibnamefont{{Hild}}},
  \bibinfo{author}{\bibfnamefont{S.}~\bibnamefont{{Chelkowski}}},
  \bibinfo{author}{\bibfnamefont{A.}~\bibnamefont{{Freise}}},
  \bibinfo{author}{\bibfnamefont{J.}~\bibnamefont{{Franc}}},
  \bibinfo{author}{\bibfnamefont{N.}~\bibnamefont{{Morgado}}},
  \bibinfo{author}{\bibfnamefont{R.}~\bibnamefont{{Flaminio}}},
  \bibnamefont{and}
  \bibinfo{author}{\bibfnamefont{R.}~\bibnamefont{{DeSalvo}}},
  \bibinfo{journal}{Classical and Quantum Gravity}
  \textbf{\bibinfo{volume}{27}}, \bibinfo{pages}{015003}
  (\bibinfo{year}{2010}), \eprint{0906.2655}.

\bibitem[{ET-()}]{ET-site}
\emph{\bibinfo{title}{{http://www.et-gw.eu/}}}.

\bibitem[{\citenamefont{{Huerta} and {Gair}}(2011)}]{eliu-ET}
\bibinfo{author}{\bibfnamefont{E.~A.} \bibnamefont{{Huerta}}} \bibnamefont{and}
  \bibinfo{author}{\bibfnamefont{J.~R.} \bibnamefont{{Gair}}},
  \bibinfo{journal}{\prd} \textbf{\bibinfo{volume}{83}}, \bibinfo{eid}{044021}
  (\bibinfo{year}{2011}), \eprint{1011.0421}.

\bibitem[{\citenamefont{{Abadie}
  et~al.}(2010{\natexlab{c}})\citenamefont{{Abadie}, {Abbott}, {Abbott},
  {Abernathy}, {Accadia}, {Acernese}, {Adams}, {Adhikari}, {Ajith}, {Allen}
  et~al.}}]{Abadie2010}
\bibinfo{author}{\bibfnamefont{J.}~\bibnamefont{{Abadie}}},
  \bibinfo{author}{\bibfnamefont{B.~P.} \bibnamefont{{Abbott}}},
  \bibinfo{author}{\bibfnamefont{R.}~\bibnamefont{{Abbott}}},
  \bibinfo{author}{\bibfnamefont{M.}~\bibnamefont{{Abernathy}}},
  \bibinfo{author}{\bibfnamefont{T.}~\bibnamefont{{Accadia}}},
  \bibinfo{author}{\bibfnamefont{F.}~\bibnamefont{{Acernese}}},
  \bibinfo{author}{\bibfnamefont{C.}~\bibnamefont{{Adams}}},
  \bibinfo{author}{\bibfnamefont{R.}~\bibnamefont{{Adhikari}}},
  \bibinfo{author}{\bibfnamefont{P.}~\bibnamefont{{Ajith}}},
  \bibinfo{author}{\bibfnamefont{B.}~\bibnamefont{{Allen}}},
  \bibnamefont{et~al.}, \bibinfo{journal}{\prd} \textbf{\bibinfo{volume}{82}},
  \bibinfo{pages}{102001} (\bibinfo{year}{2010}{\natexlab{c}}).

\bibitem[{\citenamefont{{O'Shaughnessy}
  et~al.}(2010)\citenamefont{{O'Shaughnessy}, {Kalogera}, and
  {Belczynski}}}]{oshaughnessy2010-binary-compact}
\bibinfo{author}{\bibfnamefont{R.}~\bibnamefont{{O'Shaughnessy}}},
  \bibinfo{author}{\bibfnamefont{V.}~\bibnamefont{{Kalogera}}},
  \bibnamefont{and}
  \bibinfo{author}{\bibfnamefont{K.}~\bibnamefont{{Belczynski}}},
  \bibinfo{journal}{\apj} \textbf{\bibinfo{volume}{716}}, \bibinfo{pages}{615}
  (\bibinfo{year}{2010}), \eprint{0908.3635}.

\bibitem[{\citenamefont{{Nutzman} et~al.}(2004)\citenamefont{{Nutzman},
  {Kalogera}, {Finn}, {Hendrickson}, and {Belczynski}}}]{Nutzman2004}
\bibinfo{author}{\bibfnamefont{P.}~\bibnamefont{{Nutzman}}},
  \bibinfo{author}{\bibfnamefont{V.}~\bibnamefont{{Kalogera}}},
  \bibinfo{author}{\bibfnamefont{L.~S.} \bibnamefont{{Finn}}},
  \bibinfo{author}{\bibfnamefont{C.}~\bibnamefont{{Hendrickson}}},
  \bibnamefont{and}
  \bibinfo{author}{\bibfnamefont{K.}~\bibnamefont{{Belczynski}}},
  \bibinfo{journal}{\apj} \textbf{\bibinfo{volume}{612}}, \bibinfo{pages}{364}
  (\bibinfo{year}{2004}), \eprint{arXiv:astro-ph/0402091}.

\bibitem[{\citenamefont{{Regimbau} et~al.}(2012)\citenamefont{{Regimbau},
  {Dent}, {Del Pozzo}, {Giampanis}, {Li}, {Robinson}, {Van Den Broeck},
  {Meacher}, {Rodriguez}, {Sathyaprakash} et~al.}}]{regimbau2012}
\bibinfo{author}{\bibfnamefont{T.}~\bibnamefont{{Regimbau}}},
  \bibinfo{author}{\bibfnamefont{T.}~\bibnamefont{{Dent}}},
  \bibinfo{author}{\bibfnamefont{W.}~\bibnamefont{{Del Pozzo}}},
  \bibinfo{author}{\bibfnamefont{S.}~\bibnamefont{{Giampanis}}},
  \bibinfo{author}{\bibfnamefont{T.~G.~F.} \bibnamefont{{Li}}},
  \bibinfo{author}{\bibfnamefont{C.}~\bibnamefont{{Robinson}}},
  \bibinfo{author}{\bibfnamefont{C.}~\bibnamefont{{Van Den Broeck}}},
  \bibinfo{author}{\bibfnamefont{D.}~\bibnamefont{{Meacher}}},
  \bibinfo{author}{\bibfnamefont{C.}~\bibnamefont{{Rodriguez}}},
  \bibinfo{author}{\bibfnamefont{B.~S.} \bibnamefont{{Sathyaprakash}}},
  \bibnamefont{et~al.}, \bibinfo{journal}{ArXiv e-prints}
  (\bibinfo{year}{2012}), \eprint{1201.3563}.

\bibitem[{adv(2009)}]{advirgo-noise}
\emph{\bibinfo{title}{{Advanced Virgo Baseline Design}}}
  (\bibinfo{year}{2009}),
  \bibinfo{note}{https://pub3.ego-gw.it/itf/tds/file.php?callFile=VIR-
  0027A-09.pdf.}

\bibitem[{\citenamefont{{Sathyaprakash} and {Schutz}}(2009)}]{sathy2009}
\bibinfo{author}{\bibfnamefont{B.~S.} \bibnamefont{{Sathyaprakash}}}
  \bibnamefont{and} \bibinfo{author}{\bibfnamefont{B.~F.}
  \bibnamefont{{Schutz}}}, \bibinfo{journal}{Living Reviews in Relativity}
  \textbf{\bibinfo{volume}{12}}, \bibinfo{pages}{2} (\bibinfo{year}{2009}),
  \eprint{0903.0338}.

\bibitem[{\citenamefont{{Schutz}}(2011)}]{schutz2011}
\bibinfo{author}{\bibfnamefont{B.~F.} \bibnamefont{{Schutz}}},
  \bibinfo{journal}{Classical and Quantum Gravity}
  \textbf{\bibinfo{volume}{28}}, \bibinfo{pages}{125023}
  (\bibinfo{year}{2011}), \eprint{1102.5421}.

\bibitem[{\citenamefont{{Veitch} et~al.}(2012)\citenamefont{{Veitch}, {Mandel},
  {Aylott}, {Farr}, {Raymond}, {Rodriguez}, {van der Sluys}, {Kalogera}, and
  {Vecchio}}}]{veitch2012}
\bibinfo{author}{\bibfnamefont{J.}~\bibnamefont{{Veitch}}},
  \bibinfo{author}{\bibfnamefont{I.}~\bibnamefont{{Mandel}}},
  \bibinfo{author}{\bibfnamefont{B.}~\bibnamefont{{Aylott}}},
  \bibinfo{author}{\bibfnamefont{B.}~\bibnamefont{{Farr}}},
  \bibinfo{author}{\bibfnamefont{V.}~\bibnamefont{{Raymond}}},
  \bibinfo{author}{\bibfnamefont{C.}~\bibnamefont{{Rodriguez}}},
  \bibinfo{author}{\bibfnamefont{M.}~\bibnamefont{{van der Sluys}}},
  \bibinfo{author}{\bibfnamefont{V.}~\bibnamefont{{Kalogera}}},
  \bibnamefont{and}
  \bibinfo{author}{\bibfnamefont{A.}~\bibnamefont{{Vecchio}}},
  \bibinfo{journal}{\prd} \textbf{\bibinfo{volume}{85}}, \bibinfo{eid}{104045}
  (\bibinfo{year}{2012}), \eprint{1201.1195}.

\bibitem[{\citenamefont{{Kiziltan} et~al.}(2010)\citenamefont{{Kiziltan},
  {Kottas}, and {Thorsett}}}]{kiziltan2010}
\bibinfo{author}{\bibfnamefont{B.}~\bibnamefont{{Kiziltan}}},
  \bibinfo{author}{\bibfnamefont{A.}~\bibnamefont{{Kottas}}}, \bibnamefont{and}
  \bibinfo{author}{\bibfnamefont{S.~E.} \bibnamefont{{Thorsett}}},
  \bibinfo{journal}{ArXiv e-prints}  (\bibinfo{year}{2010}),
  \eprint{1011.4291}.

\bibitem[{\citenamefont{{Valentim} et~al.}(2011)\citenamefont{{Valentim},
  {Rangel}, and {Horvath}}}]{valentim2011}
\bibinfo{author}{\bibfnamefont{R.}~\bibnamefont{{Valentim}}},
  \bibinfo{author}{\bibfnamefont{E.}~\bibnamefont{{Rangel}}}, \bibnamefont{and}
  \bibinfo{author}{\bibfnamefont{J.~E.} \bibnamefont{{Horvath}}},
  \bibinfo{journal}{\mnras} \textbf{\bibinfo{volume}{414}},
  \bibinfo{pages}{1427} (\bibinfo{year}{2011}), \eprint{1101.4872}.

\bibitem[{\citenamefont{{Belczynski} et~al.}(2008)\citenamefont{{Belczynski},
  {Kalogera}, {Rasio}, {Taam}, {Zezas}, {Bulik}, {Maccarone}, and
  {Ivanova}}}]{Belczynski:2008}
\bibinfo{author}{\bibfnamefont{K.}~\bibnamefont{{Belczynski}}},
  \bibinfo{author}{\bibfnamefont{V.}~\bibnamefont{{Kalogera}}},
  \bibinfo{author}{\bibfnamefont{F.~A.} \bibnamefont{{Rasio}}},
  \bibinfo{author}{\bibfnamefont{R.~E.} \bibnamefont{{Taam}}},
  \bibinfo{author}{\bibfnamefont{A.}~\bibnamefont{{Zezas}}},
  \bibinfo{author}{\bibfnamefont{T.}~\bibnamefont{{Bulik}}},
  \bibinfo{author}{\bibfnamefont{T.~J.} \bibnamefont{{Maccarone}}},
  \bibnamefont{and}
  \bibinfo{author}{\bibfnamefont{N.}~\bibnamefont{{Ivanova}}},
  \bibinfo{journal}{Astrophys. J. Suppl.} \textbf{\bibinfo{volume}{174}},
  \bibinfo{pages}{223} (\bibinfo{year}{2008}).

\bibitem[{\citenamefont{{Mandel} and
  {O'Shaughnessy}}(2010{\natexlab{b}})}]{MandelOshaughnessy:2010}
\bibinfo{author}{\bibfnamefont{I.}~\bibnamefont{{Mandel}}} \bibnamefont{and}
  \bibinfo{author}{\bibfnamefont{R.}~\bibnamefont{{O'Shaughnessy}}},
  \bibinfo{journal}{Classical and Quantum Gravity}
  \textbf{\bibinfo{volume}{27}}, \bibinfo{pages}{114007}
  (\bibinfo{year}{2010}{\natexlab{b}}), \eprint{0912.1074}.

\bibitem[{\citenamefont{{Ozel} et~al.}(2012)\citenamefont{{Ozel}, {Psaltis},
  {Narayan}, and {Santos Villarreal}}}]{ozel2012}
\bibinfo{author}{\bibfnamefont{F.}~\bibnamefont{{Ozel}}},
  \bibinfo{author}{\bibfnamefont{D.}~\bibnamefont{{Psaltis}}},
  \bibinfo{author}{\bibfnamefont{R.}~\bibnamefont{{Narayan}}},
  \bibnamefont{and} \bibinfo{author}{\bibfnamefont{A.}~\bibnamefont{{Santos
  Villarreal}}}, \bibinfo{journal}{ArXiv e-prints}  (\bibinfo{year}{2012}),
  \eprint{1201.1006}.

\bibitem[{\citenamefont{{O'Shaughnessy}
  et~al.}(2008)\citenamefont{{O'Shaughnessy}, {Belczynski}, and
  {Kalogera}}}]{oshaughnessy2008}
\bibinfo{author}{\bibfnamefont{R.}~\bibnamefont{{O'Shaughnessy}}},
  \bibinfo{author}{\bibfnamefont{K.}~\bibnamefont{{Belczynski}}},
  \bibnamefont{and}
  \bibinfo{author}{\bibfnamefont{V.}~\bibnamefont{{Kalogera}}},
  \bibinfo{journal}{\apj} \textbf{\bibinfo{volume}{675}}, \bibinfo{pages}{566}
  (\bibinfo{year}{2008}), \eprint{0706.4139}.

\bibitem[{\citenamefont{{de Freitas Pacheco}}(1997)}]{pacheco1997}
\bibinfo{author}{\bibfnamefont{J.~A.} \bibnamefont{{de Freitas Pacheco}}},
  \bibinfo{journal}{Astroparticle Physics} \textbf{\bibinfo{volume}{8}},
  \bibinfo{pages}{21} (\bibinfo{year}{1997}).

\bibitem[{\citenamefont{{Totani}}(1997)}]{totani1997}
\bibinfo{author}{\bibfnamefont{T.}~\bibnamefont{{Totani}}},
  \bibinfo{journal}{Astrophys. J.} \textbf{\bibinfo{volume}{486}},
  \bibinfo{pages}{L71+} (\bibinfo{year}{1997}),
  \eprint{arXiv:astro-ph/9707051}.

\bibitem[{\citenamefont{{Schneider} et~al.}(2001)\citenamefont{{Schneider},
  {Ferrari}, {Matarrese}, and {Portegies Zwart}}}]{schneider2001}
\bibinfo{author}{\bibfnamefont{R.}~\bibnamefont{{Schneider}}},
  \bibinfo{author}{\bibfnamefont{V.}~\bibnamefont{{Ferrari}}},
  \bibinfo{author}{\bibfnamefont{S.}~\bibnamefont{{Matarrese}}},
  \bibnamefont{and} \bibinfo{author}{\bibfnamefont{S.~F.}
  \bibnamefont{{Portegies Zwart}}}, \bibinfo{journal}{\mnras}
  \textbf{\bibinfo{volume}{324}}, \bibinfo{pages}{797} (\bibinfo{year}{2001}),
  \eprint{arXiv:astro-ph/0002055}.

\bibitem[{\citenamefont{{Piran}}(1992)}]{piran1992}
\bibinfo{author}{\bibfnamefont{T.}~\bibnamefont{{Piran}}},
  \bibinfo{journal}{\apjl} \textbf{\bibinfo{volume}{389}}, \bibinfo{pages}{L45}
  (\bibinfo{year}{1992}).

\bibitem[{\citenamefont{{Abt}}(1983)}]{abt1983}
\bibinfo{author}{\bibfnamefont{H.~A.} \bibnamefont{{Abt}}},
  \bibinfo{journal}{Annu. Rev. Astron. Astrophys.}
  \textbf{\bibinfo{volume}{21}}, \bibinfo{pages}{343} (\bibinfo{year}{1983}).

\bibitem[{\citenamefont{{Champion} et~al.}(2004)\citenamefont{{Champion},
  {Lorimer}, {McLaughlin}, {Cordes}, {Arzoumanian}, {Weisberg}, and
  {Taylor}}}]{champion2004}
\bibinfo{author}{\bibfnamefont{D.~J.} \bibnamefont{{Champion}}},
  \bibinfo{author}{\bibfnamefont{D.~R.} \bibnamefont{{Lorimer}}},
  \bibinfo{author}{\bibfnamefont{M.~A.} \bibnamefont{{McLaughlin}}},
  \bibinfo{author}{\bibfnamefont{J.~M.} \bibnamefont{{Cordes}}},
  \bibinfo{author}{\bibfnamefont{Z.}~\bibnamefont{{Arzoumanian}}},
  \bibinfo{author}{\bibfnamefont{J.~M.} \bibnamefont{{Weisberg}}},
  \bibnamefont{and} \bibinfo{author}{\bibfnamefont{J.~H.}
  \bibnamefont{{Taylor}}}, \bibinfo{journal}{\mnras}
  \textbf{\bibinfo{volume}{350}}, \bibinfo{pages}{L61} (\bibinfo{year}{2004}),
  \eprint{arXiv:astro-ph/0403553}.

\bibitem[{\citenamefont{{Guetta} and {Piran}}(2005)}]{guetta2005}
\bibinfo{author}{\bibfnamefont{D.}~\bibnamefont{{Guetta}}} \bibnamefont{and}
  \bibinfo{author}{\bibfnamefont{T.}~\bibnamefont{{Piran}}},
  \bibinfo{journal}{\aap} \textbf{\bibinfo{volume}{435}}, \bibinfo{pages}{421}
  (\bibinfo{year}{2005}), \eprint{arXiv:astro-ph/0407429}.

\bibitem[{\citenamefont{{Guetta} and {Piran}}(2006)}]{guetta2006}
\bibinfo{author}{\bibfnamefont{D.}~\bibnamefont{{Guetta}}} \bibnamefont{and}
  \bibinfo{author}{\bibfnamefont{T.}~\bibnamefont{{Piran}}},
  \bibinfo{journal}{\aap} \textbf{\bibinfo{volume}{453}}, \bibinfo{pages}{823}
  (\bibinfo{year}{2006}), \eprint{arXiv:astro-ph/0511239}.

\bibitem[{\citenamefont{{Belczynski} et~al.}(2006)\citenamefont{{Belczynski},
  {Perna}, {Bulik}, {Kalogera}, {Ivanova}, and {Lamb}}}]{belczynski2006}
\bibinfo{author}{\bibfnamefont{K.}~\bibnamefont{{Belczynski}}},
  \bibinfo{author}{\bibfnamefont{R.}~\bibnamefont{{Perna}}},
  \bibinfo{author}{\bibfnamefont{T.}~\bibnamefont{{Bulik}}},
  \bibinfo{author}{\bibfnamefont{V.}~\bibnamefont{{Kalogera}}},
  \bibinfo{author}{\bibfnamefont{N.}~\bibnamefont{{Ivanova}}},
  \bibnamefont{and} \bibinfo{author}{\bibfnamefont{D.~Q.}
  \bibnamefont{{Lamb}}}, \bibinfo{journal}{\apj}
  \textbf{\bibinfo{volume}{648}}, \bibinfo{pages}{1110} (\bibinfo{year}{2006}),
  \eprint{arXiv:astro-ph/0601458}.

\bibitem[{\citenamefont{{Belczynski} et~al.}(2007)\citenamefont{{Belczynski},
  {Stanek}, and {Fryer}}}]{belczynski-twin-2007}
\bibinfo{author}{\bibfnamefont{K.}~\bibnamefont{{Belczynski}}},
  \bibinfo{author}{\bibfnamefont{K.~Z.} \bibnamefont{{Stanek}}},
  \bibnamefont{and} \bibinfo{author}{\bibfnamefont{C.~L.}
  \bibnamefont{{Fryer}}}, \bibinfo{journal}{ArXiv e-prints}
  (\bibinfo{year}{2007}), \eprint{0712.3309}.

\bibitem[{\citenamefont{{Dominik} et~al.}(2012)\citenamefont{{Dominik},
  {Belczynski}, {Fryer}, {Holz}, {Berti}, {Bulik}, {Mandel}, and
  {O'Shaughnessy}}}]{dominik2012}
\bibinfo{author}{\bibfnamefont{M.}~\bibnamefont{{Dominik}}},
  \bibinfo{author}{\bibfnamefont{K.}~\bibnamefont{{Belczynski}}},
  \bibinfo{author}{\bibfnamefont{C.}~\bibnamefont{{Fryer}}},
  \bibinfo{author}{\bibfnamefont{D.}~\bibnamefont{{Holz}}},
  \bibinfo{author}{\bibfnamefont{E.}~\bibnamefont{{Berti}}},
  \bibinfo{author}{\bibfnamefont{T.}~\bibnamefont{{Bulik}}},
  \bibinfo{author}{\bibfnamefont{I.}~\bibnamefont{{Mandel}}}, \bibnamefont{and}
  \bibinfo{author}{\bibfnamefont{R.}~\bibnamefont{{O'Shaughnessy}}},
  \bibinfo{journal}{ArXiv e-prints}  (\bibinfo{year}{2012}),
  \eprint{1202.4901}.

\bibitem[{\citenamefont{{Belczynski} et~al.}(2010)\citenamefont{{Belczynski},
  {Holz}, {Fryer}, {Berger}, {Hartmann}, and {O'Shea}}}]{belczynski2010}
\bibinfo{author}{\bibfnamefont{K.}~\bibnamefont{{Belczynski}}},
  \bibinfo{author}{\bibfnamefont{D.~E.} \bibnamefont{{Holz}}},
  \bibinfo{author}{\bibfnamefont{C.~L.} \bibnamefont{{Fryer}}},
  \bibinfo{author}{\bibfnamefont{E.}~\bibnamefont{{Berger}}},
  \bibinfo{author}{\bibfnamefont{D.~H.} \bibnamefont{{Hartmann}}},
  \bibnamefont{and} \bibinfo{author}{\bibfnamefont{B.}~\bibnamefont{{O'Shea}}},
  \bibinfo{journal}{\apj} \textbf{\bibinfo{volume}{708}}, \bibinfo{pages}{117}
  (\bibinfo{year}{2010}), \eprint{0812.2470}.

\bibitem[{\citenamefont{{Porciani} and {Madau}}(2001)}]{porciani2001}
\bibinfo{author}{\bibfnamefont{C.}~\bibnamefont{{Porciani}}} \bibnamefont{and}
  \bibinfo{author}{\bibfnamefont{P.}~\bibnamefont{{Madau}}},
  \bibinfo{journal}{\apj} \textbf{\bibinfo{volume}{548}}, \bibinfo{pages}{522}
  (\bibinfo{year}{2001}), \eprint{arXiv:astro-ph/0008294}.

\bibitem[{\citenamefont{{Bouwens} et~al.}(2009)\citenamefont{{Bouwens},
  {Illingworth}, {Franx}, {Chary}, {Meurer}, {Conselice}, {Ford}, {Giavalisco},
  and {van Dokkum}}}]{bouwens2009}
\bibinfo{author}{\bibfnamefont{R.~J.} \bibnamefont{{Bouwens}}},
  \bibinfo{author}{\bibfnamefont{G.~D.} \bibnamefont{{Illingworth}}},
  \bibinfo{author}{\bibfnamefont{M.}~\bibnamefont{{Franx}}},
  \bibinfo{author}{\bibfnamefont{R.-R.} \bibnamefont{{Chary}}},
  \bibinfo{author}{\bibfnamefont{G.~R.} \bibnamefont{{Meurer}}},
  \bibinfo{author}{\bibfnamefont{C.~J.} \bibnamefont{{Conselice}}},
  \bibinfo{author}{\bibfnamefont{H.}~\bibnamefont{{Ford}}},
  \bibinfo{author}{\bibfnamefont{M.}~\bibnamefont{{Giavalisco}}},
  \bibnamefont{and} \bibinfo{author}{\bibfnamefont{P.}~\bibnamefont{{van
  Dokkum}}}, \bibinfo{journal}{\apj} \textbf{\bibinfo{volume}{705}},
  \bibinfo{pages}{936} (\bibinfo{year}{2009}), \eprint{0909.4074}.

\bibitem[{\citenamefont{{Gonz{\'a}lez}
  et~al.}(2010)\citenamefont{{Gonz{\'a}lez}, {Labb{\'e}}, {Bouwens},
  {Illingworth}, {Franx}, {Kriek}, and {Brammer}}}]{gonzalez2010}
\bibinfo{author}{\bibfnamefont{V.}~\bibnamefont{{Gonz{\'a}lez}}},
  \bibinfo{author}{\bibfnamefont{I.}~\bibnamefont{{Labb{\'e}}}},
  \bibinfo{author}{\bibfnamefont{R.~J.} \bibnamefont{{Bouwens}}},
  \bibinfo{author}{\bibfnamefont{G.}~\bibnamefont{{Illingworth}}},
  \bibinfo{author}{\bibfnamefont{M.}~\bibnamefont{{Franx}}},
  \bibinfo{author}{\bibfnamefont{M.}~\bibnamefont{{Kriek}}}, \bibnamefont{and}
  \bibinfo{author}{\bibfnamefont{G.~B.} \bibnamefont{{Brammer}}},
  \bibinfo{journal}{\apj} \textbf{\bibinfo{volume}{713}}, \bibinfo{pages}{115}
  (\bibinfo{year}{2010}), \eprint{0909.3517}.

\bibitem[{\citenamefont{{Cucchiara} et~al.}(2011)\citenamefont{{Cucchiara},
  {Levan}, {Fox}, {Tanvir}, {Ukwatta}, {Berger}, {Kr{\"u}hler}, {K{\"u}pc{\"u}
  Yolda{\c s}}, {Wu}, {Toma} et~al.}}]{cucchiara2011}
\bibinfo{author}{\bibfnamefont{A.}~\bibnamefont{{Cucchiara}}},
  \bibinfo{author}{\bibfnamefont{A.~J.} \bibnamefont{{Levan}}},
  \bibinfo{author}{\bibfnamefont{D.~B.} \bibnamefont{{Fox}}},
  \bibinfo{author}{\bibfnamefont{N.~R.} \bibnamefont{{Tanvir}}},
  \bibinfo{author}{\bibfnamefont{T.~N.} \bibnamefont{{Ukwatta}}},
  \bibinfo{author}{\bibfnamefont{E.}~\bibnamefont{{Berger}}},
  \bibinfo{author}{\bibfnamefont{T.}~\bibnamefont{{Kr{\"u}hler}}},
  \bibinfo{author}{\bibfnamefont{A.}~\bibnamefont{{K{\"u}pc{\"u} Yolda{\c
  s}}}}, \bibinfo{author}{\bibfnamefont{X.~F.} \bibnamefont{{Wu}}},
  \bibinfo{author}{\bibfnamefont{K.}~\bibnamefont{{Toma}}},
  \bibnamefont{et~al.}, \bibinfo{journal}{\apj} \textbf{\bibinfo{volume}{736}},
  \bibinfo{pages}{7} (\bibinfo{year}{2011}), \eprint{1105.4915}.

\bibitem[{\citenamefont{{Bouwens} et~al.}(2011)\citenamefont{{Bouwens},
  {Illingworth}, {Labbe}, {Oesch}, {Trenti}, {Carollo}, {van Dokkum}, {Franx},
  {Stiavelli}, {Gonz{\'a}lez} et~al.}}]{bouwens2011}
\bibinfo{author}{\bibfnamefont{R.~J.} \bibnamefont{{Bouwens}}},
  \bibinfo{author}{\bibfnamefont{G.~D.} \bibnamefont{{Illingworth}}},
  \bibinfo{author}{\bibfnamefont{I.}~\bibnamefont{{Labbe}}},
  \bibinfo{author}{\bibfnamefont{P.~A.} \bibnamefont{{Oesch}}},
  \bibinfo{author}{\bibfnamefont{M.}~\bibnamefont{{Trenti}}},
  \bibinfo{author}{\bibfnamefont{C.~M.} \bibnamefont{{Carollo}}},
  \bibinfo{author}{\bibfnamefont{P.~G.} \bibnamefont{{van Dokkum}}},
  \bibinfo{author}{\bibfnamefont{M.}~\bibnamefont{{Franx}}},
  \bibinfo{author}{\bibfnamefont{M.}~\bibnamefont{{Stiavelli}}},
  \bibinfo{author}{\bibfnamefont{V.}~\bibnamefont{{Gonz{\'a}lez}}},
  \bibnamefont{et~al.}, \bibinfo{journal}{\nat} \textbf{\bibinfo{volume}{469}},
  \bibinfo{pages}{504} (\bibinfo{year}{2011}), \eprint{0912.4263}.

\bibitem[{\citenamefont{{Chevallier} and {Polarski}}(2001)}]{chevallier2000}
\bibinfo{author}{\bibfnamefont{M.}~\bibnamefont{{Chevallier}}}
  \bibnamefont{and}
  \bibinfo{author}{\bibfnamefont{D.}~\bibnamefont{{Polarski}}},
  \bibinfo{journal}{International Journal of Modern Physics D}
  \textbf{\bibinfo{volume}{10}}, \bibinfo{pages}{213} (\bibinfo{year}{2001}),
  \eprint{arXiv:gr-qc/0009008}.

\bibitem[{\citenamefont{{Shafieloo} et~al.}(2009)\citenamefont{{Shafieloo},
  {Sahni}, and {Starobinsky}}}]{shafieloo2009}
\bibinfo{author}{\bibfnamefont{A.}~\bibnamefont{{Shafieloo}}},
  \bibinfo{author}{\bibfnamefont{V.}~\bibnamefont{{Sahni}}}, \bibnamefont{and}
  \bibinfo{author}{\bibfnamefont{A.~A.} \bibnamefont{{Starobinsky}}},
  \bibinfo{journal}{\prd} \textbf{\bibinfo{volume}{80}}, \bibinfo{eid}{101301}
  (\bibinfo{year}{2009}), \eprint{0903.5141}.

\bibitem[{\citenamefont{{Linder}}(2003)}]{linder2003}
\bibinfo{author}{\bibfnamefont{E.~V.} \bibnamefont{{Linder}}},
  \bibinfo{journal}{Physical Review Letters} \textbf{\bibinfo{volume}{90}},
  \bibinfo{eid}{091301} (\bibinfo{year}{2003}),
  \eprint{arXiv:astro-ph/0208512}.

\bibitem[{\citenamefont{{Albrecht} et~al.}(2009)\citenamefont{{Albrecht},
  {Amendola}, {Bernstein}, {Clowe}, {Eisenstein}, {Guzzo}, {Hirata}, {Huterer},
  {Kirshner}, {Kolb} et~al.}}]{detf-findings-2009}
\bibinfo{author}{\bibfnamefont{A.}~\bibnamefont{{Albrecht}}},
  \bibinfo{author}{\bibfnamefont{L.}~\bibnamefont{{Amendola}}},
  \bibinfo{author}{\bibfnamefont{G.}~\bibnamefont{{Bernstein}}},
  \bibinfo{author}{\bibfnamefont{D.}~\bibnamefont{{Clowe}}},
  \bibinfo{author}{\bibfnamefont{D.}~\bibnamefont{{Eisenstein}}},
  \bibinfo{author}{\bibfnamefont{L.}~\bibnamefont{{Guzzo}}},
  \bibinfo{author}{\bibfnamefont{C.}~\bibnamefont{{Hirata}}},
  \bibinfo{author}{\bibfnamefont{D.}~\bibnamefont{{Huterer}}},
  \bibinfo{author}{\bibfnamefont{R.}~\bibnamefont{{Kirshner}}},
  \bibinfo{author}{\bibfnamefont{E.}~\bibnamefont{{Kolb}}},
  \bibnamefont{et~al.}, \bibinfo{journal}{ArXiv e-prints}
  (\bibinfo{year}{2009}), \eprint{0901.0721}.

\bibitem[{\citenamefont{{Jarosik} et~al.}(2011)\citenamefont{{Jarosik},
  {Bennett}, {Dunkley}, {Gold}, {Greason}, {Halpern}, {Hill}, {Hinshaw},
  {Kogut}, {Komatsu} et~al.}}]{wmap_plus}
\bibinfo{author}{\bibfnamefont{N.}~\bibnamefont{{Jarosik}}},
  \bibinfo{author}{\bibfnamefont{C.~L.} \bibnamefont{{Bennett}}},
  \bibinfo{author}{\bibfnamefont{J.}~\bibnamefont{{Dunkley}}},
  \bibinfo{author}{\bibfnamefont{B.}~\bibnamefont{{Gold}}},
  \bibinfo{author}{\bibfnamefont{M.~R.} \bibnamefont{{Greason}}},
  \bibinfo{author}{\bibfnamefont{M.}~\bibnamefont{{Halpern}}},
  \bibinfo{author}{\bibfnamefont{R.~S.} \bibnamefont{{Hill}}},
  \bibinfo{author}{\bibfnamefont{G.}~\bibnamefont{{Hinshaw}}},
  \bibinfo{author}{\bibfnamefont{A.}~\bibnamefont{{Kogut}}},
  \bibinfo{author}{\bibfnamefont{E.}~\bibnamefont{{Komatsu}}},
  \bibnamefont{et~al.}, \bibinfo{journal}{\apjs}
  \textbf{\bibinfo{volume}{192}}, \bibinfo{pages}{14} (\bibinfo{year}{2011}),
  \eprint{1001.4744}.

\bibitem[{\citenamefont{Haario et~al.}(1999)\citenamefont{Haario, Saksman, and
  Tamminen}}]{haario1999}
\bibinfo{author}{\bibfnamefont{H.}~\bibnamefont{Haario}},
  \bibinfo{author}{\bibfnamefont{E.}~\bibnamefont{Saksman}}, \bibnamefont{and}
  \bibinfo{author}{\bibfnamefont{J.}~\bibnamefont{Tamminen}},
  \bibinfo{journal}{Computational Statistics} \textbf{\bibinfo{volume}{14}},
  \bibinfo{pages}{375} (\bibinfo{year}{1999}),
  \urlprefix\url{http://www.springerlink.com/index/10.1007/s001800050022}.

\bibitem[{\citenamefont{Haario et~al.}(2001)\citenamefont{Haario, Saksman, and
  Tamminen}}]{haario2001}
\bibinfo{author}{\bibfnamefont{H.}~\bibnamefont{Haario}},
  \bibinfo{author}{\bibfnamefont{E.}~\bibnamefont{Saksman}}, \bibnamefont{and}
  \bibinfo{author}{\bibfnamefont{J.}~\bibnamefont{Tamminen}},
  \bibinfo{journal}{Official Journal for the Bernoulli Society of Mathematical
  Statistics and Probability} \textbf{\bibinfo{volume}{7}}, \bibinfo{pages}{pp.
  223} (\bibinfo{year}{2001}), ISSN \bibinfo{issn}{13507265},
  \urlprefix\url{http://www.jstor.org/stable/3318737}.

\bibitem[{\citenamefont{{Dunkley} et~al.}(2005)\citenamefont{{Dunkley},
  {Bucher}, {Ferreira}, {Moodley}, and {Skordis}}}]{dunkley2005}
\bibinfo{author}{\bibfnamefont{J.}~\bibnamefont{{Dunkley}}},
  \bibinfo{author}{\bibfnamefont{M.}~\bibnamefont{{Bucher}}},
  \bibinfo{author}{\bibfnamefont{P.~G.} \bibnamefont{{Ferreira}}},
  \bibinfo{author}{\bibfnamefont{K.}~\bibnamefont{{Moodley}}},
  \bibnamefont{and}
  \bibinfo{author}{\bibfnamefont{C.}~\bibnamefont{{Skordis}}},
  \bibinfo{journal}{\mnras} \textbf{\bibinfo{volume}{356}},
  \bibinfo{pages}{925} (\bibinfo{year}{2005}), \eprint{arXiv:astro-ph/0405462}.

\bibitem[{\citenamefont{Gelman et~al.}(1996)\citenamefont{Gelman, Roberts, and
  Gilks}}]{gelman1995}
\bibinfo{author}{\bibfnamefont{A.}~\bibnamefont{Gelman}},
  \bibinfo{author}{\bibfnamefont{G.}~\bibnamefont{Roberts}}, \bibnamefont{and}
  \bibinfo{author}{\bibfnamefont{W.}~\bibnamefont{Gilks}}, in
  \emph{\bibinfo{booktitle}{Bayesian Statistics}}, edited by
  \bibinfo{editor}{\bibfnamefont{J.~M.} \bibnamefont{Bernado}}
  \bibnamefont{et~al.} (\bibinfo{publisher}{Oxford University Press, Oxford},
  \bibinfo{year}{1996}), vol.~\bibinfo{volume}{5}, p. \bibinfo{pages}{599}.

\bibitem[{bri()}]{bridle2011}
\emph{\bibinfo{title}{{\textsc{CosmoloGUI} package}}},
  \bibinfo{note}{http://www.sarahbridle.net/ cosmologui/}.

\bibitem[{\citenamefont{{Wang}}(2008)}]{wang-parametrise-2008}
\bibinfo{author}{\bibfnamefont{Y.}~\bibnamefont{{Wang}}},
  \bibinfo{journal}{\prd} \textbf{\bibinfo{volume}{77}}, \bibinfo{eid}{123525}
  (\bibinfo{year}{2008}), \eprint{0803.4295}.

\bibitem[{\citenamefont{{Holz} and {Linder}}(2005)}]{holz-linder-2005}
\bibinfo{author}{\bibfnamefont{D.~E.} \bibnamefont{{Holz}}} \bibnamefont{and}
  \bibinfo{author}{\bibfnamefont{E.~V.} \bibnamefont{{Linder}}},
  \bibinfo{journal}{\apj} \textbf{\bibinfo{volume}{631}}, \bibinfo{pages}{678}
  (\bibinfo{year}{2005}), \eprint{arXiv:astro-ph/0412173}.

\bibitem[{\citenamefont{{Kocsis} et~al.}(2006)\citenamefont{{Kocsis}, {Frei},
  {Haiman}, and {Menou}}}]{kocsis2006}
\bibinfo{author}{\bibfnamefont{B.}~\bibnamefont{{Kocsis}}},
  \bibinfo{author}{\bibfnamefont{Z.}~\bibnamefont{{Frei}}},
  \bibinfo{author}{\bibfnamefont{Z.}~\bibnamefont{{Haiman}}}, \bibnamefont{and}
  \bibinfo{author}{\bibfnamefont{K.}~\bibnamefont{{Menou}}},
  \bibinfo{journal}{\apj} \textbf{\bibinfo{volume}{637}}, \bibinfo{pages}{27}
  (\bibinfo{year}{2006}), \eprint{arXiv:astro-ph/0505394}.

\bibitem[{\citenamefont{{Hirata} et~al.}(2010)\citenamefont{{Hirata}, {Holz},
  and {Cutler}}}]{hirata-2010}
\bibinfo{author}{\bibfnamefont{C.~M.} \bibnamefont{{Hirata}}},
  \bibinfo{author}{\bibfnamefont{D.~E.} \bibnamefont{{Holz}}},
  \bibnamefont{and} \bibinfo{author}{\bibfnamefont{C.}~\bibnamefont{{Cutler}}},
  \bibinfo{journal}{\prd} \textbf{\bibinfo{volume}{81}}, \bibinfo{eid}{124046}
  (\bibinfo{year}{2010}), \eprint{1004.3988}.

\bibitem[{\citenamefont{{Hilbert} et~al.}(2011)\citenamefont{{Hilbert}, {Gair},
  and {King}}}]{hilbert-gair-king-2011}
\bibinfo{author}{\bibfnamefont{S.}~\bibnamefont{{Hilbert}}},
  \bibinfo{author}{\bibfnamefont{J.~R.} \bibnamefont{{Gair}}},
  \bibnamefont{and} \bibinfo{author}{\bibfnamefont{L.~J.}
  \bibnamefont{{King}}}, \bibinfo{journal}{\mnras}
  \textbf{\bibinfo{volume}{412}}, \bibinfo{pages}{1023} (\bibinfo{year}{2011}),
  \eprint{1007.2468}.

\bibitem[{\citenamefont{{Nishizawa}
  et~al.}(2012{\natexlab{b}})\citenamefont{{Nishizawa}, {Yagi}, {Taruya}, and
  {Tanaka}}}]{nishizawa2012}
\bibinfo{author}{\bibfnamefont{A.}~\bibnamefont{{Nishizawa}}},
  \bibinfo{author}{\bibfnamefont{K.}~\bibnamefont{{Yagi}}},
  \bibinfo{author}{\bibfnamefont{A.}~\bibnamefont{{Taruya}}}, \bibnamefont{and}
  \bibinfo{author}{\bibfnamefont{T.}~\bibnamefont{{Tanaka}}},
  \bibinfo{journal}{Journal of Physics Conference Series}
  \textbf{\bibinfo{volume}{363}}, \bibinfo{pages}{012052}
  (\bibinfo{year}{2012}{\natexlab{b}}), \eprint{1204.2877}.

\bibitem[{\citenamefont{{Vitale} et~al.}(2012)\citenamefont{{Vitale}, {Del
  Pozzo}, {Li}, {Van Den Broeck}, {Mandel}, {Aylott}, and
  {Veitch}}}]{calibration-errors-2011}
\bibinfo{author}{\bibfnamefont{S.}~\bibnamefont{{Vitale}}},
  \bibinfo{author}{\bibfnamefont{W.}~\bibnamefont{{Del Pozzo}}},
  \bibinfo{author}{\bibfnamefont{T.~G.~F.} \bibnamefont{{Li}}},
  \bibinfo{author}{\bibfnamefont{C.}~\bibnamefont{{Van Den Broeck}}},
  \bibinfo{author}{\bibfnamefont{I.}~\bibnamefont{{Mandel}}},
  \bibinfo{author}{\bibfnamefont{B.}~\bibnamefont{{Aylott}}}, \bibnamefont{and}
  \bibinfo{author}{\bibfnamefont{J.}~\bibnamefont{{Veitch}}},
  \bibinfo{journal}{\prd} \textbf{\bibinfo{volume}{85}}, \bibinfo{eid}{064034}
  (\bibinfo{year}{2012}), \eprint{1111.3044}.

\bibitem[{\citenamefont{{Gair} et~al.}(2010)\citenamefont{{Gair}, {Tang}, and
  {Volonteri}}}]{gair2010}
\bibinfo{author}{\bibfnamefont{J.~R.} \bibnamefont{{Gair}}},
  \bibinfo{author}{\bibfnamefont{C.}~\bibnamefont{{Tang}}}, \bibnamefont{and}
  \bibinfo{author}{\bibfnamefont{M.}~\bibnamefont{{Volonteri}}},
  \bibinfo{journal}{\prd} \textbf{\bibinfo{volume}{81}},
  \bibinfo{pages}{104014} (\bibinfo{year}{2010}), \eprint{1004.1921}.

\bibitem[{\citenamefont{{Wambsganss} et~al.}(1996)\citenamefont{{Wambsganss},
  {Cen}, {Xu}, and {Ostriker}}}]{wambsganss-1996}
\bibinfo{author}{\bibfnamefont{J.}~\bibnamefont{{Wambsganss}}},
  \bibinfo{author}{\bibfnamefont{R.}~\bibnamefont{{Cen}}},
  \bibinfo{author}{\bibfnamefont{G.}~\bibnamefont{{Xu}}}, \bibnamefont{and}
  \bibinfo{author}{\bibfnamefont{J.~P.} \bibnamefont{{Ostriker}}},
  \bibinfo{journal}{ArXiv Astrophysics e-prints}  (\bibinfo{year}{1996}),
  \eprint{arXiv:astro-ph/9607084}.

\bibitem[{\citenamefont{{Duquennoy} and {Mayor}}(1991)}]{duquennoy1991}
\bibinfo{author}{\bibfnamefont{A.}~\bibnamefont{{Duquennoy}}} \bibnamefont{and}
  \bibinfo{author}{\bibfnamefont{M.}~\bibnamefont{{Mayor}}},
  \bibinfo{journal}{\aap} \textbf{\bibinfo{volume}{248}}, \bibinfo{pages}{485}
  (\bibinfo{year}{1991}).

\bibitem[{\citenamefont{{Raghavan} et~al.}(2010)\citenamefont{{Raghavan},
  {McAlister}, {Henry}, {Latham}, {Marcy}, {Mason}, {Gies}, {White}, and {ten
  Brummelaar}}}]{raghavan2010}
\bibinfo{author}{\bibfnamefont{D.}~\bibnamefont{{Raghavan}}},
  \bibinfo{author}{\bibfnamefont{H.~A.} \bibnamefont{{McAlister}}},
  \bibinfo{author}{\bibfnamefont{T.~J.} \bibnamefont{{Henry}}},
  \bibinfo{author}{\bibfnamefont{D.~W.} \bibnamefont{{Latham}}},
  \bibinfo{author}{\bibfnamefont{G.~W.} \bibnamefont{{Marcy}}},
  \bibinfo{author}{\bibfnamefont{B.~D.} \bibnamefont{{Mason}}},
  \bibinfo{author}{\bibfnamefont{D.~R.} \bibnamefont{{Gies}}},
  \bibinfo{author}{\bibfnamefont{R.~J.} \bibnamefont{{White}}},
  \bibnamefont{and} \bibinfo{author}{\bibfnamefont{T.~A.} \bibnamefont{{ten
  Brummelaar}}}, \bibinfo{journal}{\apjs} \textbf{\bibinfo{volume}{190}},
  \bibinfo{pages}{1} (\bibinfo{year}{2010}), \eprint{1007.0414}.

\bibitem[{\citenamefont{{Belczy{\'n}ski} and
  {Kalogera}}(2001)}]{belczynski2000-dce}
\bibinfo{author}{\bibfnamefont{K.}~\bibnamefont{{Belczy{\'n}ski}}}
  \bibnamefont{and}
  \bibinfo{author}{\bibfnamefont{V.}~\bibnamefont{{Kalogera}}},
  \bibinfo{journal}{\apjl} \textbf{\bibinfo{volume}{550}},
  \bibinfo{pages}{L183} (\bibinfo{year}{2001}),
  \eprint{arXiv:astro-ph/0012172}.

\bibitem[{\citenamefont{{Belczynski}
  et~al.}(2002{\natexlab{a}})\citenamefont{{Belczynski}, {Bulik}, and
  {Kalogera}}}]{belczynski2002-mergersites}
\bibinfo{author}{\bibfnamefont{K.}~\bibnamefont{{Belczynski}}},
  \bibinfo{author}{\bibfnamefont{T.}~\bibnamefont{{Bulik}}}, \bibnamefont{and}
  \bibinfo{author}{\bibfnamefont{V.}~\bibnamefont{{Kalogera}}},
  \bibinfo{journal}{\apjl} \textbf{\bibinfo{volume}{571}},
  \bibinfo{pages}{L147} (\bibinfo{year}{2002}{\natexlab{a}}),
  \eprint{arXiv:astro-ph/0204416}.

\bibitem[{\citenamefont{{Ando}}(2004)}]{ando2004}
\bibinfo{author}{\bibfnamefont{S.}~\bibnamefont{{Ando}}},
  \bibinfo{journal}{\jcap} \textbf{\bibinfo{volume}{6}}, \bibinfo{pages}{7}
  (\bibinfo{year}{2004}), \eprint{arXiv:astro-ph/0405411}.

\bibitem[{\citenamefont{{Os{\l}owski} et~al.}(2011)\citenamefont{{Os{\l}owski},
  {Bulik}, {Gondek-Rosi{\'n}ska}, and {Belczy{\'n}ski}}}]{oslowski2011}
\bibinfo{author}{\bibfnamefont{S.}~\bibnamefont{{Os{\l}owski}}},
  \bibinfo{author}{\bibfnamefont{T.}~\bibnamefont{{Bulik}}},
  \bibinfo{author}{\bibfnamefont{D.}~\bibnamefont{{Gondek-Rosi{\'n}ska}}},
  \bibnamefont{and}
  \bibinfo{author}{\bibfnamefont{K.}~\bibnamefont{{Belczy{\'n}ski}}},
  \bibinfo{journal}{\mnras} \textbf{\bibinfo{volume}{413}},
  \bibinfo{pages}{461} (\bibinfo{year}{2011}), \eprint{0903.3538}.

\bibitem[{\citenamefont{{Bhattacharya} and {van den
  Heuvel}}(1991)}]{bhatta-1991}
\bibinfo{author}{\bibfnamefont{D.}~\bibnamefont{{Bhattacharya}}}
  \bibnamefont{and} \bibinfo{author}{\bibfnamefont{E.~P.~J.} \bibnamefont{{van
  den Heuvel}}}, \bibinfo{journal}{\physrep} \textbf{\bibinfo{volume}{203}},
  \bibinfo{pages}{1} (\bibinfo{year}{1991}).

\bibitem[{\citenamefont{{Belczynski}
  et~al.}(2002{\natexlab{b}})\citenamefont{{Belczynski}, {Kalogera}, and
  {Bulik}}}]{belczynski2002-comprehensive}
\bibinfo{author}{\bibfnamefont{K.}~\bibnamefont{{Belczynski}}},
  \bibinfo{author}{\bibfnamefont{V.}~\bibnamefont{{Kalogera}}},
  \bibnamefont{and} \bibinfo{author}{\bibfnamefont{T.}~\bibnamefont{{Bulik}}},
  \bibinfo{journal}{\apj} \textbf{\bibinfo{volume}{572}}, \bibinfo{pages}{407}
  (\bibinfo{year}{2002}{\natexlab{b}}), \eprint{arXiv:astro-ph/0111452}.

\bibitem[{\citenamefont{{Bouwens} et~al.}(2007)\citenamefont{{Bouwens},
  {Illingworth}, {Franx}, and {Ford}}}]{bouwens2007}
\bibinfo{author}{\bibfnamefont{R.~J.} \bibnamefont{{Bouwens}}},
  \bibinfo{author}{\bibfnamefont{G.~D.} \bibnamefont{{Illingworth}}},
  \bibinfo{author}{\bibfnamefont{M.}~\bibnamefont{{Franx}}}, \bibnamefont{and}
  \bibinfo{author}{\bibfnamefont{H.}~\bibnamefont{{Ford}}},
  \bibinfo{journal}{\apj} \textbf{\bibinfo{volume}{670}}, \bibinfo{pages}{928}
  (\bibinfo{year}{2007}), \eprint{0707.2080}.

\bibitem[{\citenamefont{{Virgili} et~al.}(2011)\citenamefont{{Virgili},
  {Zhang}, {O'Brien}, and {Troja}}}]{virgili2011}
\bibinfo{author}{\bibfnamefont{F.~J.} \bibnamefont{{Virgili}}},
  \bibinfo{author}{\bibfnamefont{B.}~\bibnamefont{{Zhang}}},
  \bibinfo{author}{\bibfnamefont{P.}~\bibnamefont{{O'Brien}}},
  \bibnamefont{and} \bibinfo{author}{\bibfnamefont{E.}~\bibnamefont{{Troja}}},
  \bibinfo{journal}{\apj} \textbf{\bibinfo{volume}{727}}, \bibinfo{pages}{109}
  (\bibinfo{year}{2011}), \eprint{0909.1850}.

\end{thebibliography}

\end{document}